\documentclass[11pt,fleqn]{article}
\usepackage{amsfonts}
\usepackage{amsmath}
\usepackage{pifont}
\usepackage{amssymb}
\usepackage{a4wide}
\usepackage{subcaption}
\usepackage[font=small]{caption}
\usepackage{float}
\usepackage[multiple]{footmisc}
\usepackage{threeparttable}
\usepackage{graphicx}
\usepackage{authblk}
\DeclareMathOperator*{\plim}{plim}
\usepackage{lscape}
\usepackage{mathpazo}
\usepackage[hidelinks]{hyperref}
\usepackage{chngcntr}
\usepackage[round]{natbib}
\usepackage{rotating}
\numberwithin{equation}{section}
\usepackage{color}
\usepackage{setspace}
\usepackage{dcolumn}
\usepackage{booktabs}
\usepackage{diagbox}
\usepackage{multirow}

\usepackage[ruled,vlined]{algorithm2e}
\allowdisplaybreaks

\newcommand{\F}{\mathbf{F}}

\DeclareMathOperator*{\argmin}{arg\,min}

\usepackage{enumitem}
\usepackage{siunitx}[=v2] 
\newcolumntype{d}[1]{D{.}{.}{#1}}

\def\*#1{\mathbf{#1}}
\def\+#1{\boldsymbol{#1}}

\newtheorem{lemma}{Lemma}
\newtheorem{corollary}{Corollary}
\usepackage[margin=0.75in,paper=letterpaper]{geometry}
\begin{document}
\setlength{\baselineskip}{0.5cm}
\title{\bf On Selection of Cross-Section Averages in Non-stationary Environments }
{
  \author[1]{Jan Ditzen}
		\author[2]{Ovidijus Stauskas\thanks{Corr: \href{ovidijus.stauskas@bi.no}{ovidijus.stauskas@bi.no. B3-015, BI Norwegian Business School, Department of Economics, Nydalsveien 37}.}}
		\affil[1]{ Free University of Bozen-Bolzano}
  \affil[2]{BI Norwegian Business School, Department of Economics}
		\maketitle
	}
\maketitle
\begin{abstract}
    \noindent Information criteria (ICs) have been widely used in factor models to estimate an unknown number of latent factors. It has recently been shown that ICs perform well in Common Correlated Effects (CCE) and related settings when selecting a set of cross-section averages (CAs) sufficient for the factor space under stationary factors. As CAs can proxy non-stationary factors, it is tempting to claim an excellent performance of ICs under general factors, too. We show formally and in simulations that they remain consistent, but the more persistent factors are, the poorer they perform in small samples, which goes against the sentiment in the CCE/CAs literature.  
\end{abstract}

\noindent MOS subject classification: 91B84

\section{Introduction}
Consider a set of $K$ variables (stacked over time $t=1,\ldots,T$) that admits a factor structure for cross-sectional units $i=1,\ldots,N$:
\begin{align}
    &\*Z_{i}=\*F\*C_i+\*U_i, \label{eq1}
    \end{align}
where $\*Z_{i}=[\*z_{i,1},\ldots,\*z_{i,T}]'\in \mathbb{R}^{T\times K}$, $\*U_i$ is an error term, while $\*F\in\mathbb{R}^{T\times m}$ are latent common factors, where $m$ denotes the number of factors, and  $\*C_i\in \mathbb{R}^{m\times K }$ is the loading matrix. (\ref{eq1}) nests several settings. The most popular one is that of \textit{interactive effects} (see \citealp{bai2009panel}). Then $\*Z_i=[\*y_i, \*X_i]\in \mathbb{R}^{T\times (k+1)}$, where $\*y_i=\*X_i\+\beta+\*F\+\gamma_i+\+\varepsilon_i$ gives a model with unobserved heterogeneity. If $\*X_i=\*F\+\Gamma_i+\*V_i\in \mathbb{R}^{T\times k}$, we have the Common Correlated Effects (CCE) setting of \cite{Pesaran2006} with $\*V_i, \+\varepsilon_i$ ($\+\Gamma_i, \+\gamma_i$) representing idiosyncratics (loadings). (\ref{eq1}) goes beyond interactive effects and CCE. \cite{stauskas2025handling} consider Distinct Correlated Effects (DCE), where $\*y_i$ and $\*X_i$ load on distinct sets of factors.  \cite{massacci2024forecasting} use $\*Z_i=\*X_i$ to model a set of co-moving regressors in a forecasting equation. \\

\noindent It is possible to estimate unobserved factors in several ways, including the Principal Components (PC) method (see \citealp{bai2006confidence}) or diversified projections (see \citealp{fan2022learning}). \cite{Pesaran2006} suggests a simple and elegant way to proxy $\*F$ up to a linear transformation by taking cross-section averages (CAs): $\overline{\*Z}=\frac{1}{N}\sum_{i=1}^N\*Z_i=\widehat{\*F}=\*F\overline{\*C}+o_p(1)$, because CAs of the idiosyncratics are negligible under a wide range of empirically relevant assumptions (see e.g. \citealp{Pesaran2011a}). This is enough to estimate $\*F$ in DCE and forecasting settings. In interactive effects, the CCE estimator of $\+\beta$ is then simply the least squares (LS) estimator augmented with $\overline{\*Z}=\widehat{\*F}$ as an additional regressor. 
\noindent  \\

\noindent A substantial advantage of CAs is their ability to estimate $\*F$ irrespective of their time series properties. Indeed, the CCE estimator enjoys popularity, because it is consistent and asymptotically normal under very general $\*F$ in large $N,T$ settings without any modifications (see e.g. \citealp{Kapetanios2011}, \citealp{westerlund2018cce}, \citealp{westerlund2018asymptotic}, or \citealp{stauskas2022complete}). This property is valid if the rank of $\overline{\*C}$ is equal to $m$, which means that the number of factors cannot exceed the effective number of CAs. However, there are reasons to ensure that the number of CAs matches $m$. In CCE, if $m<k+1$ and $TN^{-1}$ is bounded, CCE produces an asymptotic bias whose analytical correction is infeasible (see \citealp{Karabiyik2017}). This situation is common in macro datasets (see \citealp{de2024cross}). In general, too many CAs might even lead to a model that over-fits the data. In contrast, inclusion of too few generally leads to inconsistent estimates of the model
parameters (see \citealp{Juodis2022CCER}, for a discussion). Beyond CCE, \cite{stauskas2025handling} show that DCE only uses $\overline{\*Z}=\overline{\*X}$, but $m=k$ must hold. In forecasting, the same is needed to avoid only conservative confidence intervals around factor-augmented forecasts (see \citealp{karabiyik2021forecasting}). In response to this, \cite{margaritella2023using} (MW) were the first to demonstrate that the information criteria (ICs) of \cite{Bai2002} and \cite{bai2009_supp} from the PC literature can be applied in the pure CCE setting under \textit{stationary factors} to identify an optimal set of CAs from $\overline{\*Z}$. \cite{de2024cross} (DVS) show similar results when CAs are selected from $\overline{\*X}$, which is more applicable in the DCE and forecasting settings, but the procedure has a similar theoretical basis.\footnote{Even in basic CCE setting, it is advised to omit $\overline{\*y}$ and use $\overline{\*X}$ only to avoid computational issues (see \citealp{karavias2023structural}).}$^,$\footnote{PC techniques are limited in CAs setting, as they focus on $m$ by detecting the largest eigenvalues of the data matrix. They cannot detect sets of CAs as they do not have a natural ordering. With IC, we learn $m$ from the cardinality of the selected set. } \\

\noindent MW stress that an assumption of stationary $\*F$ is required only to simplify the proofs, hinting at a much greater generality of IC. As CAs proxy a general factor structure, it is natural to evaluate this statement and re-visit the ICs proposed in both MW and DVS due to their wide applicability and similar theoretical foundation. The subtlety arises here because IC is minimized by grid-search at different combinations of CAs. Inevitably, there is a need to understand the asymptotic behavior of IC evaluated at a combination inconsistent for $\*F$ that is non-stationary, which is a new undertaking in the CAs literature. Therefore, this study can be seen as the CAs counterpart of \cite{bai2004estimating}, where classical PC results of \cite{Bai2002} are evaluated against pure unit root factors. Instead, we use a mildly integrated process by \cite{magdalinos2009limit} to model $\*F$, which allows us to experiment with varying degrees of factor persistence. We demonstrate formally and in simulations that, while ICs remain consistent as $(N,T)\to \infty$, highly persistent factors negatively affect their small sample performance. We discuss differences between our results and those in \cite{bai2004estimating}, and also explore penalty functions adapted to non-stationary $\*F$, as suggested in the latter study. Since they make the performance of our ICs even worse, we argue that practitioners should not mechanically take recommendations from the PC literature if the goal is to select CAs. More importantly, the ICs of DVS and MW (with the original penalties) should be applied in relatively large samples if the presence of non-stationary factors is suspected.

\section{Econometric Setup}
\indent We focus on the IC of DVS to analyze our theoretical results, however, the conclusions apply to MW, as well. While we provide numerical experiments on both DVS and MW for comparison, theoretical arguments regarding the latter are similar, and we relegate them to Section 3.3 of the Supplementary material. To operationalize our analysis, we introduce $M$ which is a set of column indices of $\overline{\*X}$, and $\*q_{M}\in \mathbb{R}^{k\times g}$ picks the corresponding $g$ averages in practice. That is, $\overline{\*X}\*q_M=\widehat{\F}_M$ defines a selection of $g$ out of $k$ CAs. Consequently, let $M_{0}$ denote the true set of averages from $\overline{\*X}$ such that $\mathrm{rank}(\overline{\+\Gamma} \*q_{M_0})=m$, and $|M_0|=m$, where $|M|$ denotes the cardinality of an arbitrary set $M$. IC under consideration is given by 
\begin{align}
    &\mathrm{IC}(M)= \ln \det \left( \overline{\*Q}_M  \right) +g\cdot k\cdot p_{N,T},\label{DVS}
\end{align}
 where $\ln(.)$ denotes the natural logarithm and $\mathrm{det}(\*A)$ is a determinant of any square matrix $\*A$.  $\overline{\*Q}_M=\frac{1}{NT}\sum_{i=1}^N\*X_i'\*M_{\widehat{\*F}_M}\*X_i$,   $\*M_\*A=\*I-\*A(\*A'\*A)^+\*A'$ is a projection matrix, $\*A^+$ is the Moore-Penrose inverse, and $p_{N,T}$ is a penalty term.\footnote{In case of MW, $\overline{\*Q}_M=  \frac{1}{NT} \sum_{i=1}^{N} \widehat{\+\nu}_i' \mathbf{M}_{\widehat{\*F}_M}  \widehat{\+\nu}_i $ (scalar), where $\widehat{\+\nu}_i=\*y_i-\*X_i\widehat{\+\beta}$ and $\widehat{\+\beta}$ is obtained using the CCE estimator under all available $k+1$ CAs. Then $\mathrm{IC}(M)=\ln(\overline{\*Q}_M)+g\cdot p_{N,T}$.} Examples of feasible and most popular penalties are given by
\begin{align}
    p_{N,T,1}=
        \frac{N+T}{NT}\ln\left(\frac{NT}{N+T}\right),\hspace{2mm}
          p_{N,T,2}=\frac{N+T}{NT}\ln\left(C_{N,T}^2\right) 
    \label{penalties}
\end{align}
for $C_{N,T}=\mathrm{min}(\sqrt{N},\sqrt{T})$ (see more in Section 5 of \citealp{Bai2002}). Note that (\ref{DVS}) is function of (a version of) the denominator of the CCE estimator, which is robust to general unknown factors as long as CAs are (rotationally) consistent for $\*F$ (see Theorem 1 in \citealp{westerlund2018cce}). \\

\noindent Let $\overline{M}$ be the set of indices of all available CAs that, so that $|\overline{M}|=k$. Then 
\begin{align}
\widehat{M}\hspace{1mm}=\argmin_{M\subseteq \overline{M}\hspace{1mm}} \mathrm{IC}(M),
\end{align}
 where $|\widehat{M}|=g$ provides the estimator of the number of factors as a consequence. In DVS (and MW), it is demonstrated that $\mathbb{P}\left(\mathrm{IC}(M)-\mathrm{IC}(M_0)<0\right)\to0$ (therefore, $\mathbb{P}(\widehat{M}=M_0)\to1$) as $(N,T)\to \infty$. This result means that some other $M$ does not minimize IC asymptotically and theoretically justifies its use in the case of CAs. These findings are based on stationary $\*F$, which has been considerably relaxed in the CCE/CAs literature. Therefore, our natural goal is to examine (\ref{DVS}) in a way that departs from this restrictive assumption. Throughout our analysis, we employ the following set of assumptions. \\

\noindent \textbf{Assumption 1.} $\{\*f_t \}$ is a mildly integrated process as defined in \cite{magdalinos2009limit}, such that 
\begin{align*}
   \*f_t = \*R_{f,T}\*f_{t-1} + \*u_{f,t}, \quad \*R_{f,T} = \*I_{m} - \*GT^{-\tau}, \quad \*G=\mathrm{diag}(g_1,\ldots, g_m), \hspace{2mm} g_j\in (0,2),
\end{align*}
where $\*u_{f,t}$ is a zero-mean linear process. 
\bigskip

\noindent \textbf{Assumption 2.} Let $\*e_{i,t}=(\varepsilon_{i,t}, \*v_{i,t}')'\in \mathbb{R}^{k+1}$. Then
\begin{itemize}
  \item[(a)] 
  \begin{itemize}
      \item[(i)] If $\tau \in (0,1)$, then $\{\*e_{i,t}\}$ is a martingale difference sequence, $\mathbb{E}(\*e_{i,t})=\*0_{(k+1)\times 1}$, $\mathbb{E}(\*e_{i,t}\*e_{i,t}')=\+\Sigma_{ee,i,t}$, $\lim_{N,T\to \infty}\frac{1}{NT}\sum_{i=1}^N\sum_{t=1}^T\+\Sigma_{ee,i,t}=\+\Sigma_{ee}=\mathrm{diag}(\sigma^2, \+\Sigma_\*v)$ positive definite and $\mathbb{E}(\|\*e_{i,t} \|^4)<\infty$. 
      \item[(ii)]  If $\tau =0$, then we let $\*e_{i,t}= \mathbf{K}_i(L)\boldsymbol{\epsilon}_{i,t} = \sum_{j=0}^{\infty}\mathbf{K}_{i,j}\boldsymbol{\epsilon}_{i,t-j}$, where $\boldsymbol{\epsilon}_{i,t}$ is independent across $t$ with $\mathbb{E}(\boldsymbol{\epsilon}_{i,t})=\mathbf{0}_{(k+1)\times 1}$, $\mathbb{E}(\boldsymbol{\epsilon}_{i,t}\boldsymbol{\epsilon}_{i,t}') =\boldsymbol{\Sigma}_{\epsilon\epsilon,i,t}$ positive definite, $\mathbb{E}(\|\boldsymbol{\epsilon}_{i,t}\|^4)<\infty$, and $\sum_{j=0}^{\infty}j^{1/2}\| \mathbf{K}_{i,j} \|<\infty$. Also, $\+\xi_t=\mathrm{vec}\left(N\overline{\*e}_t\overline{\*e}_t'-\mathbb{E}\left(N\overline{\*e}_t\overline{\*e}_t' \right) \right)$ is strong mixing with coefficients of size $-bd(b-d)$ with $b>4$ and $b>d>2$, $\mathbb{E}(\|\+\xi_t\|^b)<\infty$, and $\lim_{T\to\infty}T^{-1}\sum_{t=1}^{T}\sum_{s=1}^{T}\mathbb{E}(\+\xi_t\+\xi_s')$ is positive definite. Here and throughout, $\left\|\*A \right\|=\sqrt{\mathrm{trace}(\*A'\*A)}$ is Frobenius norm. 
      \item[(iii)] In both cases, $\varepsilon_{i,t}$ and $\*v_{j,s}$ are independent for all $i,j,t,s$.
  \end{itemize}

  \item[(b)] We have $\left\|\frac{1}{NT}\sum_{i=1}^N\sum_{j=1}^N\sum_{t=1}^{ T}\mathbb{E}\left(\*e_{i,t}\*e_{
j,t}'\right) - \+\Sigma_{ee}\right\|=o(1)$ as  $(N,T)\to \infty$ for a positive definite matrix $\+\Sigma_{ee}$.
\end{itemize}

\bigskip

\noindent \textbf{Assumption 3.} $\mathbf{f}_t$ and $\mathbf{e}_{i,s}$ are independent for all $t$, $s$ and $i$.

\bigskip

\noindent \textbf{Assumption 4.} $\*C_i$ is a deterministic matrix, such that $\|\*C_i\|<\infty$ and $\frac{1}{N}\sum_{i=1}^N\*C_i\*C_i'\to \+\Sigma_\*C$ positive definite. Also, $\overline{\*C}=[\overline{\*C}_m, \overline{\*C}_{-m}]$, where $\overline{\*C}_{-m}\in \mathbb{R}^{m\times (k+1-m)}$ and $\overline{\*C}_m=\overline{\*C}\*q_{M_0}\in\mathbb{R}^{m \times m}$ for a unique $M_0$ is full rank for all $N$, including $N\to \infty$. If $m=k+1$, then $\overline{\*C}=\overline{\*C}_m$.
\bigskip

\noindent Assumption 1 (a) treats the factors very flexibly and offers comparative statics, as $\tau \to 1$ increases persistence of the process. We are not interested in a specific model for $\*F$, but \cite{magdalinos2009limit} allow us to vary $\tau$ and give theoretical guarantees, such as $\frac{1}{T^{1+\tau}}\*F'\*F\to_p\+\Sigma_{\*F}$, which is a constant positive definite matrix. The current specification ensures that we return to the usual stationarity conditions under $\tau=0$, as the eigenvalues of $\*I_m-\*G$ lie within a unit circle, so that the process can be inverted and admit a representation of $MA(\infty)$. Assumption 2 (a) is split into two parts. If the idiosyncratics $\{\*e_{i,t}\}$ are correlated over time, for $\*E_i=[\*e_{i,1},\ldots, \*e_{i,T}]'$, we have that 
 $\left\|\frac{1}{T^{1+\tau}}\*F'\*E_i\right\|=O_p(T^{-\tau/2})$ (see Lemma 3.1 in \citealp{magdalinos2009limit}), which is too slow to demonstrate that the factor estimation error is negligible. In order not to obscure the very effect of factor persistence on IC, we restrict the correlation of idiosyncratics and show that $\left\|\frac{1}{T^{1+\tau}}\*F'\*E_i\right\|=O_p(T^{-(1+\tau)/2})$ as needed (see our auxiliary results in the Supplementary material, and a comparative rate requirement when $\tau=0$ in e.g. \citealp{Karabiyik2017}). We present simulations under serial correlation in the Supplement, which do not show any negative effect on IC. The rest of the assumptions are standard, as they ensure weak cross-section dependence of the idiosyncratics (Assumption 2 (b)), mutual independence of factors and the error terms (Assumption 3; see also \citealp{Pesaran2006}), and informativeness of the loadings (rank condition in Assumption 4). The loadings are deterministic, but they can also admit a random coefficient model (see e.g. \citealp{de2021bias}). In our assumptions, we cover $k+1$ variables in order to accommodate MW, as well. \\

 \noindent Let $\mathrm{d}\overline{\*Q}_{M,M_0}=\ln\mathrm{det}(\overline{\*Q}_M)-\ln\mathrm{det}(\overline{\*Q}_{M_0})=\ln\det\left(\*I_k+T^{\tau}\+\Omega_{M,M_0}\right)$, where $\+\Omega_{M,M_0}:=T^{-\tau}[\overline{\*Q}_M-\overline{\*Q}_{M_0}]\overline{\*Q}_{M_0}^{-1}$. $\mathrm{d}\overline{\*Q}_{M,M_0}$  is important since it helps describing the behavior of $\mathrm{IC}(M)$ in the neighborhood of $\mathrm{IC}(M_0)$ under $M_0\subset M$ (over-specification) and $M\subset M_0$ (under-specification). In addition, it prescribes properties of $p_{N,T}$, which can be seen from
 \begin{align}\label{IC-IC0}
     \mathrm{IC}(M)-\mathrm{IC}(M_0)=\mathrm{d}\overline{\*Q}_{M,M_0}+k\cdot(g-m)p_{N,T}.
 \end{align}
 To illustrate,  \cite{bai2004estimating} (in PC setting)
 shows that under $M_0\subset M$, (an equivalent of) $\mathrm{d}\overline{\*Q}_{M,M_0}$ is $O_p(1)$ when factors are non-stationary, and so the contribution of the extra estimates to the sum of squared residuals is non-negligible. This leads to $p_{N,T}\to \infty$ to heavily penalize redundant estimates so that $\mathrm{IC}(M)-\mathrm{IC}(M_0)>0$. However, when factors are stationary, \cite{Bai2002} (and DVS) detect $\mathrm{d}\overline{\*Q}_{M,M_0}=O_p(C_{N,T}^{-2})$, which means that $p_{N,T}\to 0$ at a slower rate to penalize lightly to ensure $\mathrm{IC}(M)-\mathrm{IC}(M_0)>0$ asymptotically. Proposition 1 below can be seen as the CAs equivalent of Lemma A3 and A4 of \cite{bai2004estimating}.\footnote{ Strictly speaking, under-specification happens when $M\subset M_0$, $M_0 \cap M\neq \emptyset$ but neither is a weak subset of each other, and when $M\cap M_0=\emptyset$. We only analyze the case of $M\subset M_0$ for brevity. As \cite{margaritella2023using} pointed out, analysis would lead to the same conclusion in all cases.}\\

 \noindent \textbf{Proposition 1.} \label{Propos1}\textit{Under Assumptions 1-4 as $(N,T)\to \infty$, we have that: 
 \begin{align}
    & \textit{(a)}\hspace{2mm} \textit{Under}\hspace{1mm} M_0\subset M, \hspace{1mm} \mathrm{d}\overline{\*Q}_{M,M_0}=O_p(C_{N,T}^{-2});\\
    & \textit{(b)}\hspace{2mm} \textit{Under}\hspace{1mm} M\subset M_0,  \hspace{1mm} \mathrm{d}\overline{\*Q}_{M,M_0}=\tau b\ln(T)+\sum_{j:\lambda>0}\ln\left(\lambda_j(\+\Omega_{M.M_0}^0)\right)+R_{N,T},\label{exp_dQ}
 \end{align}
where $\+\Omega_{M.M_0}^0=\plim_{(N,T)\to\infty}\+\Omega_{M.M_0}$ (positive semi-definite, non-zero), $|R_{N,T}|=o_p(1)$ is the remainder, $\lambda_j(\*A)$ is the $j$-th eigenvalue of $\*A$, while $b$ is a number of strictly positive eigenvalues.}
\bigskip 

\noindent \textit{\textbf{Proof: } Section 3 of the Supplementary material.
}
\\

\noindent In (a), the CAs are able to approximate all $m$ factors under Assumption 4. Then $\overline{\*Q}_M$ is exactly (log-determinant of) the denominator of the CCE estimator. It uses the fact that under $M_0\subset M$ we have $\*F=\overline{\*X}(\overline{\+\Gamma}\*q_{M})^++o_p(1)$ for fixed $T$, and by inserting this into $\*X_i$, we can show that $\ln\mathrm{det}(\overline{\*Q}_M)=\ln\mathrm{det}\left(\+\Sigma_\*v\right)+o_p(1)$,
whose argument is a positive definite matrix under general factors (see also \citealp{westerlund2018cce}). Given that $\ln\mathrm{det}(\overline{\*Q}_{M_0})$ admits the same asymptotic representation, it is natural that $\mathrm{d}\overline{\*Q}_{M,M_0}$ is negligible. Moreover, the rate is already detected by DVS, MW and \cite{Bai2002} (for PC) under stationary $\*F$, and it remains identical here. This rate is instructive and determines the properties of penalty $p_{N,T}$, which is the key motivation behind the functional forms in (\ref{penalties}). Specifically, because redundant $g-m$ CAs have an asymptotically negligible contribution to the sum of squared residuals, over-specification does not need to be heavily penalized, and so (\ref{IC-IC0}) becomes
\begin{align}
    \left(\mathrm{IC}(M)-\mathrm{IC}(M_0)\right)/p_{N,T}\to_p k\cdot(g-m)>0
\end{align}
as $(N, T)\to \infty$ if $p_{N,T}\to 0$ and $p_{N,T}C_{N,T}^2\to \infty$. Note that this stands in sharp contrast to findings in Lemma A4 of \cite{bai2004estimating}, where excess factor estimates contribute non-trivially. Overall, the usual consistency of CAs under general unknown factors prevails when $M_0\subset M$ (see also \citealp{westerlund2018cce}, or \citealp{stauskas2022complete}). However, the result is more nuanced under $M\subset M_0$.  \\

\noindent In (b), the rank condition in Assumption 4 is not satisfied. This means that CAs are inconsistent for the $m-g$ factors. Note that $\overline{\*Q}_M$ is a quadratic form in $\*F$, and we know that $\left\|\*F'\*F\right\|=O_p(T^{1+\tau})$ as implied by Assumption 1. However, the objective function offers normalization by $T$ only as the integration order is unknown. Exactly this imbalance causes the divergence of $\mathrm{d}\overline{\*Q}_{M,M_0}$. A similar situation arises in Lemma A3 of \cite{bai2004estimating}, where the central step is to detect the sign of divergence, and we follow this route. In (\ref{exp_dQ}), $b$ is finite ($b< k$), so $\mathrm{d}\overline{\*Q}_{M,M_0}\to_p + \infty$ and therefore
\begin{align}
    \left(\mathrm{IC}(M)-\mathrm{IC}(M_0)\right)/\ln(T)\to_p \tau b>0,
\end{align}
 as $(N,T)\to \infty$, as required. Two comments are in order. Firstly, neither $\sum_{j:\lambda>0}\ln\left(\lambda_j(\+\Omega_{M.M_0}^0)\right)$ nor $R_{N,T}$ are guaranteed to be positive. In addition, in the Supplement, we show that $|R_{N,T}|=O_p(N^{-1})+O_p(N^{-1/2}T^{(\tau-1)/2})$, which means that for $\tau \approx 1$ and a small $N$, it vanishes slowly as $T\to \infty$. This can induce $\mathrm{IC}(M)-\mathrm{IC}(M_0)<0$ in small samples. Hence, it may be necessary to rely on large $N,T$ combinations so that $\tau b\ln(T)$ dominates to detect the minimum of IC outside of $M\subset M_0$ region if $\*F$ is very persistent. We explore and confirm such risks in Monte Carlo simulations in Section 3, where we see that both MW and DVS misselect (in fact, underselect), unless both $N$ and $T$ are large. Secondly, while this result is similar to Lemma A3 in \cite{bai2004estimating}, there the divergence rate is $O_p(T/\ln\ln(T))$, since the law of iterated logarithm is used to arrive at the expression analogous to (\ref{exp_dQ}) when factors have unit root. Theorem 1 leads to our consistency result.\\

\noindent \textbf{Theorem 1}. \textit{Under conditions of Proposition 1 with $p_{N,T}\to 0$ and $p_{N,T}C_{N,T}^2\to \infty$, we have as $(N,T)\to \infty$
\begin{align*}
    \mathbb{P}\left(\widehat{M}=M_0 \right)\to 1.
\end{align*}
}
\bigskip 
\noindent \textit{\textbf{Proof.} Follows from Proposition 1, because $\mathbb{P}\left(\mathrm{IC}(M)-\mathrm{IC}(M_0)<0\right)=\mathbb{P}\left(\left(\mathrm{IC}(M)-\mathrm{IC}(M_0)\right)/p_{N,T}<0\right)\to0$ under (a) and a similar statement holds under (b).}\\

\noindent The key message of Theorem 1 is that the IC of DVS (and MW) is consistent, but exhibits hybrid asymptotic properties characteristic to stationary and non-stationary factors. Consequently, two forces are in effect: $p_{N,T}$ should still be negligible, but the factor estimation error can be ``large'' if $M\subset M_0$ as indicated by the remainder in (\ref{exp_dQ}). To balance these properties in practice, it may be tempting to experiment further with the penalty. Indeed, the change of penalty is the key message of Theorem 1 in \cite{bai2004estimating} in the PC setting, where two conditions are satisfied: 1) $p_{N,T}\to \infty$, and 2) $p_{N,T}/\ln(T)\to0$ which is given in footnote 3 of the study to adapt the penalty to the logarithmic sum of squared residuals, which is exactly our setting. Importantly, this condition does not interfere with consistency in our Theorem 1, because clearly $p_{N,T}=o(\ln(T))$ holds. Therefore, we compare both types of penalties in the simulations to illustrate the outcome if practitioners simply extrapolate recommendations from the PC literature. We let $\widetilde{p}_{N,T}=\ln(T)p_{N,T}$, which follows the suggestion in (12) in \cite{bai2004estimating}, and obeys $\widetilde{p}_{N,T}\to \infty$ and $\widetilde{p}_{N,T}/\ln(T)\to0$. 
\\

\noindent \textbf{Remark 1.} \textit{Note that under stationarity, $\mathrm{d}\overline{\*Q}_{M,M_0}\to_pc>0$ when $M\subset M_0$, according to the results in DVS (and similarly in MW). Our Proposition 1 naturally accommodates this result under $\tau=0$, because, by following the same approximation steps, (\ref{exp_dQ}) becomes 
\begin{align*}
     \mathrm{d}\overline{\*Q}_{M,M_0}=\sum_{j:\lambda>0}\ln\left(1+\lambda_j(\+\Omega_{M.M_0}^0)\right)+R_{N,T}\to_pc>0
\end{align*}
as $(N,T)\to \infty$ because each summand is greater than 1 w.p. 1. Here, $\+\Omega_{M.M_0}^0=\plim_{(N,T)\to \infty}[\overline{\*Q}_M-\overline{\*Q}_{M_0}]\overline{\*Q}_{M_0}^{-1}$ (positive semi-definite, non-zero). Also, $|R_{N,T}|=O_p(N^{-1})+O_p((NT)^{-1/2})$, which is the same as in DVS (and MW) under $\tau=0$, as expected from the discussion of (\ref{exp_dQ}).}\\

\noindent \textbf{Remark 2.} \textit{Throughout, we assume that $M_0$ is unique. However, because CAs do not have a natural ordering, the problem is only set-identified. It can be shown that both ICs select the set that minimizes the asymptotic mean squared error out of all the sets that satisfy Assumption 4 (see e.g. Corollary 3.1 in \citealp{margaritella2023using}). In this study, the first order issue is ability of IC to select any $\widehat{M}$ with the cardinality of $m$. }\\

\noindent \textbf{Remark 3.} \textit{\cite{Juodis2022CCER} considers eigenvalue-based method of \cite{ahn2013eigenvalue} under stationarity, where the focus is learning $m$, but not the set of CAs. We leave analysis of this approach for the future research, but provide comparative simulation evidence under mildly integrated $\*F$ in the Supplementary material. }

\section{Simulation Study}
The data generating process of the simulation follows MW:
\begin{align}
    \mathbf{y}_i &= \mathbf{X}_i \boldsymbol{\beta}_i + \mathbf{F}\boldsymbol{\gamma}_i + \boldsymbol{\varepsilon}_i, \quad
    \mathbf{X}_i = \mathbf{F}\boldsymbol{\Gamma}_i + \mathbf{V}_i,
\end{align}
where \(\mathbf{X}_i = \left[\mathbf{x}_{i,1},...,\mathbf{x}_{i,T}\right]'\) is a $T\times k$ matrix of observable regressors, \(\boldsymbol{\beta} = \iota' 0.5\) a $k\times 1$ vector of unobserved parameters. The factor loadings are generated as $\boldsymbol{\Gamma}_i = [\*I_{m}, \*0_{m \times (k-m)}]\psi_i$, that is, a matrix $m \times k$, with $\psi_i\sim N(1,1)$ and the elements of the vector $1\times m $ $\boldsymbol{\gamma}_i$ are drawn from $N(1,1)$. In the case of errors independent over time and cross-sectionally, $\mathbf{e}_{i,t}$ is drawn from $N(\*0_{k+1},\*I_{k+1})$. In the case of errors correlated weakly across time and/or cross-sections we follow \cite{Bai2002,margaritella2023using}:
\begin{align}
&\*e_{i,t}=(\epsilon_{i,t}, \*v_{i,t}')'=\*P_\rho\*e_{i,t-1}+\*z_{\*w,i,t},\hspace{2mm} \*P_\rho=\begin{bmatrix}
    \rho & \*0_{1\times k}\\
    \*0_{k\times 1} & \rho_v\*I_k
\end{bmatrix}, \hspace{2mm} \*z_{\*w,i,t}=\begin{bmatrix}
\sqrt{\frac{m \left(1-\rho^2\right)}{1+2J\kappa^2}} \+\epsilon_t'\left( \+\iota_{N,i} + \kappa \*w_i'\right)\\
\sqrt{\frac{ \left(1-\rho_v^2\right)}{1+2J_v\kappa_v^2}} \+\Upsilon_t'\left(  \+\iota_{N,i} + \kappa_v \boldsymbol{\tilde{\*w}}_i'\right)
\end{bmatrix}
\label{eq:MC:weaklyerrors}
\end{align}
where \( \boldsymbol{\epsilon}_t\) is a $N\times 1$ stack of $\epsilon_{i,t} \sim N(0,1)$ over $i$, and $\*w_i$ is the $i$-th row ($1\times N$) of $\*W$, a weight matrix with the $J$-th off diagonal elements equal to zero. $\rho$ and $\rho_v$ control the correlation over time in $\boldsymbol{\varepsilon}_t$ and $\mathbf{V}_i$, respectively. For this simulation exercise, we set them to zero. In the Supplement we explore settings with $\rho=\rho_v=0.5$, an extension to our theory which prescribes uncorrelated innovations. Across all specifications, we allow for weak correlation across units with $\kappa = \kappa_v = 0.2$, $J=J_v=5$ and $\*w_i=\tilde{\*w}_i$. $\mathbf{\Upsilon}_t$ is a $N \times k$ matrix, which stacks $\+\upsilon_{i,t}\sim N(\*0_k,\*I_k)$ over $i$, and it is independent of $\+\epsilon_t$. The vector $\+\iota_{N,i}\in \mathbb{R}^{N\times 1}$ contains 1 in the $i$-th coordinate, and zeros elsewhere.
The factors are drawn as $\mathbf{f}_t = \*R_{f,T}  \mathbf{f}_{t-1} + \*u_{f,t}$, where
\begin{align}
    & \*R_{f,T} = \*I_{m} - \*GT^{-\tau}, 
    \quad \*G = \mathrm{diag}(g_1,...,g_{m}), g_j \sim U(0,2) \label{MC:eq:C}\quad\*u_{f,t} \sim N\left(\*0_{m},\left(\*I_{m}-\*R_{f,T}^2\right)^{1/2}\right) .
\end{align}
In the stationary factors case, the factors are drawn as a stable VAR with $\tau=0$. Two cases of non-stationary factors are considered: the non-stationary case of $\tau = 0.4$, and the case close to local-to-unity process $\tau = 0.9$. All simulations have 1000 repetitions, and we compare DVS and MW. In the Supplement, we provide additional simulation evidence of the eigenvalue growth ratio (ER) based on \cite{Juodis2022CCER} for comparison. The focus of this method is the number of factors ($m$), and it does not select the sufficient set of CAs. Nevertheless, its performance crumbles with $\tau\to 1$ and does not improve as $(N,T)\to \infty$. \\

\noindent Table \ref{tab:Table1} below presents the frequency of the correctly selected number of CAs ($g = 4$) with default penalties $p_{N,T,1}$ and $p_{N,T,2}$ from (\ref{penalties}), and the penalty from \cite{bai2004estimating}, $\widetilde{p}_{N,T}$, for the case of stationary factors ($\tau = 0$), the mildly non-stationary case ($\tau = 0.4$) and the close to local-to-unity process. The idiosyncratic terms $\*v_{i,t}$ and $\varepsilon_{i,t}$ are uncorrelated over time, but weakly correlated over units. In what follows, the low selection frequency reported is dominated by underslection.
We start with the case established in the literature, which is stationary factors and default penalties. For small $N$ and $T$ ($T=N=50$) and stationary factors, we observe a frequency between $16\%$ and $36\%$, a pattern which resembles the results in MW.\footnote{In comparison to \cite{margaritella2023using} we set the number of factors to 4 rather than 2 and hence underestimate the number of CAs more severely in the case of $T=20$ and $N>20$. If we reduce the number of factors to 2, we identify a similar number of factors as in the case of 4 factors. } As expected, consistency, illustrated here by a selection frequency of 100\%, is quickly achieved when $N$ and $T$ increase. We now turn to the case of non-stationary factors. For small $N$ and $T$, the selection frequency decreases to zero, which is driven by underselection of the number of CAs, as we show in Section 4 in the Supplement. To gain consistency, much larger combinations of $(N,T)$ are required. For the extreme case of a process close to local-to-unity, combinations of $(N,T)>300$ are integral. The behavior of both criteria with respect to increases in $N$, $T$, and $\tau$ is almost identical.  \cite{bai2004estimating} proposes to adjust the penalty by $\ln(T)$ in the presence of non-stationary factors. The lower panel of Table \ref{tab:Table1} presents the frequency of the correctly selected number of CAs with a penalty of $\widetilde{p}_{N,T} = \ln(T)\frac{N+T}{NT}\ln(\frac{NT}{N+T})$ for $\mathrm{IC}_1$ and $\widetilde{p}_{N,T} = \ln(T)\frac{N+T}{NT}\ln(C^2_{N,T})$ for $\mathrm{IC}_2$. We observe a selection frequency of 0 for all four ICs and for almost all combinations of $N$ and $T$ and $\tau$. Only in the large sample cases of $N=500$ and $T>300$ and only under stationary factors the selection frequency reaches 100\%. This strongly suggests that prescriptions from the PC literature do not directly apply to CAs, as ICs perform even worse, and selecting more CAs should not be heavily penalized.    \\

\noindent In the Supplement, we also investigate the share of misselected number of cross-sectional averages when increasing $\tau$ using $IC^{MW}_1$ and $IC^{DVS}_1$ for a small ($N=T=100$) and large sample ($N=T=500$). For the small sample, the share remains relatively flat until $\tau$ reaches levels of around $0.5$. As expected, this is much less of a problem for the large sample and we observe under-selection of CA only for a high degree of non-stationarity.

\begin{table}[H]
    \centering
    \resizebox{0.9\columnwidth}{!}{%
    \begin{tabular}{cc|cccc|cccc|cccc}
    \hline\hline
    & & \multicolumn{4}{c}{Stationary $\*F$} & \multicolumn{8}{c}{Non-Stationary $\*F$} \\
    & & \multicolumn{4}{c}{$\tau = 0$} & \multicolumn{4}{c}{$\tau = 0.4$} & \multicolumn{4}{c}{$\tau = 0.9$} \\
    $N$ & $T$ & $IC^{MW}_1$ & $IC^{MW}_2$ & $IC^{DVS}_1$ & $IC^{DVS}_2$ &$IC^{MW}_1$ & $IC^{MW}_2$ & $IC^{DVS}_1$ & $IC^{DVS}_2$&$IC^{MW}_1$ & $IC^{MW}_2$ & $IC^{DVS}_1$ & $IC^{DVS}_2$  \\ \hline %
    \multicolumn{14}{c}{Penalty $p$} \\ \hline
    50 & 50 & 31.00 & 15.90 & 36.40 & 16.10 & 7.30 & 3.60 & 3.50 & 0.30 & 0.10 & 0.00 & 0.00 & 0.00 \\
100 & 50 & 97.30 & 95.00 & 99.20 & 96.40 & 91.50 & 82.80 & 95.10 & 85.90 & 11.50 & 8.60 & 13.40 & 7.50 \\
200 & 50 & 95.50 & 94.30 & 99.70 & 98.80 & 48.60 & 45.50 & 77.70 & 69.20 & 0.20 & 0.20 & 5.50 & 4.10 \\
300 & 50 & 100.00 & 100.00 & 99.60 & 99.50 & 81.30 & 78.20 & 69.90 & 67.70 & 4.30 & 3.70 & 1.80 & 1.60 \\
500 & 50 & 100.00 & 100.00 & 100.00 & 100.00 & 99.60 & 99.40 & 96.30 & 95.10 & 21.20 & 20.10 & 11.50 & 10.10 \\ \hline
50 & 100 & 97.10 & 89.30 & 100.00 & 100.00 & 68.60 & 55.30 & 99.30 & 94.50 & 7.00 & 5.90 & 18.20 & 10.90 \\
100 & 100 & 100.00 & 99.90 & 99.90 & 99.00 & 62.50 & 49.40 & 56.30 & 39.70 & 2.70 & 1.30 & 1.70 & 0.60 \\
200 & 100 & 100.00 & 100.00 & 100.00 & 100.00 & 99.90 & 99.60 & 99.90 & 99.70 & 24.50 & 19.70 & 26.30 & 18.80 \\
300 & 100 & 100.00 & 100.00 & 100.00 & 100.00 & 100.00 & 100.00 & 100.00 & 100.00 & 39.10 & 34.30 & 46.60 & 39.30 \\
500 & 100 & 100.00 & 100.00 & 100.00 & 100.00 & 99.70 & 99.70 & 95.90 & 94.80 & 16.30 & 15.30 & 8.50 & 7.60 \\ \hline
50 & 200 & 99.70 & 99.90 & 99.90 & 99.80 & 97.70 & 96.70 & 89.00 & 79.90 & 9.80 & 8.70 & 5.10 & 3.50 \\
100 & 200 & 100.00 & 100.00 & 100.00 & 100.00 & 93.00 & 88.80 & 99.50 & 98.60 & 5.30 & 4.70 & 12.80 & 7.70 \\
200 & 200 & 100.00 & 100.00 & 100.00 & 100.00 & 100.00 & 100.00 & 100.00 & 100.00 & 40.80 & 32.70 & 60.20 & 44.50 \\
300 & 200 & 100.00 & 100.00 & 100.00 & 100.00 & 100.00 & 100.00 & 100.00 & 100.00 & 22.40 & 18.90 & 81.40 & 73.20 \\
500 & 200 & 100.00 & 100.00 & 100.00 & 100.00 & 100.00 & 100.00 & 100.00 & 100.00 & 54.10 & 49.30 & 80.80 & 75.50 \\ \hline
50 & 300 & 100.00 & 100.00 & 100.00 & 100.00 & 88.60 & 86.40 & 99.90 & 99.40 & 7.40 & 6.20 & 15.20 & 11.00 \\
100 & 300 & 98.90 & 99.80 & 100.00 & 100.00 & 98.70 & 99.40 & 100.00 & 100.00 & 21.30 & 18.60 & 27.20 & 22.40 \\
200 & 300 & 100.00 & 100.00 & 100.00 & 100.00 & 100.00 & 100.00 & 100.00 & 100.00 & 56.00 & 49.10 & 52.60 & 45.60 \\
300 & 300 & 100.00 & 100.00 & 100.00 & 100.00 & 100.00 & 100.00 & 100.00 & 100.00 & 89.30 & 82.90 & 94.40 & 89.10 \\
500 & 300 & 100.00 & 100.00 & 100.00 & 100.00 & 100.00 & 100.00 & 100.00 & 100.00 & 80.10 & 75.70 & 93.70 & 91.90 \\ \hline
50 & 500 & 100.00 & 100.00 & 100.00 & 100.00 & 90.10 & 89.00 & 100.00 & 100.00 & 3.40 & 3.10 & 11.20 & 10.20 \\
100 & 500 & 100.00 & 100.00 & 100.00 & 100.00 & 100.00 & 100.00 & 100.00 & 100.00 & 38.50 & 37.10 & 53.70 & 50.80 \\
200 & 500 & 100.00 & 100.00 & 100.00 & 100.00 & 100.00 & 100.00 & 100.00 & 100.00 & 84.10 & 81.30 & 97.30 & 95.80 \\
300 & 500 & 100.00 & 100.00 & 100.00 & 100.00 & 100.00 & 100.00 & 100.00 & 100.00 & 88.30 & 85.90 & 97.00 & 95.60 \\
500 & 500 & 100.00 & 100.00 & 100.00 & 100.00 & 100.00 & 100.00 & 100.00 & 100.00 & 92.70 & 88.20 & 99.50 & 98.70 \\ \hline \multicolumn{14}{c}{Penalty by \cite{bai2004estimating}, $\widetilde{p} = \ln(T)p $} \\ \hline
50 & 50 & 0.00 & 0.00 & 0.00 & 0.00 & 0.00 & 0.00 & 0.00 & 0.00 & 0.00 & 0.00 & 0.00 & 0.00 \\
100 & 50 & 0.00 & 0.00 & 0.00 & 0.00 & 0.00 & 0.00 & 0.00 & 0.00 & 0.00 & 0.00 & 0.00 & 0.00 \\
200 & 50 & 0.00 & 0.00 & 0.00 & 0.00 & 0.00 & 0.00 & 0.00 & 0.00 & 0.00 & 0.00 & 0.00 & 0.00 \\
300 & 50 & 0.00 & 0.00 & 0.00 & 0.00 & 0.00 & 0.00 & 0.00 & 0.00 & 0.00 & 0.00 & 0.00 & 0.00 \\
500 & 50 & 0.00 & 0.00 & 0.00 & 0.00 & 0.10 & 0.10 & 0.00 & 0.00 & 0.00 & 0.00 & 0.00 & 0.00 \\ \hline
50 & 100 & 0.00 & 0.00 & 0.00 & 0.00 & 0.00 & 0.00 & 0.00 & 0.00 & 0.00 & 0.00 & 0.00 & 0.00 \\
100 & 100 & 0.00 & 0.00 & 0.00 & 0.00 & 0.00 & 0.00 & 0.00 & 0.00 & 0.00 & 0.00 & 0.00 & 0.00 \\
200 & 100 & 0.00 & 0.00 & 0.00 & 0.00 & 0.00 & 0.00 & 0.00 & 0.00 & 0.00 & 0.00 & 0.00 & 0.00 \\
300 & 100 & 0.00 & 0.00 & 0.00 & 0.00 & 0.10 & 0.10 & 0.00 & 0.00 & 0.00 & 0.00 & 0.00 & 0.00 \\
500 & 100 & 0.10 & 0.00 & 0.00 & 0.00 & 1.20 & 0.40 & 0.00 & 0.00 & 0.10 & 0.10 & 0.00 & 0.00 \\ \hline
50 & 200 & 0.00 & 0.00 & 0.00 & 0.00 & 0.00 & 0.00 & 0.00 & 0.00 & 0.00 & 0.00 & 0.00 & 0.00 \\
100 & 200 & 0.00 & 0.00 & 0.00 & 0.00 & 0.00 & 0.00 & 0.00 & 0.00 & 0.00 & 0.00 & 0.00 & 0.00 \\
200 & 200 & 0.00 & 0.00 & 0.00 & 0.00 & 0.40 & 0.00 & 0.00 & 0.00 & 0.00 & 0.00 & 0.00 & 0.00 \\
300 & 200 & 2.40 & 0.00 & 0.00 & 0.00 & 8.80 & 2.70 & 0.00 & 0.00 & 0.10 & 0.00 & 0.00 & 0.00 \\
500 & 200 & 26.10 & 4.70 & 0.00 & 0.00 & 12.90 & 7.20 & 0.00 & 0.00 & 0.40 & 0.10 & 0.00 & 0.00 \\ \hline
50 & 300 & 0.00 & 0.00 & 0.00 & 0.00 & 0.00 & 0.00 & 0.00 & 0.00 & 0.00 & 0.00 & 0.00 & 0.00 \\
100 & 300 & 0.00 & 0.00 & 0.00 & 0.00 & 0.00 & 0.00 & 0.00 & 0.00 & 0.00 & 0.00 & 0.00 & 0.00 \\
200 & 300 & 0.00 & 0.00 & 0.00 & 0.00 & 0.00 & 0.00 & 0.00 & 0.00 & 0.10 & 0.10 & 0.00 & 0.00 \\
300 & 300 & 19.30 & 0.00 & 0.00 & 0.00 & 16.00 & 1.90 & 0.00 & 0.00 & 0.90 & 0.70 & 0.00 & 0.00 \\
500 & 300 & 99.90 & 92.90 & 0.00 & 0.00 & 72.70 & 53.60 & 1.20 & 0.00 & 2.70 & 1.30 & 0.00 & 0.00 \\ \hline
50 & 500 & 0.00 & 0.00 & 0.00 & 0.00 & 0.00 & 0.00 & 0.00 & 0.00 & 0.00 & 0.00 & 0.00 & 0.00 \\
100 & 500 & 0.00 & 0.00 & 0.00 & 0.00 & 0.00 & 0.00 & 0.00 & 0.00 & 0.00 & 0.00 & 0.00 & 0.00 \\
200 & 500 & 0.00 & 0.00 & 0.00 & 0.00 & 0.30 & 0.00 & 0.00 & 0.00 & 0.80 & 0.50 & 0.00 & 0.00 \\
300 & 500 & 15.80 & 0.10 & 0.00 & 0.00 & 14.90 & 3.10 & 0.00 & 0.00 & 0.90 & 0.70 & 0.00 & 0.00 \\
500 & 500 & 100.00 & 100.00 & 77.40 & 0.20 & 94.50 & 81.20 & 31.50 & 2.00 & 2.50 & 1.70 & 0.30 & 0.00 \\
\hline\hline
    \end{tabular}}
    \caption{Correct Selection Frequency for $g$ with $m = 4$ and $K \in \{8,9\}$. DVS criteria from \cite{de2024cross}, MW from \cite{margaritella2023using}, see \eqref{DVS}. In the upper part, for $IC^{DVS}_1$ and $IC^{MW}_1$, $p_{N,T} = \frac{N+T}{NT}\ln(\frac{NT}{N+T})$; for  $IC^{DVS}_2$ and $IC^{MW}_2$, $p_{N,T} = \frac{N+T}{NT}\ln(C^2_{N,T})$ with $C_{N,T} = \min(\sqrt{N},\sqrt{T})$. In the lower part the penalty from \cite{bai2004estimating} is used and  defined as $\widetilde{p}_{N,T}=\ln(T)p_{N,T} $. Idiosyncratics in $\*x_{i,t}$, $\*v_{i,t}$, and $\varepsilon_{i,t}$  uncorrelated over time, but weakly correlated across units, see \eqref{eq:MC:weaklyerrors}. }
    \label{tab:Table1}
\end{table}

\section{Conclusion}
\noindent This study is the first to examine the selection of an optimal set of CAs in CCE and related settings by ICs inspired by \cite{Bai2002} when latent factors are non-stationary. In particular, we use mild integration to explore varying degrees of factor persistence and demonstrate that ICs remain consistent without any modifications. However, the more persistent common factors are, the worse their small sample performance becomes, and ICs regain selection consistency only in very large samples (i.e. $(N,T)>300$, according to our experiments). Importantly, a divergent penalty suggested by \cite{bai2004estimating} to address non-stationary factors in PC makes our ICs perform even worse in the case of CAs. Therefore, our recommendation for CCE/CAs practitioners is not to automatically take PC literature prescriptions and interpret IC results with caution in the presence of highly persistent data, unless $N,T$ are substantial.

\clearpage
\section{Supplement}
\renewcommand{\theequation}{A.\arabic{equation}}

\begin{abstract}
    In this Supplementary Material, we provide the proofs of our auxiliary results and our main result in Proposition 1 (Sections 2 and 3). In addition, we demonstrate how the proofs change in the case of the IC of \cite{margaritella2023using}. In Section 4, we provide additional simulation evidence, including, e.g., correlated idiosyncratics or eigenvalue ratio method of \cite{Juodis2022CCER}. 
\end{abstract}

\subsection{Assumptions}
Throughout our analysis, we employ the following set of assumptions. \\

\noindent \textbf{Assumption 1.} $\{\*f_t \}$ is a mildly integrated process as defined in \cite{magdalinos2009limit}, such that 
\begin{align*}
   \*f_t = \*R_{f,T}\*f_{t-1} + \*u_{f,t}, \quad \*R_{f,T} = \*I_{m} - \*GT^{-\tau}, \quad \*G=\mathrm{diag}(g_1,\ldots, g_m), \hspace{2mm} g_j\in (0,2),
\end{align*}
where $\*u_{f,t}$ is a zero-mean linear process. 
\bigskip

\noindent \textbf{Assumption 2.} Let $\*e_{i,t}=(\varepsilon_{i,t}, \*v_{i,t}')'\in \mathbb{R}^{k+1}$. Then
\begin{itemize}
  \item[(a)] 
  \begin{itemize}
      \item[(i)] If $\tau \in (0,1)$, then $\{\*e_{i,t}\}$ is a martingale difference sequence with $\mathbb{E}(\*e_{i,t})=\*0_{(k+1)\times 1}$, $\mathbb{E}(\*e_{i,t}\*e_{i,t}')=\+\Sigma_{ee,i,t}$ with $\lim_{N,T\to \infty}\frac{1}{NT}\sum_{i=1}^N\sum_{t=1}^T\+\Sigma_{ee,i,t}=\+\Sigma_{ee}=\mathrm{diag}(\sigma^2, \+\Sigma_\*v)$ positive definite and $\mathbb{E}(\|\*e_{i,t} \|^4)<\infty$. 
      \item[(ii)]  If $\tau =0$, then we let $\*e_{i,t}= \mathbf{K}_i(L)\boldsymbol{\epsilon}_{i,t} = \sum_{j=0}^{\infty}\mathbf{K}_{i,j}\boldsymbol{\epsilon}_{i,t-j}$, where $\boldsymbol{\epsilon}_{i,t}$ is independent across $t$ with $\mathbb{E}(\boldsymbol{\epsilon}_{i,t})=\mathbf{0}_{(k+1)\times 1}$, $\mathbb{E}(\boldsymbol{\epsilon}_{i,t}\boldsymbol{\epsilon}_{i,t}') =\boldsymbol{\Sigma}_{\epsilon\epsilon,i,t}$ positive definite, $\mathbb{E}(\|\boldsymbol{\epsilon}_{i,t}\|^4)<\infty$, and $\sum_{j=0}^{\infty}j^{1/2}\| \mathbf{K}_{i,j} \|<\infty$. Also, $\+\xi_t=\mathrm{vec}\left(N\overline{\*e}_t\overline{\*e}_t'-\mathbb{E}\left(N\overline{\*e}_t\overline{\*e}_t' \right) \right)$ is strong mixing with coefficients of size $-bd(b-d)$ with $b>4$ and $b>d>2$, $\mathbb{E}(\|\+\xi_t\|^b)<\infty$, and $\lim_{T\to\infty}T^{-1}\sum_{t=1}^{T}\sum_{s=1}^{T}\mathbb{E}(\+\xi_t\+\xi_s')$ is positive definite. Here and throughout, $\left\|\*A \right\|=\sqrt{\mathrm{trace}(\*A'\*A)}$ is Frobenius norm. 
      \item[(iii)] In both cases, $\varepsilon_{i,t}$ and $\*v_{j,s}$ are independent for all $i,j,t,s$.
  \end{itemize}

  \item[(b)] We have $\left\|\frac{1}{NT}\sum_{i=1}^N\sum_{j=1}^N\sum_{t=1}^{ T}\mathbb{E}\left(\*e_{i,t}\*e_{
j,t}'\right) - \+\Sigma_{ee}\right\|=o(1)$ as  $(N,T)\to \infty$ for a positive definite matrix $\+\Sigma_{ee}$.
\end{itemize}

\bigskip

\noindent \textbf{Assumption 3.} $\mathbf{f}_t$ and $\mathbf{e}_{i,s}$ are independent for all $t$, $s$ and $i$.

\bigskip

\noindent \textbf{Assumption 4.} $\*C_i$ is a deterministic matrix, such that $\|\*C_i\|<\infty$ and $\frac{1}{N}\sum_{i=1}^N\*C_i\*C_i'\to \+\Sigma_\*C$ positive definite. Also, $\overline{\*C}=[\overline{\*C}_m, \overline{\*C}_{-m}]$, where $\overline{\*C}_{-m}\in \mathbb{R}^{m\times (K-m)}$ and $\overline{\*C}_m=\overline{\*C}\*q_{M_0}\in\mathbb{R}^{m \times m}$ for a unique $M_0$ is full rank for all $N$, including $N\to \infty$. If $m=K$, then $\overline{\*C}=\overline{\*C}_m$.
\subsection{Auxiliary Results }
\noindent \begin{lemma}\label{LemmaA1} \textit{Under Assumptions 1 - 4, we have that 
\begin{align*}
    \left\|\frac{1}{T^{1+\tau}}\*F'\overline{\*E} \right\|=O_p(N^{-1/2}T^{-\tau/2})
\end{align*}
 as $(N,T)\to \infty$}.\\
\end{lemma}
\noindent \textbf{Proof.} To begin with, we can write $\frac{1}{T^{1+\tau}}\*F'\overline{\*E}=\frac{1}{T^{1+\tau}}\sum_{t=1}^T\*f_t\overline{\*e}_t'$. Then for some positive $\epsilon$,
\begin{align}
     \mathbb{P}\left(\left\|\frac{1}{T^{1+\tau}}\sum_{t=1}^T\*f_t\overline{\*e}_t' \right\|>\epsilon \right)&\leq \epsilon^{-1}\mathbb{E}\left(\left\| \frac{1}{T^{1+\tau}}\sum_{t=1}^T\*f_t\overline{\*e}_t'\right\| \right)\notag\\
     &=\frac{1}{\epsilon \sqrt{N}}\mathbb{E}\left(\left\| \frac{1}{T^{1+\tau}}\sum_{t=1}^T\*f_t(\sqrt{N}\overline{\*e}_t')\right\| \right)\notag\\
     &\leq \frac{1}{\epsilon \sqrt{N}}\mathbb{E}\left( \frac{1}{T^{1+\tau}}\sum_{t=1}^T\left\|\*f_t\right\|\left\|(\sqrt{N}\overline{\*e}_t)\right\| \right)\notag\\
     &= \frac{1}{\epsilon \sqrt{N}} \frac{1}{T^{1+\tau}}\sum_{t=1}^T\mathbb{E}(\left\|\*f_t\right\|)\mathbb{E}\left(\left\|(\sqrt{N}\overline{\*e}_t)\right\|\right) \notag\\
     &\leq \frac{1}{\epsilon \sqrt{N}T^{\tau/2}} \left(\frac{1}{T^{1+\tau}}\sum_{t=1}^T\mathbb{E}(\left\|\*f_t\right\|)^2\right)^{1/2}\underbrace{\left(\frac{1}{T}\sum_{t=1}^T\mathbb{E}\left(\left\|(\sqrt{N}\overline{\*e}_t)\right\|\right)^2\right)^{1/2}}_{O(1)}\notag\\
     &=O(N^{-1/2}T^{-\tau/2}),
\end{align}
because 
\begin{align}
    \frac{1}{T^{1+\tau}}\sum_{t=1}^T\mathbb{E}(\left\|\*f_t\right\|)^2&\leq \frac{1}{T^{1+\tau}}\sum_{t=1}^T\mathbb{E}\left[\left(\sqrt{(\mathrm{tr}(\*f_t\*f_t'))} \right)^2\right]\notag\\
    &=\frac{1}{T^{1+\tau}}\sum_{t=1}^T\mathbb{E}(\mathrm{tr}[\*f_t\*f_t'])=O(1)
\end{align}
by the concavity of function appearing in Jensen's inequality. Note that this result gives a different rate than for similar terms in, for example, \cite{pitarakis2023direct}, which is brought down exactly by time dependence in $\{\overline{\*e}_t\}$.  Also note that independence between the factors and idiosyncratics is not strictly necessary, and we use it for simplicity. Indeed, 
\begin{align}
     \mathbb{P}\left(\left\|\frac{1}{T^{1+\tau}}\sum_{t=1}^T\*f_t\overline{\*e}_t' \right\|>\epsilon \right)&\leq \frac{1}{\epsilon \sqrt{N}}\mathbb{E}\left( \frac{1}{T^{1+\tau}}\sum_{t=1}^T\left\|\*f_t\right\|\left\|(\sqrt{N}\overline{\*e}_t)\right\| \right)\notag\\
     &\leq \frac{1}{\epsilon \sqrt{N}} \frac{1}{T^{1+\tau}}\sum_{t=1}^T\mathbb{E}\left( \left\|\*f_t \right\|^2\right)^{1/2}\mathbb{E}\left(\left\| \sqrt{N}\overline{\*e}_t\right\|^2\right)^{1/2}\notag\\
     &\leq \frac{\sup_t\mathbb{E}\left(\left\| \sqrt{N} \overline{\*e}_t\right\|^2\right)^{1/2}}{\epsilon \sqrt{N}T^{\tau/2}}\sup_t \mathbb{E}\left( \left\|T^{-\tau/2}\*f_t \right\|^2\right)^{1/2}\notag\\
     &=O(N^{-1/2}T^{-\tau/2}),
\end{align}
because $\sup_t \mathbb{E}\left( \left\|T^{-\tau/2}\*f_t \right\|^2\right)^{1/2}=O(1)$ by Lemma 3.1 in \cite{magdalinos2009limit}. 
\\

\noindent \begin{corollary}\label{CorollaryA1} \textit{Under Assumptions 1 - 4, but $\*v_{i,t}$ and $\varepsilon_{i,t}$ are uncorrelated over time, we have that 
\begin{align*}
    \left\|\frac{1}{T^{1+\tau}}\*F'\overline{\*E} \right\|=O_p(N^{-1/2}T^{-(1+\tau)/2})
\end{align*}
 as $(N,T)\to \infty$}.\\
\end{corollary}
\noindent \textbf{Proof.} By assumption, we then have that $\overline{\*e}_t$ is uncorrelated over time. Then by Markov's inequality we obtain 
\begin{align}
    \mathbb{P}\left(\left\|\frac{1}{T^{1+\tau}}\sum_{t=1}^T\*f_t\overline{\*e}_t' \right\|>\epsilon \right)&\leq \epsilon^{-2}\mathbb{E}\left(\left\|\frac{1}{T^{1+\tau}}\sum_{t=1}^T\*f_t\overline{\*e}_t' \right\|^2 \right)\notag\\
    &=\epsilon^{-2} \frac{1}{T^{1+\tau}}\mathbb{E}\left(\mathrm{tr}\left[\frac{1}{T^{1+\tau}}\sum_{t=1}^T\sum_{s=1}^T\*f_t\overline{\*e}_t'\overline{\*e}_s\*f_s' \right] \right)\notag\\
    &= \epsilon^{-2} \frac{1}{T^{1+\tau}} \left(\mathrm{tr}\left[\frac{1}{T^{1+\tau}}\sum_{t=1}^T\sum_{s=1}^T\mathbb{E}\left(\*f_s'\*f_t \overline{\*e}_t'\overline{\*e}_s\right) \right] \right)\notag\\
    &= \epsilon^{-2} \frac{1}{T^{1+\tau}} \left(\mathrm{tr}\left[\frac{1}{T^{1+\tau}}\sum_{t=1}^T\sum_{s=1}^T\mathbb{E}\left(\*f_s'\*f_t \right)\mathbb{E}\left(\overline{\*e}_t'\overline{\*e}_s\right) \right] \right)\notag\\
    &= \epsilon^{-2} \frac{1}{T^{1+\tau}} \left(\mathrm{tr}\left[\frac{1}{T^{1+\tau}}\sum_{t=1}^T\sum_{s=1}^T\mathbb{E}\left(\*f_t'\*f_t \right)\mathbb{E}\left(\mathbb{E}\left(\overline{\*e}_t'\overline{\*e}_s|\mathcal{F}_{(t-1)\lor (s-1)}\right)\right)\right]\right)\notag\\
    &= \epsilon^{-2} \frac{1}{T^{1+\tau}} \left(\mathrm{tr}\left[\frac{1}{T^{1+\tau}}\sum_{t=1}^T\mathbb{E}\left(\*f_t'\*f_t \right)\mathbb{E}\left(\overline{\*e}_t'\overline{\*e}_t\right) \right] \right)\notag\\
    &=\epsilon^{-2} \frac{1}{NT^{1+\tau}} \left(\mathrm{tr}\left[\frac{1}{T^{1+\tau}}\sum_{t=1}^T\mathbb{E}\left(\*f_t'\*f_t \right)\mathbb{E}\left(N\overline{\*e}_t'\overline{\*e}_t\right) \right] \right)\notag\\
    &\leq \epsilon^{-2} \frac{\sup_{t}\mathbb{E}\left(N\overline{\*e}_t'\overline{\*e}_t\right)}{NT^{1+\tau}} \left(\mathrm{tr}\left[\frac{1}{T^{1+\tau}}\sum_{t=1}^T\mathbb{E}\left(\*f_t'\*f_t \right) \right] \right)\notag\\
    &\leq O(N^{-1}T^{-1-\tau})\times \underbrace{\mathrm{tr}\left[ \frac{1}{T^{1+\tau}}\sum_{t=1}^T\mathbb{E}\left(\*f_t\*f_t' \right)\right]}_{O(1)}\notag\\
    &=  O(N^{-1}T^{-1-\tau}),
\end{align}
which implies that $\left\|\frac{1}{T^{1+\tau}}\*F'\overline{\*E} \right\|=O_p(N^{-1/2}T^{-(1+\tau)/2})$ if $\overline{\*e}_t$ is uncorrelated over time. Note that 
\begin{align}
    \mathbb{E}\left(N\overline{\*e}_t'\overline{\*e}_t\right)=\mathbb{E}\left(\frac{1}{N}\sum_{i=1}^N\sum_{j=1}^N\*e_{i,t}'\*e_{j,t} \right) =\frac{1}{N}\sum_{i=1}^N\sum_{j=1}^N\mathrm{tr}\left[\mathbb{E}\left(\*e_{i,t}\*e_{j,t}' \right) \right]=O(1),
\end{align}
uniformly in $t$ by our assumptions. \\

\noindent \noindent \begin{lemma}\label{LemmaA2} \textit{Under Assumptions 1 - 4, we have that 
\begin{align*}
    \left\|\frac{1}{T^{1+\tau}}\*F'\*E_i \right\|=O_p(T^{-\tau/2})
\end{align*}
 as $T\to \infty$}.\\
\end{lemma}
 \noindent \textbf{Proof}. Similarly to Lemma \ref{LemmaA1}, we have that by Markov's and Cauchy-Schwarz inequalities
 \begin{align}
     \mathbb{P}\left(\left\|\frac{1}{T^{1+\tau}}\sum_{t=1}^T\*f_t\*e_{i,t}' \right\|>\epsilon \right)&\leq \epsilon^{-1}\mathbb{E}\left(\left\| \frac{1}{T^{1+\tau}}\sum_{t=1}^T\*f_t\*e_{i,t}'\right\| \right)\notag\\
     &=\frac{1}{\epsilon }\mathbb{E}\left(\left\| \frac{1}{T^{1+\tau}}\sum_{t=1}^T\*f_t\*e_{i,t}'\right\| \right)\notag\\
     &\leq \frac{1}{\epsilon }\mathbb{E}\left( \frac{1}{T^{1+\tau}}\sum_{t=1}^T\left\|\*f_t\right\|\left\|\*e_{i,t}\right\| \right)\notag\\
     &= \frac{1}{\epsilon } \frac{1}{T^{1+\tau}}\sum_{t=1}^T\mathbb{E}(\left\|\*f_t\right\|)\mathbb{E}\left(\left\|\*e_{i,t}\right\|\right) \notag\\
     &\leq \frac{1}{\epsilon T^{\tau/2}} \left(\frac{1}{T^{1+\tau}}\sum_{t=1}^T\mathbb{E}(\left\|\*f_t\right\|)^2\right)^{1/2}\underbrace{\left(\frac{1}{T}\sum_{t=1}^T\mathbb{E}\left(\left\|\*e_{i,t}\right\|\right)^2\right)^{1/2}}_{O(1)}\notag\\
     &=O(T^{-\tau/2}),
\end{align}
as expected. \\

\noindent \begin{corollary}\label{CorollaryA2} \textit{Under Assumptions 1 - 4, but $\*v_{i,t}$ and $\varepsilon_{i,t}$ are uncorrelated over time, we have that 
\begin{align*}
    \left\|\frac{1}{T^{1+\tau}}\*F'\*E_i \right\|=O_p(T^{-(1+\tau)/2})
\end{align*}
 as $T\to \infty$}.\\
\end{corollary}
\noindent \textbf{Proof.} We use an approach similar to the one in Corollary \ref{CorollaryA1}. Then by Markov's inequality we obtain 
\begin{align}
    \mathbb{P}\left(\left\|\frac{1}{T^{1+\tau}}\sum_{t=1}^T\*f_t\*e_{i,t}' \right\|>\epsilon \right)&\leq \epsilon^{-2}\mathbb{E}\left(\left\|\frac{1}{T^{1+\tau}}\sum_{t=1}^T\*f_t\*e_{i,t}' \right\|^2 \right)\notag\\
    &=\epsilon^{-2} \frac{1}{T^{1+\tau}}\mathbb{E}\left(\mathrm{tr}\left[\frac{1}{T^{1+\tau}}\sum_{t=1}^T\sum_{s=1}^T\*f_t\*e_{i,t}'\*e_{i,s}\*f_s' \right] \right)\notag\\
    &= \epsilon^{-2} \frac{1}{T^{1+\tau}} \left(\mathrm{tr}\left[\frac{1}{T^{1+\tau}}\sum_{t=1}^T\sum_{s=1}^T\mathbb{E}\left(\*f_s'\*f_t \*e_{i,t}'\*e_{i,s}\right) \right] \right)\notag\\
    &= \epsilon^{-2} \frac{1}{T^{1+\tau}} \left(\mathrm{tr}\left[\frac{1}{T^{1+\tau}}\sum_{t=1}^T\sum_{s=1}^T\mathbb{E}\left(\*f_s'\*f_t \right)\mathbb{E}\left(\*e_{i,t}'\*e_{i,s}\right) \right] \right)\notag\\
    &=\epsilon^{-2} \frac{1}{T^{1+\tau}} \left(\mathrm{tr}\left[\frac{1}{T^{1+\tau}}\sum_{t=1}^T\sum_{s=1}^T\mathbb{E}\left(\*f_t'\*f_t \right)\mathbb{E}\left(\mathbb{E}\left(\*e_{i,t}'\*e_{i,s}|\mathcal{F}_{(t-1)\lor (s-1)}\right)\right)\right]\right)\notag\\
    &= \epsilon^{-2} \frac{1}{T^{1+\tau}} \left(\mathrm{tr}\left[\frac{1}{T^{1+\tau}}\sum_{t=1}^T\mathbb{E}\left(\*f_t'\*f_t \right)\mathbb{E}\left(\*e_{i,t}'\*e_{i,t}\right) \right] \right)\notag\\
    &=\epsilon^{-2} \frac{1}{T^{1+\tau}} \left(\mathrm{tr}\left[\frac{1}{T^{1+\tau}}\sum_{t=1}^T\mathbb{E}\left(\*f_t'\*f_t \right)\mathbb{E}\left(\*e_{i,t}'\*e_{i,t}\right) \right] \right)\notag\\
    &\leq \epsilon^{-2} \frac{\sup_{t}\mathbb{E}\left(\*e_{i,t}'\*e_{i,t}\right)}{T^{1+\tau}} \left(\mathrm{tr}\left[\frac{1}{T^{1+\tau}}\sum_{t=1}^T\mathbb{E}\left(\*f_t'\*f_t \right) \right] \right)\notag\\
    &\leq O(T^{-1-\tau})\times \underbrace{\mathrm{tr}\left[ \frac{1}{T^{1+\tau}}\sum_{t=1}^T\mathbb{E}\left(\*f_t\*f_t' \right)\right]}_{O(1)}\notag\\
    &=  O(T^{-1-\tau}),
\end{align}
which implies that $\left\|\frac{1}{T^{1+\tau}}\*F'\*E_i \right\|=O_p(T^{-(1+\tau)/2})$.
\subsection{Proposition 1 and the Proof}
\textbf{Proposition 1.} \label{Propos1}\textit{Under Assumptions 1-4 as $(N,T)\to \infty$, we have that: 
 \begin{align}
    & \textit{(a)}\hspace{2mm} \textit{Under}\hspace{1mm} M_0\subset M, \hspace{1mm} \mathrm{d}\overline{\*Q}_{M,M_0}=O_p(C_{N,T}^{-2});\\
    & \textit{(b)}\hspace{2mm} \textit{Under}\hspace{1mm} M\subset M_0,  \hspace{1mm} \mathrm{d}\overline{\*Q}_{M,M_0}=\tau b\ln(T)+\sum_{j:\lambda>0}\ln\left(\lambda_j(\+\Omega_{M.M_0}^0)\right)+R_{N,T},\label{exp_dQ}
 \end{align}
where $\+\Omega_{M.M_0}^0=\plim_{(N,T)\to\infty}\+\Omega_{M.M_0}$ (positive semi-definite, non-zero), $|R_{N,T}|=o_p(1)$ is the remainder, $\lambda_j(\*A)$ is the $j$-th eigenvalue of $\*A$, while $b$ is a number of strictly positive eigenvalues.}
\\

\noindent \textbf{Proof}. In the proof, we focus on the IC of \cite{de2024cross} ($j=DVS$). Therefore, we will have $K=k$ throughout. The argument for the IC of \cite{margaritella2023using} ($j=MW$) is very similar, and we only need to take into account that we are working with the usual CCE setup, where $\widehat{\+\beta}=\+\beta+o_p(1)$ under the full set of $k+1$ CAs. We will comment on this at the end of the proof. For notational simplicity, we will use $\*q_M\in \mathbb{R}^{k\times g}$ without the subscript $\*x$. Also, because IC of DVS covers Examples 2 and 3 (see the main text), we denote factors as $\*F_\*x$. \\

\noindent For brevity, let $\overline{\*Q}_M=\frac{1}{NT}\sum_{i=1}^N\*X_i'\*M_{\widehat{\*F}_{M_{\*x}}}\*X_i$, where $\widehat{\*F}_{M_{\*x}}=\overline{\*X}\*q_M=(\*F_\*x\overline{\+\Gamma}+\overline{\*V})\*q_M$ for the selector matrix $\*q_M\in \mathbb{R}^{k\times g}$. Then, for $M_0$ being the true subset of CAs such that $g=m$, let us characterize difference of two Information Criteria: 
\begin{align}\label{det_exp_IC}
	\mathrm{IC}^{DVS}(M)-\mathrm{IC}^{DVS}(M_0)&=\ln \mathrm{det}\left( \overline{\*Q}_{M}\right)-\ln \mathrm{det}\left( \overline{\*Q}_{M_0}\right)+k(g-m)p_{N,T}\notag\\
	&=\ln \left[\frac{\mathrm{det}( \overline{\*Q}_{M})}{\mathrm{det}( \overline{\*Q}_{M_0})}  \right]+k(g-m)p_{N,T}\notag\\
	&=\ln \left[ \det(\overline{\*Q}_{M})\mathrm{det}(\overline{\*Q}_{M_0}^{-1})\right]+k(g-m)p_{N,T}\notag\\
	&=\ln\mathrm{det}\left( \overline{\*Q}_{M}\overline{\*Q}_{M_0}^{-1}\right)+k(g-m)p_{N,T}\notag\\
	&=\ln\mathrm{det}\left(\*I_k+\overline{\*Q}_{M}\overline{\*Q}_{M_0}^{-1}-\*I_k \right)+k(g-m)p_{N,T}\notag\\
	&=\ln\mathrm{det}\left(\*I_k+\overline{\*Q}_{M}\overline{\*Q}_{M_0}^{-1}-\overline{\*Q}_{M_0}\overline{\*Q}_{M_0}^{-1} \right)+k(g-m)p_{N,T}\notag\\
	&=\ln\mathrm{det}\left(\*I_k+\left[\overline{\*Q}_{M}-\overline{\*Q}_{M_0}\right]\overline{\*Q}_{M_0}^{-1}\right)+k(g-m)p_{N,T}\notag\\
    &=\ln\mathrm{det}\left(\*I_k+T^{\tau}\left[T^{-\tau}\left(\overline{\*Q}_{M}-\overline{\*Q}_{M_0}\right)\right]\overline{\*Q}_{M_0}^{-1}\right)+k(g-m)p_{N,T},
\end{align}
where we used $\ln(a)-\ln(b)=\ln\left(\frac{a}{b}\right)$ and $\mathrm{det}(\*A\*B)=\mathrm{det}(\*A)\mathrm{det}(\*B)$. This implies that we will examine the asymptotic behavior of $T^{-\tau}\left(\overline{\*Q}_{M}-\overline{\*Q}_{M_0}\right)$. Its behavior is determined by whether $M_0\subset M$ (over-specification) or $M \subset M_0$ (under-specification). In the latter case, under-specification happens when $M\subset M_0$, $M_0 \cap M\neq \emptyset$ but neither is a weak subset of each other, and when $M\cap M_0=\emptyset$. We only analyze the case of $M\subset M_0$ as in \cite{margaritella2023using} or \cite{de2024cross}, because other cases will lead to the same asymptotic conclusions. 
\subsubsection{Case of $ M\subset M_0$ }
By following \cite{de2024cross}, we have
\begin{align}\label{I-II-III}	T^{-\tau}\left(\overline{\*Q}_{M}-\overline{\*Q}_{M_0}\right)&=T^{-\tau}\left[\overline{\*Q}_{M}-\widehat{\*Q}_{M,\*M_{\*F_{\*x}\overline{\+\Gamma}\*q_{M}}}\right] - T^{-\tau}\left[ \overline{\*Q}_{M_0}-\widehat{\*Q}_{M_0,\*M_{\*F_{\*x}\overline{\+\Gamma}\*q_{M_0}}}\right] 
	\notag\\
    &+T^{-\tau}\left[ \widehat{\*Q}_{M,\*M_{\*F_{\*x}\overline{\+\Gamma}\*q_M}}
    -\widehat{\*Q}_{M_0,\*M_{\*F_{\*x}\overline{\+\Gamma}\*q_{M_0}}}\right]\notag\\
    &=\*I-\mathbf{II}+\mathbf{III}. 
\end{align}
where, for instance, $\widehat{\*Q}_{M_0,\*M_{\*F_{\*x}\overline{\+\Gamma}\*q_{M}}}$ means that it is evaluated at $\*F_\*x\overline{\+\Gamma}\*q_{M}$. We begin with 
\begin{align}
    \*I &=\frac{1}{NT^{1+\tau}}\sum_{i=1}^N\*X_i'(\*M_{\widehat{\*F}_{M_{\*x}}}-\*M_{\*F_{\*x}\overline{\+\Gamma}\*q_M})\*X_i\notag\\
	&=\frac{1}{NT^{1+\tau}}\sum_{i=1}^N\*V_i'(\*M_{\widehat{\*F}_{M_{\*x}}}-\*M_{\*F_{\*x}\overline{\+\Gamma}\*q_M})\*V_i+\frac{1}{NT^{1+\tau}}\sum_{i=1}^N\+\Gamma_i'\*F_\*x'(\*M_{\widehat{\*F}_{M_{\*x}}}-\*M_{\*F_{\*x}\overline{\+\Gamma}\*q_M})\*F_\*x\+\Gamma_i\notag\\
	&+\frac{1}{NT^{1+\tau}}\sum_{i=1}^N\*V_i'(\*M_{\widehat{\*F}_{M_{\*x}}}-\*M_{\*F_{\*x}\overline{\+\Gamma}\*q_M})\*F_\*x\+\Gamma_i + \frac{1}{NT^{1+\tau}}\sum_{i=1}^N\+\Gamma_i'\*F_\*x'(\*M_{\widehat{\*F}_{M_{\*x}}}-\*M_{\*F_{\*x}\overline{\+\Gamma}\*q_M})\*V_i\notag\\
    &=\*I_a+\*I_b+\*I_c+\*I_d.
\end{align}
where the expansion $\*M_{\widehat{\*F}_{M_{\*x}}}-\*M_{\*F_{\*x}\overline{\+\Gamma}\*q_M}$ is the key.  Note how 
\begin{align}
   T^{-(1+\tau)}\widehat{\*F}_{M_{\*x}}'\widehat{\*F}_{M_{\*x}}&= \*q_M'\overline{\+\Gamma}'T^{-(1+\tau)}\*F_\*x'\*F_\*x\overline{\+\Gamma}\*q_M + \*q_M'\overline{\+\Gamma}'T^{-(1+\tau)}\*F_\*x'\overline{\*V}\*q_M+\*q_M'T^{-(1+\tau)}\overline{\*V}'\*F_\*x\overline{\+\Gamma}\*q_M\notag\\
    &+\*q_M'TT^{-(1+\tau)}T^{-1}\overline{\*V}'\overline{\*V}\*q_M\notag\\
    &= \*q_M'\overline{\+\Gamma}'T^{-(1+\tau)}\*F_\*x'\*F_\*x\overline{\+\Gamma}\*q_M +O_p(N^{-1}T^{-\tau})+O_p(N^{-1/2}T^{-(1+\tau)/2})\notag\\
    &=\*q_M'\overline{\+\Gamma}'T^{-(1+\tau)}\*F_\*x'\*F_\*x\overline{\+\Gamma}\*q_M + O_p(\xi^{-1}_{N,T,\tau})
\end{align}
by Corollary \ref{CorollaryA2}, where we defined $\xi_{N,T,\tau}^{-1}=N^{-1}T^{-\tau}+N^{-1/2}T^{-(1+\tau)/2}$. The remainder has a slightly different rate than in \cite{de2024cross} (see e.g. 2.184 in their Online Supplement), because the idiosyncratic component is still stationary, and so
\begin{align}
    \left\|T^{-(1+\tau)}\overline{\*V}'\overline{\*V} \right\|= N^{-1}T^{-\tau}\left\|NT^{-1}\overline{\*V}'\overline{\*V} \right\|=O_p(N^{-1}T^{-\tau}).
\end{align}
Moreover, note that 
\begin{align}
    N\times \xi_{N,T,\tau}^{-1}&=N\times  (O_p(N^{-1}T^{-\tau})+O_p(N^{-1/2}T^{-(1+\tau)/2}))\notag\\
    &=O_p(T^{-\tau})+O_p(\sqrt{N}T^{-1/2}T^{-\tau/2})=O_p(1)
\end{align}
 without further restrictions (but negligible under $TN^{-1}=O(1)$. Also, let 
 \begin{align}
     T^\tau\times \xi_{N,T,\tau}^{-1}&= T^\tau \times (O_p(N^{-1}T^{-\tau})+O_p(N^{-1/2}T^{-(1+\tau)/2}))\notag\\
      &=O_p(N^{-1})+O_p(N^{-1/2}T^{(\tau-1)/2})\notag\\
      &=o_p(1))=R_{N,T}
 \end{align}
  without further restrictions as $\tau \in (0,1)$. Note that the latter will be the dominant component in the further analysis. Both of these results will be important when determining the orders of many terms. \\
  
  \noindent Next, in case of $M\subset M_0$, we have that $\mathrm{rank}\left(  \frac{1}{T^{1+\tau}}\widehat{\*F}_{M_{\*x}}'\widehat{\*F}_{M_{\*x}}\right)=m$ even asymptotically, and hence 
\begin{align}
     \left(T^{-(1+\tau)}\widehat{\*F}_{M_{\*x}}'\widehat{\*F}_{M_{\*x}}\right)^+=\left(\*q_M'\overline{\+\Gamma}'T^{-(1+\tau)}\*F_\*x'\*F_\*x\overline{\+\Gamma}\*q_M\right)^+ + O_p(\xi_{N,T,\tau}^{-1}),
\end{align}
because 
\begin{align}
    &\left\| \left(T^{-(1+\tau)}\widehat{\*F}_{M_{\*x}}'\widehat{\*F}_{M_{\*x}}\right)^+- \left(\*q_M'\overline{\+\Gamma}'T^{-(1+\tau)}\*F_\*x'\*F_\*x\overline{\+\Gamma}\*q_M\right)^+\right\|\notag\\
    &=\left\|  \left(T^{-(1+\tau)}\widehat{\*F}_{M_{\*x}}'\widehat{\*F}_{M_{\*x}}\right)^+\left(\*q_M'\overline{\+\Gamma}'T^{-(1+\tau)}\*F_\*x'\*F_\*x\overline{\+\Gamma}\*q_M-T^{-(1+\tau)}\widehat{\*F}_{M_{\*x}}'\widehat{\*F}_{M_{\*x}}\right) \left(\*q_M'\overline{\+\Gamma}'T^{-(1+\tau)}\*F_\*x'\*F_\*x\overline{\+\Gamma}\*q_M\right)^+\right\|\notag\\
    &\leq \left\| \left(T^{-(1+\tau)}\widehat{\*F}_{M_{\*x}}'\widehat{\*F}_{M_{\*x}}\right)^+\right\| \left\| T^{-(1+\tau)}\widehat{\*F}_{M_{\*x}}'\widehat{\*F}_{M_{\*x}}-\*q_M'\overline{\+\Gamma}'T^{-(1+\tau)}\*F_\*x'\*F_\*x\overline{\+\Gamma}\*q_M\right\|\left\|\left(\*q_M'\overline{\+\Gamma}'T^{-(1+\tau)}\*F_\*x'\*F_\*x\overline{\+\Gamma}\*q_M\right)^+ \right\|\notag\\
    &=O_p(\xi_{N,T,\tau}^{-1}).
\end{align}
This leads to the expansion
\begin{align}\label{IC_expansion}
	\*M_{\*F_{\*x}\overline{\+\Gamma}\*q_M}-\*M_{\widehat{\*F}_{M_{\*x}}}&=\overline{\*V}\*q_M(T^{-(1+\tau)}\widehat{\*F}_{M_{\*x}}'\widehat{\*F}_{M_{\*x}})^+ T^{-(1+\tau)}\*q_M'\overline{\*V}' + \overline{\*V}\*q_M(T^{-(1+\tau)}\widehat{\*F}_{M_{\*x}}'\widehat{\*F}_{M_{\*x}})^+ T^{-(1+\tau)}\*q_M'\overline{\+\Gamma}'\*F_\*x'\notag\\
	&+\*F_\*x\overline{\+\Gamma}\*q_M(T^{-(1+\tau)}\widehat{\*F}_{M_{\*x}}'\widehat{\*F}_{M_{\*x}})^+ T^{-(1+\tau)}\*q_M'\overline{\*V}'\notag\\
	&+\*F_\*x\overline{\+\Gamma}\*q_M \left[(T^{-(1+\tau)}\widehat{\*F}_{M_{\*x}}'\widehat{\*F}_{M_{\*x}})^+ - (T^{-(1+\tau)}\*q_M'\overline{\+\Gamma}'\*F_\*x'\*F_\*x\overline{\+\Gamma}\*q_M)^+  \right]T^{-(1+\tau)}\*q_M'\overline{\+\Gamma}'\*F_\*x',
\end{align}
which now takes into account the mildly integrated factors. Then, we can continue with 
\begin{align}
    \left\|\*I_a \right\|&=\left\| \frac{1}{NT^{1+\tau}}\sum_{i=1}^N\*V_i'(\*M_{\widehat{\*F}_{M_{\*x}}}-\*M_{\*F_{\*x}\overline{\+\Gamma}\*q_M})\*V_i\right\|\notag\\
    &\leq \left\| \frac{1}{NT^{1+\tau}}\sum_{i=1}^N\*V_i'\overline{\*V}\*q_M(T^{-(1+\tau)}\widehat{\*F}_{M_{\*x}}'\widehat{\*F}_{M_{\*x}})^+ T^{-(1+\tau)}\*q_M'\overline{\*V}'\*V_i\right\|\notag\\
    &+\left\|\frac{1}{NT^{1+\tau}}\sum_{i=1}^N\*V_i'\overline{\*V}\*q_M(T^{-(1+\tau)}\widehat{\*F}_{M_{\*x}}'\widehat{\*F}_{M_{\*x}})^+ T^{-(1+\tau)}\*q_M'\overline{\+\Gamma}'\*F_\*x'\*V_i\right\|\notag\\
    &+\left\|\frac{1}{NT^{1+\tau}}\sum_{i=1}^N\*V_i'\*F_\*x\overline{\+\Gamma}\*q_M(T^{-(1+\tau)}\widehat{\*F}_{M_{\*x}}'\widehat{\*F}_{M_{\*x}})^+ T^{-(1+\tau)}\*q_M'\overline{\*V}'\*V_i \right\| \notag\\
    &+\left\| \frac{1}{NT^{1+\tau}}\sum_{i=1}^N\*V_i'\*F_\*x\overline{\+\Gamma}\*q_M \left[(T^{-(1+\tau)}\widehat{\*F}_{M_{\*x}}'\widehat{\*F}_{M_{\*x}})^+ - (T^{-(1+\tau)}\*q_M'\overline{\+\Gamma}'\*F_\*x'\*F_\*x\overline{\+\Gamma}\*q_M)^+  \right]T^{-(1+\tau)}\*q_M'\overline{\+\Gamma}'\*F_\*x'\*V_i \right\| \notag\\
    &\leq \underbrace{T^{-2\tau}\left\| (T^{-(1+\tau)}\widehat{\*F}_{M_{\*x}}'\widehat{\*F}_{M_{\*x}})^+\right\|\frac{1}{N}\sum_{i=1}^N \left\|T^{-1}\overline{\*V}'\*V_i \right\|^2\left\|\*q_M \right\|^2}_{O_p(N^{-1}T^{-2\tau})+O_p(N^{-1/2}T^{-(4\tau+1)/2})}\notag\\
    &+ 2\underbrace{T^{-\tau}\left\| (T^{-(1+\tau)}\widehat{\*F}_{M_{\*x}}'\widehat{\*F}_{M_{\*x}})^+\right\| \frac{1}{N}\sum_{i=1}^N\left\|\*q_M \right\|^2\left\|T^{-1}\overline{\*V}'\*V_i \right\|\left\|\overline{\+\Gamma} \right\|\left\|T^{-(1+\tau)}\*F_\*x'\*V_i \right\|}_{O_p(N^{-1}T^{-(1+2\tau)/2})+O_p(N^{-1/2}T^{-(2+3\tau)/2})}\notag\\
    &+ \underbrace{T^{-1-\tau}\left\|(T^{-(1+\tau)}\widehat{\*F}_{M_{\*x}}'\widehat{\*F}_{M_{\*x}})^+ - (T^{-(1+\tau)}\*q_M'\overline{\+\Gamma}'\*F_\*x'\*F_\*x\overline{\+\Gamma}\*q_M)^+ \right\|\frac{1}{N}\sum_{i=1}^N\left\|\*q_M \right\|^2\left\| \overline{\+\Gamma}\right\|^2\left\|T^{-(1+\tau)/2}\*F_\*x'\*V_i \right\|^2}_{O_p(T^{-(1+\tau)}\xi_{N,T,\tau}^{-1})}\notag\\
    &= o_p(\xi_{N,T,\tau}^{-1})=o_p(T^{-\tau}),
\end{align}
where each of the term is of the lower order than $O_p(\xi_{N,T,\tau}^{-1})$. Next, we have
\begin{align}\label{1b}
    \left\|\*I_b \right\| &= \left\| \frac{1}{NT^{1+\tau}}\sum_{i=1}^N\+\Gamma_i'\*F_\*x'(\*M_{\widehat{\*F}_{M_{\*x}}}-\*M_{\*F_{\*x}\overline{\+\Gamma}\*q_M})\*F_\*x\+\Gamma_i\right\|\notag\\
    &\leq  \left\| \frac{1}{NT^{1+\tau}}\sum_{i=1}^N\+\Gamma_i'\*F_\*x'\overline{\*V}\*q_M(T^{-(1+\tau)}\widehat{\*F}_{M_{\*x}}'\widehat{\*F}_{M_{\*x}})^+ T^{-(1+\tau)}\*q_M'\overline{\*V}'\*F_\*x\+\Gamma_i\right\|\notag\\
    &+\left\|\frac{1}{NT^{1+\tau}}\sum_{i=1}^N\+\Gamma_i'\*F_\*x'\overline{\*V}\*q_M(T^{-(1+\tau)}\widehat{\*F}_{M_{\*x}}'\widehat{\*F}_{M_{\*x}})^+ T^{-(1+\tau)}\*q_M'\overline{\+\Gamma}'\*F_\*x'\*F_\*x\+\Gamma_i\right\|\notag\\
    &+\left\|\frac{1}{NT^{1+\tau}}\sum_{i=1}^N\+\Gamma_i'\*F_\*x'\*F_\*x\overline{\+\Gamma}\*q_M(T^{-(1+\tau)}\widehat{\*F}_{M_{\*x}}'\widehat{\*F}_{M_{\*x}})^+ T^{-(1+\tau)}\*q_M'\overline{\*V}'\*F_\*x\+\Gamma_i \right\| \notag\\
    &+\left\| \frac{1}{NT^{1+\tau}}\sum_{i=1}^N\+\Gamma_i'\*F_\*x'\*F_\*x\overline{\+\Gamma}\*q_M \left[(T^{-(1+\tau)}\widehat{\*F}_{M_{\*x}}'\widehat{\*F}_{M_{\*x}})^+ - (T^{-(1+\tau)}\*q_M'\overline{\+\Gamma}'\*F_\*x'\*F_\*x\overline{\+\Gamma}\*q_M)^+  \right]T^{-(1+\tau)}\*q_M'\overline{\+\Gamma}'\*F_\*x'\*F_\*x\+\Gamma_i \right\| \notag\\
    &\leq \underbrace{N^{-1}T^{-1-\tau}\left\| (T^{-(1+\tau)}\widehat{\*F}_{M_{\*x}}'\widehat{\*F}_{M_{\*x}})^+\right\|\frac{1}{N}\sum_{i=1}^N\left\|T^{-(1+\tau)/2}\sqrt{N}\*F_\*x\overline{\*V} \right\|^2\left\|\*q_M \right\|^2\left\| \+\Gamma_i\right\|^2}_{O_p(N^{-1}T^{-1-\tau})}\notag\\
    &+\underbrace{2T^{-(\tau+1)/2}N^{-1/2}\left\| T^{-(1+\tau)}\*F_\*x'\*F_\*x\right\|\left\| (T^{-(1+\tau)}\widehat{\*F}_{M_{\*x}}'\widehat{\*F}_{M_{\*x}})^+\right\| \frac{1}{N}\sum_{i=1}^N\left\|T^{-(1+\tau)/2}\sqrt{N}\*F_\*x'\overline{\*V} \right\| \left\| \*q_M\right\|^2 \left\|\+\Gamma_i \right\|^2\left\| \overline{\+\Gamma}\right\|}_{O_p(T^{-(\tau+1)/2}N^{-1/2})}\notag\\
    &+\underbrace{\left\|(T^{-(1+\tau)}\widehat{\*F}_{M_{\*x}}'\widehat{\*F}_{M_{\*x}})^+ - (T^{-(1+\tau)}\*q_M'\overline{\+\Gamma}'\*F_\*x'\*F_\*x\overline{\+\Gamma}\*q_M)^+  \right\|\left\| T^{-(1+\tau)}\*F_\*x'\*F_\*x\right\|^2\left\| \*q_M\right\|^2\left\|\overline{\+\Gamma} \right\|^2 \frac{1}{N}\sum_{i=1}^N \left\|\+\Gamma_i \right\|^2}_{O_p( \xi_{N,T,\tau}^{-1})}\notag\\
    &= O_p(\xi_{N,T,\tau}^{-1})=o_p(T^{-\tau}),
\end{align}
which remains a slowly decaying term even if scaled by $T^{\tau}$, because it will be dominated by $T^{(\tau-1)/2}=o(1)$ stemming from the second term, since $\tau \in (0,1)$. We let $C_{N,T}=\mathrm{min}(\sqrt{N},\sqrt{T})$, and we move on to 
\begin{align}
    \left\| \*I_c\right\|&=\left\|\frac{1}{NT^{1+\tau}}\sum_{i=1}^N\*V_i'(\*M_{\widehat{\*F}_{M_{\*x}}}-\*M_{\*F_{\*x}\overline{\+\Gamma}\*q_M})\*F_\*x\+\Gamma_i \right\|\notag\\
    &  =\left\| \frac{1}{NT^{1+\tau}}\sum_{i=1}^N\*V_i'\overline{\*V}\*q_M(T^{-(1+\tau)}\widehat{\*F}_{M_{\*x}}'\widehat{\*F}_{M_{\*x}})^+ T^{-(1+\tau)}\*q_M'\overline{\*V}'\*F_\*x\+\Gamma_i\right\|\notag\\
    &+\left\|\frac{1}{NT^{1+\tau}}\sum_{i=1}^N\*V_i'\overline{\*V}\*q_M(T^{-(1+\tau)}\widehat{\*F}_{M_{\*x}}'\widehat{\*F}_{M_{\*x}})^+ T^{-(1+\tau)}\*q_M'\overline{\+\Gamma}'\*F_\*x'\*F_\*x\+\Gamma_i\right\|\notag\\
    &+\left\|\frac{1}{NT^{1+\tau}}\sum_{i=1}^N\*V_i'\*F_\*x\overline{\+\Gamma}\*q_M(T^{-(1+\tau)}\widehat{\*F}_{M_{\*x}}'\widehat{\*F}_{M_{\*x}})^+ T^{-(1+\tau)}\*q_M'\overline{\*V}'\*F_\*x\+\Gamma_i \right\| \notag\\
    &+\left\| \frac{1}{NT^{1+\tau}}\sum_{i=1}^N\*V_i'\*F_\*x\overline{\+\Gamma}\*q_M \left[(T^{-(1+\tau)}\widehat{\*F}_{M_{\*x}}'\widehat{\*F}_{M_{\*x}})^+ - (T^{-(1+\tau)}\*q_M'\overline{\+\Gamma}'\*F_\*x'\*F_\*x\overline{\+\Gamma}\*q_M)^+  \right]T^{-(1+\tau)}\*q_M'\overline{\+\Gamma}'\*F_\*x'\*F_\*x\+\Gamma_i\right\| \notag\\
    &\leq \underbrace{T^{-(1+3\tau)/2}N^{-1/2}\left\|(T^{-(1+\tau)}\widehat{\*F}_{M_{\*x}}'\widehat{\*F}_{M_{\*x}})^+ \right\|\frac{1}{N}\sum_{i=1}^N\left\|T^{-1}\*V_i'\overline{\*V} \right\|\left\| \*q_M\right\|^2 \left\|\+\Gamma_i \right\|\left\|T^{-(1+\tau)/2}\sqrt{N}\overline{\*V}'\*F_\*x \right\|}_{O_p(T^{-(1+3\tau)/2}N^{-1/2}C_{N,T}^{-2})}\notag\\
    &+ \underbrace{ T^{-\tau}\left\|(T^{-(1+\tau)}\widehat{\*F}_{M_{\*x}}'\widehat{\*F}_{M_{\*x}})^+ \right\|\frac{1}{N}\sum_{i=1}^N\left\|T^{-1}\*V_i'\overline{\*V} \right\|\left\| \*q_M\right\|^2 \left\|\+\Gamma_i \right\|\left\|T^{-(1+\tau)}\*F_\*x'\*F_\*x \right\|\left\| \overline{\+\Gamma}\right\|}_{O_p(T^{-\tau}C_{N,T}^{-2})}\notag\\
    &+ \underbrace{N^{-1/2}T^{-(1+\tau)}\left\|(T^{-(1+\tau)}\widehat{\*F}_{M_{\*x}}'\widehat{\*F}_{M_{\*x}})^+ \right\|\frac{1}{N}\sum_{i=1}^N\left\|T^{-(1+\tau)/2}\*V_i'\*F_\*x \right\| \left\|\sqrt{N}T^{-(1+\tau)/2}\overline{\*V}'\*F_\*x \right\|\left\| \*q_M\right\|^2\left\|\overline{\+\Gamma} \right\|\left\| \+\Gamma_i\right\|}_{O_p(N^{-1/2}T^{-1-\tau})}\notag\\
    &+ \underbrace{T^{-(\tau+1)/2}\left\|(T^{-(1+\tau)}\widehat{\*F}_{M_{\*x}}'\widehat{\*F}_{M_{\*x}})^+ - (T^{-(1+\tau)}\*q_M'\overline{\+\Gamma}'\*F_\*x'\*F_\*x\overline{\+\Gamma}\*q_M)^+ \right\|\frac{1}{N}\sum_{i=1}^N\left\|T^{-(1+\tau)}\*F_\*x'\*F_\*x \right\|\left\|T^{-(1+\tau)/2}\*F_\*x'\*V_i \right\|}_{O_p(T^{-(\tau+1)/2}\xi_{N,T,\tau}^{-1})}\notag\\
    &\times \underbrace{\left\|\overline{\+\Gamma} \right\|^2\left\|\*q_M \right\|^2 \left\|\+\Gamma_i \right\|}_{O_p(1)}\notag\\
    &=O_p(T^{-\tau}C_{N,T}^{-2})+O_p(N^{-1/2}T^{-1-\tau})=o_p(T^{-\tau}),
\end{align}
which is negligible. Ultimately, 
\begin{align}
    \left\|\*I_d \right\| &= \left\|  \frac{1}{NT^{1+\tau}}\sum_{i=1}^N\+\Gamma_i'\*F_\*x'(\*M_{\widehat{\*F}_{M_{\*x}}}-\*M_{\*F_{\*x}\overline{\+\Gamma}\*q_M})\*V_i \right\|\notag\\
    &\leq \left\| \frac{1}{NT^{1+\tau}}\sum_{i=1}^N\+\Gamma_i'\*F_\*x'\overline{\*V}\*q_M(T^{-(1+\tau)}\widehat{\*F}_{M_{\*x}}'\widehat{\*F}_{M_{\*x}})^+ T^{-(1+\tau)}\*q_M'\overline{\*V}'\*V_i\right\|\notag\\
    &+\left\|\frac{1}{NT^{1+\tau}}\sum_{i=1}^N\+\Gamma_i'\*F_\*x'\overline{\*V}\*q_M(T^{-(1+\tau)}\widehat{\*F}_{M_{\*x}}'\widehat{\*F}_{M_{\*x}})^+ T^{-(1+\tau)}\*q_M'\overline{\+\Gamma}'\*F_\*x'\*V_i\right\|\notag\\
    &+\left\|\frac{1}{NT^{1+\tau}}\sum_{i=1}^N\+\Gamma_i'\*F_\*x'\*F_\*x\overline{\+\Gamma}\*q_M(T^{-(1+\tau)}\widehat{\*F}_{M_{\*x}}'\widehat{\*F}_{M_{\*x}})^+ T^{-(1+\tau)}\*q_M'\overline{\*V}'\*V_i \right\| \notag\\
    &+\left\| \frac{1}{NT^{1+\tau}}\sum_{i=1}^N\+\Gamma_i'\*F_\*x'\*F_\*x\overline{\+\Gamma}\*q_M \left[(T^{-(1+\tau)}\widehat{\*F}_{M_{\*x}}'\widehat{\*F}_{M_{\*x}})^+ - (T^{-(1+\tau)}\*q_M'\overline{\+\Gamma}'\*F_\*x'\*F_\*x\overline{\+\Gamma}\*q_M)^+  \right]T^{-(1+\tau)}\*q_M'\overline{\+\Gamma}'\*F_\*x'\*V_i \right\| \notag\\
    &\leq \underbrace{T^{-(1+3\tau)/2}N^{-1/2}\left\|(T^{-(1+\tau)}\widehat{\*F}_{M_{\*x}}'\widehat{\*F}_{M_{\*x}})^+ \right\|\frac{1}{N}\sum_{i=1}^N\left\|T^{-1}\*V_i'\overline{\*V} \right\|\left\| \*q_M\right\|^2 \left\|\+\Gamma_i \right\|\left\|T^{-(1+\tau)/2}\sqrt{N}\overline{\*V}'\*F_\*x \right\|}_{O_p(T^{-(1+3\tau)/2}N^{-1/2}C_{N,T}^{-2})}\notag\\
    &+ \underbrace{N^{-1/2}T^{-1-\tau}\left\|(T^{-(1+\tau)}\widehat{\*F}_{M_{\*x}}'\widehat{\*F}_{M_{\*x}})^+ \right\|\frac{1}{N}\sum_{i=1}^N\left\|T^{-(1+\tau)/2}\*V_i'\*F_\*x \right\| \left\|\sqrt{N}T^{-(1+\tau)/2}\overline{\*V}'\*F_\*x \right\|\left\| \*q_M\right\|^2\left\|\overline{\+\Gamma} \right\|\left\| \+\Gamma_i\right\|}_{O_p(N^{-1/2}T^{-1-\tau})}\notag\\
    &+ \underbrace{T^{-\tau}\left\|(T^{-(1+\tau)}\widehat{\*F}_{M_{\*x}}'\widehat{\*F}_{M_{\*x}})^+ \right\|\frac{1}{N}\sum_{i=1}^N\left\|T^{-1}\*V_i'\overline{\*V} \right\|\left\| \*q_M\right\|^2 \left\|\+\Gamma_i \right\|\left\|T^{-(1+\tau)}\*F_\*x'\*F_\*x \right\|\left\| \overline{\+\Gamma}\right\|}_{O_p(T^{-\tau}C_{N,T}^{-2})}\notag\\
    &+  \underbrace{T^{-(\tau+1)/2}\left\|(T^{-(1+\tau)}\widehat{\*F}_{M_{\*x}}'\widehat{\*F}_{M_{\*x}})^+ - (T^{-(1+\tau)}\*q_M'\overline{\+\Gamma}'\*F_\*x'\*F_\*x\overline{\+\Gamma}\*q_M)^+ \right\|\frac{1}{N}\sum_{i=1}^N\left\|T^{-(1+\tau)}\*F_\*x'\*F_\*x \right\|\left\|T^{-(1+\tau)/2}\*F_\*x'\*V_i \right\|}_{O_p(T^{-(\tau+1)/2}\xi_{N,T,\tau}^{-1})}\notag\\
    &\times \underbrace{\left\|\overline{\+\Gamma} \right\|^2\left\|\*q_M \right\|^2 \left\|\+\Gamma_i \right\|}_{O_p(1)}\notag\\
    &= O_p(T^{-\tau}C_{N,T}^{-2})+O_p(N^{-1/2}T^{-1-\tau})=o_p(T^{-\tau}),
\end{align}
which has an identical asymptotic behavior to the one of $\*I_c$. Hence, overall, 
\begin{align}
    \left\| \*I\right\| =\left\|T^{-\tau}\left(\overline{\*Q}_{M}-\widehat{\*Q}_{M,\*M_{\*F_{\*x}\overline{\+\Gamma}\*q_{M}}} \right)\right\|=O_p(\xi_{N,T,\tau}^{-1})=o_p(T^{-\tau} ),
\end{align}
which is the slowest decaying term. Importantly, we also have that 
\begin{align}\label{Q_m0-Qmf0}
     \left\| \*I\*I\right\| =\left\|T^{-\tau}\left(\overline{\*Q}_{M_0}-\widehat{\*Q}_{M_0,\*M_{\*F_{\*x}\overline{\+\Gamma}\*q_{M_0}}} \right)\right\|=O_p(\xi_{N,T,\tau}^{-1})=o_p(T^{-\tau} ),
\end{align}
as in this case $m=g$, and therefore the asymptotic behavior of this term will be identical. \\

\noindent We are left to analyze $\mathbf{III}$. Because we are in the case of under-specification, we can decompose $\*F_\*x \overline{\+\Gamma}\*q_{M_0}$ into $\*F_\*x \overline{\+\Gamma}\*q_{M_0}=\*A=[\*F_\*x \overline{\+\Gamma}\*q_{M},\*F_\*x \overline{\+\Gamma}\*q_{M^C}]=[\*B,\*C]$, where $\*q_{M^C}$ is the \textit{complement} selector matrix that is $k\times (m-g)$. By utilizing this representation, we can decompose the projection matrix $\*P_\*A=\*P_\*B+\*P_{\*M_\*B\*C}$. By following this, we obtain
\begin{align}
    \mathbf{III}&=T^{-\tau}\left(\widehat{\*Q}_{M,\*M_{\*F_{\*x}\overline{\+\Gamma}\*q_M}}-\widehat{\*Q}_{M_0,\*M_{\*F_{\*x}\overline{\+\Gamma}\*q_{M_0}}}\right)\notag\\
    &= \frac{1}{NT^{1+\tau}}\sum_{i=1}^N\*X_i'(\*M_{\*F_{\*x}\overline{\+\Gamma}\*q_{M}}-\*M_{\*F_{\*x}\overline{\+\Gamma}\*q_{M_0}})\*X_i\notag\\
    &= \frac{1}{NT^{1+\tau}}\sum_{i=1}^N\*X_i'(\*P_{\*F_{\*x}\overline{\+\Gamma}\*q_{M_0}}-\*P_{\*F_{\*x}\overline{\+\Gamma}\*q_{M}})\*X_i\notag\\
    & =\frac{1}{NT^{1+\tau}}\sum_{i=1}^N\*X_i'(\*P_{\*F_{\*x}\overline{\+\Gamma}\*q_{M}}+\*P_{\*M_{\*F_{\*x}\overline{\+\Gamma}\*q_{M}}\*F_{\*x}\overline{\+\Gamma}\*q_{M^C}} - \*P_{\*F_{\*x}\overline{\+\Gamma}\*q_{M}})\*X_i\notag\\
    &= \frac{1}{NT^{1+\tau}}\sum_{i=1}^N\*X_i'\*P_{\*M_{\*F_{\*x}\overline{\+\Gamma}\*q_{M}}\*F_{\*x}\overline{\+\Gamma}\*q_{M^C}}\*X_i\notag\\
    &= \frac{1}{NT^{1+\tau}}\sum_{i=1}^N\*V_i'\*P_{\*M_{\*F_{\*x}\overline{\+\Gamma}\*q_{M}}\*F_{\*x}\overline{\+\Gamma}\*q_{M^C}}\*V_i + \frac{1}{NT^{1+\tau}}\sum_{i=1}^N\*V_i'\*P_{\*M_{\*F_{\*x}\overline{\+\Gamma}\*q_{M}}\*F_{\*x}\overline{\+\Gamma}\*q_{M^C}}\*F_\*x\+\Gamma_i\notag\\
    & + \frac{1}{NT^{1+\tau}}\sum_{i=1}^N\+\Gamma_i'\*F_\*x'\*P_{\*M_{\*F_{\*x}\overline{\+\Gamma}\*q_{M}}\*F_{\*x}\overline{\+\Gamma}\*q_{M^C}}\*V_i + \frac{1}{NT^{1+\tau}}\sum_{i=1}^N\+\Gamma_i'\*F_\*x'\*P_{\*M_{\*F_{\*x}\overline{\+\Gamma}\*q_{M}}\*F_{\*x}\overline{\+\Gamma}\*q_{M^C}}\*F_\*x\+\Gamma_i\notag\\
    &=\mathbf{III}_a+\mathbf{III}_b+\mathbf{III}_c+\mathbf{III}_d,
\end{align}
where we will start the term $\mathbf{III}_d$. Note that it explicitly looks like
\begin{align}
      \mathbf{III}_d &= \frac{1}{NT^{1+\tau}}\sum_{i=1}^N\+\Gamma_i'\*F_\*x'\*M_{\*F_{\*x}\overline{\+\Gamma}\*q_M}\*F_{\*x}\overline{\+\Gamma}\*q_{M^C}(\*q_{M^C}'\overline{\+\Gamma}'\*F_\*x'\*M_{\*F_{\*x}\overline{\+\Gamma}\*q_M}\*F_{\*x}\overline{\+\Gamma}\*q_{M^C})^+ \*q_{M^C}'\overline{\+\Gamma}'\*F_\*x'\*M_{\*F_{\*x}\overline{\+\Gamma}\*q_M}\*F_\*x\+\Gamma_i
\end{align}
with the key component $\*F_\*x'\*M_{\*F_{\*x}\overline{\+\Gamma}\*q_M}\*F_{\*x}$, where
\begin{align}
    T^{-(1+\tau)}\*F_\*x'\*M_{\*F_{\*x}\overline{\+\Gamma}\*q_M}\*F_{\*x}&= T^{-(1+\tau)}\*F_\*x'\*F_\*x -  T^{-(1+\tau)}\*F_\*x'\*F_\*x\overline{\+\Gamma}\*q_{M}(\*q_M'\overline{\+\Gamma}'\*F_\*x'\*F_\*x\overline{\+\Gamma}\*q_{M})^+\*q_M'\overline{\+\Gamma}'\*F_\*x'\*F_\*x\notag\\
    &=T^{-(1+\tau)}\*F_\*x'\*F_\*x -  T^{-(1+\tau)}\*F_\*x'\*F_\*x\overline{\+\Gamma}\*q_{M}(T^{-(1+\tau)}\*q_M'\overline{\+\Gamma}'\*F_\*x'\*F_\*x\overline{\+\Gamma}\*q_{M})^+T^{-(1+\tau)}\*q_M'\overline{\+\Gamma}'\*F_\*x'\*F_\*x\notag\\
    &\to_p \+\Sigma_{\*F_\*x}-\+\Sigma_{\*F_\*x}\overline{\+\Gamma}\*q_{M}(\*q_M'\overline{\+\Gamma}'\+\Sigma_{\*F_\*x}\overline{\+\Gamma}\*q_{M})^+\*q_M'\overline{\+\Gamma}'\+\Sigma_{\*F_\*x},
\end{align}
as $T\to \infty$, which is positive definite constant matrix (see a similar result in A. 51 in the Supplement of \citealp{margaritella2023using}). Therefore, clearly $\mathbf{III}_d$ converges to a positive semi-definite matrix, because it is a matrix-valued form. Importantly, it is not a zero matrix, because, according to Exercise 8.13 (b) in \cite{Abadir2005}, $\mathbf{III}_d=\*0_{k\times k}$ is equivalent to $\mathrm{tr}(\mathbf{III}_d)=0$ for a positive semi-definite matrix. Note that
\begin{align}
    &\mathrm{tr}\left( \mathbf{III}_d\right)=\mathrm{tr}\left( \frac{1}{NT^{1+\tau}}\sum_{i=1}^N\+\Gamma_i'\*F_\*x'\*M_{\*F_{\*x}\overline{\+\Gamma}\*q_M}\*F_{\*x}\overline{\+\Gamma}\*q_{M^C}(\*q_{M^C}'\overline{\+\Gamma}'\*F_\*x'\*M_{\*F_{\*x}\overline{\+\Gamma}\*q_M}\*F_{\*x}\overline{\+\Gamma}\*q_{M^C})^+ \*q_{M^C}'\overline{\+\Gamma}'\*F_\*x'\*M_{\*F_{\*x}\overline{\+\Gamma}\*q_M}\*F_\*x\+\Gamma_i\right)\notag\\
    &= \mathrm{tr}\left( T^{-(1+\tau)}\*F_\*x'\*M_{\*F_{\*x}\overline{\+\Gamma}\*q_M}\*F_{\*x}\overline{\+\Gamma}\*q_{M^C}(T^{-(1+\tau)}\*q_{M^C}'\overline{\+\Gamma}'\*F_\*x'\*M_{\*F_{\*x}\overline{\+\Gamma}\*q_M}\*F_{\*x}\overline{\+\Gamma}\*q_{M^C})^+ T^{-(1+\tau)}\*q_{M^C}'\overline{\+\Gamma}'\*F_\*x'\*M_{\*F_{\*x}\overline{\+\Gamma}\*q_M}\*F_\*x\left[\frac{1}{N}\sum_{i=1}^N\+\Gamma_i\+\Gamma_i'\right]\right)\notag\\
    &=\mathrm{tr}\left(\left[\frac{1}{N}\sum_{i=1}^N\+\Gamma_i\+\Gamma_i'\right] T^{-(1+\tau)}\*F_\*x'\*M_{\*F_{\*x}\overline{\+\Gamma}\*q_M}\*F_{\*x}\overline{\+\Gamma}\*q_{M^C}(T^{-(1+\tau)}\*q_{M^C}'\overline{\+\Gamma}'\*F_\*x'\*M_{\*F_{\*x}\overline{\+\Gamma}\*q_M}\*F_{\*x}\overline{\+\Gamma}\*q_{M^C})^+ T^{-(1+\tau)}\*q_{M^C}'\overline{\+\Gamma}'\*F_\*x'\*M_{\*F_{\*x}\overline{\+\Gamma}\*q_M}\*F_\*x \right) \notag\\
    &\geq \lambda_{\min}\left( \frac{1}{N}\sum_{i=1}^N\+\Gamma_i\+\Gamma_i'\right)\notag\\
    &\times \mathrm{tr}\left( T^{-(1+\tau)}\*F_\*x'\*M_{\*F_{\*x}\overline{\+\Gamma}\*q_M}\*F_{\*x}\overline{\+\Gamma}\*q_{M^C}(T^{-(1+\tau)}\*q_{M^C}'\overline{\+\Gamma}'\*F_\*x'\*M_{\*F_{\*x}\overline{\+\Gamma}\*q_M}\*F_{\*x}\overline{\+\Gamma}\*q_{M^C})^+ T^{-(1+\tau)}\*q_{M^C}'\overline{\+\Gamma}'\*F_\*x'\*M_{\*F_{\*x}\overline{\+\Gamma}\*q_M}\*F_\*x\right)\notag\\
    &\to_p c>0
\end{align}
as $(N,T)\to \infty$, for a positive constant $c$ by the results in \cite{fang1994inequalities}. This follows from the fact that second positive semi-definite matrix in the product is non-zero by Exercise 8.26 (c) in \cite{Abadir2005}. Particularly, for a positive definite $\*A\in \mathbb{R}^{p\times p}$ and $\*B\in \mathbb{R}^{p\times q}$, $\*B'\*A\*B=\*0_{q\times q}$ if and only if $\*B=\*0_{p\times q}$, which means that $\mathrm{rk}(\*B)=0$. In our case, $\*B= T^{-(1+\tau)}\*q_{M^C}'\overline{\+\Gamma}'\*F_\*x'\*M_{\*F_{\*x}\overline{\+\Gamma}\*q_M}\*F_\*x$ and we know that $\mathrm{rk}\left( T^{-(1+\tau)}\*q_{M^C}'\overline{\+\Gamma}'\*F_\*x'\*M_{\*F_{\*x}\overline{\+\Gamma}\*q_M}\*F_\*x\right)=\mathrm{rk}(\overline{\+\Gamma}{\*q}_{M^C})=m_x-g>0$ even in the limit. Also, $\lim_{N\to \infty}\frac{1}{N}\sum_{i=1}^N\+\Gamma_i\+\Gamma_i'$ is an $m_x\times m_x$ positive definite matrix by our assumptions.\\ 

\noindent Next in our analysis, we move on to $\mathbf{III}_a$, and for this we firstly analyze 
\begin{align}
    \left\|T^{-(1+\tau)/2} \*V_i'\*M_{\*F_{\*x}\overline{\+\Gamma}\*q_M}\*F_{\*x}\right\|&\leq \left\|T^{-(1+\tau)/2}\*V_i'\*F_{\*x} \right\|\notag\\
    &+ \left\|T^{-(1+\tau)/2}\*V_i'\*F_\*x\overline{\+\Gamma}\*q_{M}(T^{-(1+\tau)}\*q_M'\overline{\+\Gamma}'\*F_\*x'\*F_\*x\overline{\+\Gamma}\*q_{M})^+T^{-(1+\tau)}\*q_M'\overline{\+\Gamma}'\*F_\*x'\*V_i\right\|\notag\\
    &\leq  \underbrace{\left\|T^{-(1+\tau)/2}\*V_i'\*F_{\*x} \right\|}_{O_p(1)}\notag\\
    &+\underbrace{T^{-(1+\tau)/2}\left\|T^{-(1+\tau)/2}\*V_i'\*F_\*x \right\|^2\left\|\overline{\+\Gamma}\*q_{M} \right\|^2\left\|(T^{-(1+\tau)}\*q_M'\overline{\+\Gamma}'\*F_\*x'\*F_\*x\overline{\+\Gamma}\*q_{M})^+ \right\|}_{O_p(T^{-(1+\tau)/2})}\notag\\
    &=O_p(1).
\end{align}
This gives us
\begin{align}
     \left\|\mathbf{III}_a \right\|&= \left\| \frac{1}{NT^{1+\tau}}\sum_{i=1}^N\*V_i'\*P_{\*M_{\*F_{\*x}\overline{\+\Gamma}\*q_{M}}\*F_{\*x}\overline{\+\Gamma}\*q_{M^C}}\*V_i\right\|\notag\\
     &=\left\| \frac{1}{NT^{1+\tau}}\sum_{i=1}^N\*V_i'\*M_{\*F_{\*x}\overline{\+\Gamma}\*q_M}\*F_{\*x}\overline{\+\Gamma}\*q_{M^C}(\*q_{M^C}'\overline{\+\Gamma}'\*F_\*x'\*M_{\*F_{\*x}\overline{\+\Gamma}\*q_M}\*F_{\*x}\overline{\+\Gamma}\*q_{M^C})^+\*q_{M^C}'\overline{\+\Gamma}'\*F_\*x'\*M_{\*F_{\*x}\overline{\+\Gamma}\*q_M}\*V_i\right\|\notag\\
     &\leq T^{-1-\tau}\frac{1}{N}\sum_{i=1}^N\left\|T^{-(1+\tau)/2}\*V_i'\*M_{\*F_{\*x}\overline{\+\Gamma}\*q_M}\*F_{\*x} \right\|^2\left\| \overline{\+\Gamma}\*q_{M^C}\right\|^2\left\| (T^{-(1+\tau)}\*q_{M^C}'\overline{\+\Gamma}'\*F_\*x'\*M_{\*F_{\*x}\overline{\+\Gamma}\*q_M}\*F_{\*x}\overline{\+\Gamma}\*q_{M^C})^+\right\|\notag\\
     &=O_p(T^{-1-\tau})=o_p(T^{-\tau}),
\end{align}
which would not be negligible under correlated idiosyncratics (see Lemma \ref{LemmaA2}). The remaining two terms are vanishing, as well. In particular, 
\begin{align}
    \left\|\mathbf{III}_b \right\|&=\left\|  \frac{1}{NT^{1+\tau}}\sum_{i=1}^N\*V_i'\*P_{\*M_{\*F_{\*x}\overline{\+\Gamma}\*q_{M}}\*F_{\*x}\overline{\+\Gamma}\*q_{M^C}}\*F_\*x\+\Gamma_i\right\|\notag\\
    &=\left\|  \frac{1}{NT^{1+\tau}}\sum_{i=1}^N\*V_i'\*M_{\*F_{\*x}\overline{\+\Gamma}\*q_M}\*F_{\*x}\overline{\+\Gamma}\*q_{M^C}(\*q_{M^C}'\overline{\+\Gamma}'\*F_\*x'\*M_{\*F_{\*x}\overline{\+\Gamma}\*q_M}\*F_{\*x}\overline{\+\Gamma}\*q_{M^C})^+\*q_{M^C}'\overline{\+\Gamma}'\*F_\*x'\*M_{\*F_{\*x}\overline{\+\Gamma}\*q_M}\*F_\*x\+\Gamma_i\right\|\notag\\
    &\leq T^{-(\tau+1)/2}\frac{1}{N}\sum_{i=1}^N\left\|T^{-(1+\tau)/2} \*V_i'\*M_{\*F_{\*x}\overline{\+\Gamma}\*q_M}\*F_{\*x}\right\|\left\|\overline{\+\Gamma}\*q_M \right\|^2\left\|T^{-(1+\tau)} \*F_\*x'\*M_{\*F_{\*x}\overline{\+\Gamma}\*q_M}\*F_\*x\right\|\notag\\
    &\times \left\|(\*q_{M^C}'\overline{\+\Gamma}'T^{-(1+\tau)}\*F_\*x'\*M_{\*F_{\*x}\overline{\+\Gamma}\*q_M}\*F_{\*x}\overline{\+\Gamma}\*q_{M^C})^+ \right\|\left\|\+\Gamma_i \right\|\notag\\
    &=O_p(T^{-(\tau+1)/2})=o_p(T^{-\tau}),
\end{align}
for $\tau \in (0,1)$, and \begin{align}
    \left\|\mathbf{III}_c \right\|&=\left\| \frac{1}{NT^{1+\tau}}\sum_{i=1}^N\+\Gamma_i'\*F_\*x'\*P_{\*M_{\*F_{\*x}\overline{\+\Gamma}\*q_{M}}\*F_{\*x}\overline{\+\Gamma}\*q_{M^C}}\*V_i\right\|\notag\\
    &= \left\|\frac{1}{NT^{1+\tau}}\sum_{i=1}^N\+\Gamma_i'\*F_\*x'\*M_{\*F_{\*x}\overline{\+\Gamma}\*q_M}\*F_{\*x}\overline{\+\Gamma}\*q_{M^C}(\*q_{M^C}'\overline{\+\Gamma}'\*F_\*x'\*M_{\*F_{\*x}\overline{\+\Gamma}\*q_M}\*F_{\*x}\overline{\+\Gamma}\*q_{M^C})^+\*q_{M^C}'\overline{\+\Gamma}'\*F_\*x'\*M_{\*F_{\*x}\overline{\+\Gamma}\*q_M}\*V_i\right\|\notag\\
    &\leq T^{-(\tau+1)/2}\frac{1}{N}\sum_{i=1}^N\left\|T^{-(1+\tau)/2} \*V_i'\*M_{\*F_{\*x}\overline{\+\Gamma}\*q_M}\*F_{\*x}\right\|\left\|\overline{\+\Gamma}\*q_M \right\|^2\left\|T^{-(1+\tau)} \*F_\*x'\*M_{\*F_{\*x}\overline{\+\Gamma}\*q_M}\*F_\*x\right\|\notag\\
    &\times \left\|(\*q_{M^C}'\overline{\+\Gamma}'T^{-(1+\tau)}\*F_\*x'\*M_{\*F_{\*x}\overline{\+\Gamma}\*q_M}\*F_{\*x}\overline{\+\Gamma}\*q_{M^C})^+ \right\|\left\|\+\Gamma_i \right\|\notag\\
    &=O_p(T^{-(\tau+1)/2})=o_p(T^{-\tau}),
\end{align}
which is just a transpose of $\mathbf{III}_b$. Note again, that both of the terms would slowly diverge under the conditions of Lemma \ref{LemmaA2}. Therefore, by combining our interim results, we obtain 
\begin{align}\label{eventual_exp}
   T^{-\tau}\left( \overline{\*Q}_{M}-\overline{\*Q}_{M_0}\right)&=  T^{-\tau}\left[\overline{\*Q}_{M}-\widehat{\*Q}_{M,\*M_{\*F_{\*x}\overline{\+\Gamma}\*q_{M}}}\right] -  T^{-\tau}\left[ \overline{\*Q}_{M_0}-\widehat{\*Q}_{M_0,\*M_{\*F_{\*x}\overline{\+\Gamma}\*q_{M_0}}}\right] 
	\notag\\
    &+ T^{-\tau}\left[ \widehat{\*Q}_{M,\*M_{\*F_{\*x}\overline{\+\Gamma}\*q_M}}-\widehat{\*Q}_{M_0,\*M_{\*F_{\*x}\overline{\+\Gamma}\*q_{M_0}}}\right]\notag\\
    &=\mathbf{III}_d+o_p(T^{-\tau}),
\end{align}
which is a positive semi-definite matrix and $o_p(T^{-\tau})=O_p(\xi_{N,T,\tau}^{-1})$. Further, note how by using the fact that $\overline{\*Q}_{M_0}=\widehat{\*Q}_{M_0,\*M_{\*F_{\*x}\overline{\+\Gamma}\*q_{M_0}}}+o_p(1)$ from (\ref{Q_m0-Qmf0}), we have that 
\begin{align}\label{Q_0}
   \overline{\*Q}_{M_0}&=\widehat{\*Q}_{M_0,\*M_{\*F_{\*x}\overline{\+\Gamma}\*q_{M_0}}}+o_p(1)\notag\\
   &= \widehat{\*Q}_{M_0,\*M_{\*F_{\*x}}}+o_p(1)\notag\\
   &= \frac{1}{T}\frac{1}{N}\sum_{i=1}^N\*X_i'\*M_{\*F_\*x}\*X_i+o_p( 1)\notag\\
   &= \frac{1}{T}\frac{1}{N}\sum_{i=1}^N\*V_i'\*M_{\*F_\*x}\*V_i+o_p( 1)\notag\\
   &= \frac{1}{N}\sum_{i=1}^NT^{-1}\*V_i'\*V_i - T^{-1}\frac{1}{N}\sum_{i=1}^NT^{-(1+\tau)/2}\*V_i'\*F_\*x(T^{-(1+\tau)}\*F_\*x'\*F_\*x)^+T^{-(1+\tau)/2}\*F_\*x'\*V_i+o_p( 1)\notag\\
   &= \frac{1}{N}\sum_{i=1}^NT^{-1}\*V_i'\*V_i  + o_p( 1)\notag\\
   &\to_p \+\Sigma_\*v,
\end{align}
because the idiosyncratics are stationary. This is a positive definite matrix. In what follows, we use Theorem 2 and Exercise 1 in Section 4 of \cite{magnus2019matrix}. In particular, 
\begin{align}
    \ln\mathrm{det}(\*A+\*H)=\ln\mathrm{det}(\*A)+\mathrm{tr}(\*A^{-1}\*H)+O\left(\left\| \*A^{-1}\*H\right\|^2\right),
\end{align}
for a positive definite matrix $\*A$ and a symmetric perturbation $\*H$. Further, to simplify notation, let $\+\Omega_{M,M_0}:=T^{-\tau}[\overline{\*Q}_M-\overline{\*Q}_{M_0}]\overline{\*Q}_{M_0}^{-1}$ and $\+\Omega_{M,M_0}^0:=\mathbf{III}_d\times\overline{\*Q}_{M_0}^{-1}$, which we know is a positive semi-definite matrix. Next, let $\mathrm{d}\+\Omega_{M,M_0}=\+\Omega_{M,M_0}-\+\Omega_{M,M_0}^0$, where we know that $\left\|\mathrm{d}\+\Omega_{M,M_0} \right\|=O_p(\xi_{N,T,\tau}^{-1})=o_p(T^{-\tau})$. In our setting, $\*A=\*I_k+T^\tau \+\Omega_{M,M_0}$ and $\*H=\mathrm{d}\+\Omega_{M,M_0}$. Therefore, 
\begin{align}\label{dQ_exp}
   \mathrm{d}\overline{\*Q}_{M,M_0}=\ln\mathrm{det}\left(\*I_k+T^\tau \+\Omega_{M,M_0}\right)&=\ln\mathrm{det}\left(\*I_k+T^\tau \+\Omega_{M,M_0}^0+\mathrm{d}\+\Omega_{M,M_0}\right)\notag\\
   &=\ln\mathrm{det}\left(\*I_k+T^\tau \+\Omega_{M,M_0}^0\right)\notag\\
   &+\ln\mathrm{det}\left(\*I_k+T^\tau \+\Omega_{M,M_0}^0+T^{\tau}\mathrm{d}\+\Omega_{M,M_0}\right)\notag\\
   &-\ln\mathrm{det}\left(\*I_k+T^\tau \+\Omega_{M,M_0}^0\right)\notag\\
   &=\ln\mathrm{det}\left(\*I_k+T^\tau \+\Omega_{M,M_0}^0\right)+\mathrm{tr}\left[\left(\*I_k+T^\tau \+\Omega_{M,M_0}^0 \right)^{-1}T^{\tau} \mathrm{d}\+\Omega_{M,M_0}\right]\notag\\
   &+O_p\left(\left\| \left(\*I_k+T^\tau \+\Omega_{M,M_0}^0 \right)^{-1}T^{\tau} \mathrm{d}\+\Omega_{M,M_0}\right\|^2 \right)\notag\\
   &= \ln\mathrm{det}\left(\*I_k+T^\tau \+\Omega_{M,M_0}^0\right)+O_p(T^{\tau}\xi_{N,T,\tau}^{-1}).
\end{align}
Here, we used the following important facts. Let $\lambda_{NT,j}\geq 0$ be an eigenvalue of $\+\Omega_{M,M_0}^0$ for $j=1,\ldots,k$ with the probability limit $\lambda >0$ or 0 immediately. As this positive semi-definite matrix is nonzero even in the limit, we know that not all eigenvalues are zero. Then we know that the eigenvalues of $\*B=\*I_k+c\*A$ are given by $1+c\lambda_j$ for a constant $c$, where $\lambda_j$ is an eigenvalue of generic square matrix $\*A$. In particular, for the $j$-th eigenvector of $\*A$ ($\+\nu_j$), we have
\begin{align}
    \*B\+\nu_j=(\*I_k+c\*A)\+\nu_j=\+\nu_j+c\*A\+\nu_j=1\times \+\nu+c\lambda_j\+\nu_j =(1+c\+\lambda_j)\+\nu_j.
\end{align}
Next, we know that eigenvalues of $\*B^{-1}=(\*I_k+c\*A)^{-1}$ are given by $\frac{1}{1+c\lambda_j}$. Which means that eigenvalues of $ \left(\*I_k+T^\tau \+\Omega_{M,M_0}^0 \right)^{-1}$ are given by $\frac{1}{1+T^{\tau}\lambda_{NT,j}}$. Note that we can diagonalize this matrix and obtain
\begin{align}
    \left(\*I_k+T^\tau \+\Omega_{M,M_0}^0 \right)^{-1}=\+\Upsilon\left(\underbrace{\mathrm{diag}\left[\frac{1}{1+T^{\tau}\lambda_{NT,1}},\ldots, \frac{1}{1+T^{\tau}\lambda_{NT,k}}\right]}_{=\+\Lambda_{N,T}}\right)\+\Upsilon',
\end{align}
where $\+\Upsilon$ is an orthonormal vector. Let $b$ denote the number of eigenvalues of $\+\Omega_{M,M_0}^0$ that is strictly positive. Then $k-b$ is zero eigenvalues, which means that we will have $k-b$ eigenvalues of $ \left(\*I_k+T^\tau \+\Omega_{M,M_0}^0 \right)^{-1}$ that are equal to 1. Then by using the trace properties, we obtain 
\begin{align}
    \left\| \left(\*I_k+T^\tau \+\Omega_{M,M_0}^0 \right)^{-1}\right\|=\sqrt{\mathrm{tr}\left(\+\Upsilon\+\Lambda_{N,T}\+\Upsilon'\+\Upsilon\+\Lambda_{N,T}\+\Upsilon' \right)}&=\sqrt{\mathrm{tr}\left(\+\Upsilon\+\Lambda_{N,T}^2\+\Upsilon' \right)}\notag\\
    &=\sqrt{\mathrm{tr}\left(\+\Upsilon' \+\Upsilon\+\Lambda_{N,T}^2\right)}\notag\\
    &= \sqrt{\sum_{j=1}^k\frac{1}{(1+T^{\tau}\lambda_{NT,j})^2}}\notag\\
    &\to_p\sqrt{k-b},
\end{align}
because $\lambda_{NT,j}=0$ or $\lambda_{NT,j}\to_p\lambda_j>0$, where the latter case produces $b$ zero entries into the total sum, and the former case gives $k-b$ units. Therefore, $ \left\| \left(\*I_k+T^\tau \+\Omega_{M,M_0}^0 \right)^{-1}\right\|=O_p(1)$. Further, 
\begin{align}\label{norm_differential}
   \left| \mathrm{tr}\left[\left(\*I_k+T^\tau \+\Omega_{M,M_0}^0 \right)^{-1}T^{\tau} \mathrm{d}\+\Omega_{M,M_0}\right]\right|&\leq \left\|\left(\*I_k+T^\tau \+\Omega_{M,M_0}^0 \right)^{-1}\right\| T^\tau \left\|\mathrm{d}\+\Omega_{M,M_0} \right\|\notag\\
   &=O_p(T^{\tau}\xi_{N,T,\tau}^{-1})=O_p(N^{-1})+O_p(N^{-1/2}T^{(\tau-1)/2})\notag\\
   &=R_{N,T}. 
   \end{align}
   Therefore, combining (\ref{dQ_exp}) and (\ref{norm_differential}), we get 
\begin{align}
    \mathrm{IC}^{DVS}(M)- \mathrm{IC}^{DVS}(M_0)&=\ln\mathrm{det}\left(\*I_k+T^{\tau}\+\Omega_{M,M_0}\right)+k(g-m)p_{N,T}\notag\\
    &=\ln\mathrm{det}\left(\*I_k+T^\tau\left(\+\Omega_{M,M_0}^0+\mathrm{d}\+\Omega_{M,M_0}\right) \right)+k(g-m)p_{N,T}\notag\\
    &=\ln\mathrm{det}\left(\*I_k+T^\tau\+\Omega_{M,M_0}^0+T^{\tau}\mathrm{d}\+\Omega_{M,M_0} \right) + k(g-m)p_{N,T}\notag\\
    &=\ln\mathrm{det}\left(\*I_k+T^\tau\+\Omega_{M,M_0}^0 \right) +k(g-m)p_{N,T}+ R_{N,T}.
\end{align}
Clearly, the difference diverges to $+\infty$, because $\+\Omega_{M,M_0}^0$ converges to a nonzero positive semi-definite matrix. To complete the proof, it is useful to derive the rate at which the divergence occurs. Again, recall that $\lambda_{NT,j}\geq 0$ is an eigenvalue of $\+\Omega_{M,M_0}^0$ for $j=1,\ldots,k$ with the probability limit $\lambda > 0$ or zero immediately. As this positive semi-definite matrix is nonzero even in the limit, we know that not all eigenvalues are zero. Recall that $b$ is the number of positive eigenvalues. Therefore, using $\ln(1+x)=x+O(x^2)$ for a small $x$ and $p_{N,T}/\ln(T)=o(1)$, we obtain the following:  
\begin{align}
     ( \mathrm{IC}^{DVS}(M)&- \mathrm{IC}^{DVS}(M_0))/\ln(T)=\left[\ln\mathrm{det}\left(\*I_k+T^\tau\+\Omega_{M,M_0}^0 \right)\right] /\ln(T)+ o_p(1)\notag\\
     &=\left[ \ln \prod_{j:\lambda>0}(1+T^\tau\lambda_{NT,j})\right]/\ln(T)+o_p(1)\notag\\
     &=\left[\sum_{j:\lambda>0}\ln(1+T^\tau \lambda_{NT,j}) \right]/\ln(T)+o_p(1)\notag\\
     &=\left[\sum_{j:\lambda>0} \ln(T^\tau \lambda_{NT,j}) \right]/\ln(T) + \left[\sum_{j:\lambda>0}[\ln(1+T^\tau \lambda_{NT,j})-\ln(T^\tau \lambda_{NT,j}) ]\right]/\ln(T) +o_p(1)\notag\\
     &= \frac{\tau b\ln(T)}{\ln(T)} + \frac{1}{\ln(T)}\sum_{j:\lambda>0}\ln(\lambda_{NT,j})+ \frac{1}{\ln(T)}\sum_{j:\lambda>0}\ln \left( 1+\frac{1}{T^\tau \lambda_{NT,j}}\right) + o_p(1)\notag\\
     &=\tau b + \frac{1}{\ln(T)}\sum_{j:\lambda>0}\ln(\lambda_{NT,j})+ \frac{1}{\ln(T)}\sum_{j:\lambda>0}\left(\frac{1}{T^\tau \lambda_{NT,j}} + O(T^{-2\tau})\right) + o_p(1)\notag\\
     &\to_p \tau b >0,
\end{align}
because $\lambda_{NT,j}\to_p \lambda_j>0$ as $(N,T)\to \infty$. This overall implies that 
\begin{align}  \mathbb{P}\left(\mathrm{IC}^{DVS}(M)- \mathrm{IC}^{DVS}(M_0) <0\right)=\mathbb{P}\left(\frac{\mathrm{IC}^{DVS}(M)- \mathrm{IC}^{DVS}(M_0)}{\ln(T)} <0\right)\to 0
\end{align}
as $(N,T)\to \infty$. 
\subsubsection{Case of  $ M_0\subset M$}
We now move to the case when $M_0\subset M$ (over-specification). To analyze this case we will introduce additional notation. We decompose the selector matrix $\*q_M=[\*q_{M_0}, \*q_{M_0^C}]$, where $\*q_{M_0}\in\mathbb{R}^{k\times m}$ and $\*q_{M_0^C}\in\mathbb{R}^{k\times (g-m)}$, where the latter selector corresponds to $g-m$ excess (over-selected) averages. By using this, we can further decompose 
\begin{align}
	\overline{\+\Gamma}\*q_M=\begin{bmatrix} \overline{\+\Gamma}\*q_{M_0}, \overline{\+\Gamma}\*q_{M_0^C}\end{bmatrix}=\begin{bmatrix} \overline{\+\Gamma}_{M_0}, \overline{\+\Gamma}_{M_0^C}\end{bmatrix}=\overline{\+\Gamma}_M
\end{align}
for short-hard notation. Then, we have 
\begin{align}\label{partial_rep}
\widehat{\*F}_{M_{\*x}}=\overline{\*X}\*q_M=\*F_\*x\overline{\+\Gamma}\*q_M+\overline{\*V}\*q_M=\*F_\*x\begin{bmatrix}\overline{\+\Gamma}_{M_0}, \overline{\+\Gamma}_{M_0^C} \end{bmatrix} + \begin{bmatrix}\overline{\*V}_{M_0}, \overline{\*V}_{M_0^C} \end{bmatrix},
\end{align}
where $\overline{\*V}\*q_M=\overline{\*V}_M$ is partitioned accordingly. Similarly to the case of using the total available set of CAs ($k$), we introduce in spirit of \cite{Karabiyik2017} the following rotation matrix
\begin{align}
	\overline{\*H}_M=\begin{bmatrix} \overline{\+\Gamma}_{M_0}^{-1} & -\overline{\+\Gamma}_{M_0}^{-1}\overline{\+\Gamma}_{M_0^C}\\
		\*0_{(g-m)\times m} & \*I_{g-m}\end{bmatrix}=\begin{bmatrix} \overline{\*H}_{M_0}, \overline{\*H}_{M_0^C} \end{bmatrix},
\end{align}
with the obvious definitions of $\overline{\*H}_{M_0}$ and $\overline{\*H}_{M_0^C}$ and $\mathrm{rank}(\overline{\*H}_M)=g$. In what follows, the post-multiplication by this matrix leads to 
\begin{align}
	\widehat{\*F}_{M_{\*x}}\overline{\*H}_M= \*F_\*x\overline{\+\Gamma}\*q_M\overline{\*H}_M+\overline{\*V}\*q_M\overline{\*H}_M= \begin{bmatrix}\*F_\*x, \*0_{T\times (g-m)} \end{bmatrix} + \begin{bmatrix} \overline{\*V}\*q_M\overline{\*H}_{M_0}, \overline{\*V}\*q_M\overline{\*H}_{M_0^C} \end{bmatrix}.
\end{align}
Finally, we introduce $\*D_M=\mathrm{diag}(\*I_{m}, \sqrt{N}\*I_{g-m})$, such that 
\begin{align}\label{rotated_F_IC}
	\widehat{\*F}^0_{M_{\*x}}=\*F_\*x\overline{\+\Gamma}\*q_M\overline{\*H}_M\*D_M+\overline{\*V}\*q_M\*D_M &= \begin{bmatrix}\*F_\*x, \*0_{T\times (g-m)} \end{bmatrix} + \begin{bmatrix} \overline{\*V}\*q_M\overline{\*H}_{M_0}, \sqrt{N}\overline{\*V}\*q_M\overline{\*H}_{M_0^C} \end{bmatrix}\notag\\
	&=\begin{bmatrix}\*F_\*x, \*0_{T\times (g-m)} \end{bmatrix} + \begin{bmatrix}\overline{\*V}_{M_1}^0,\overline{\*V}_{M_2}^0 \end{bmatrix}\notag\\
	&=\*F_\*x^0+ \begin{bmatrix}\overline{\*V}_{M_1}^0,\overline{\*V}_{M_2}^0 \end{bmatrix}\notag\\
    &=\*F_\*x^0+\overline{\*V}_{M}^0,
\end{align}
where $\left\| \overline{\*V}_{M_1}^0\right\|=O_p(N^{-1/2})$, but $\left\| \overline{\*V}_{M_2}^0\right\|=O_p(1)$, such that the last $g-m$ columns are non-degenerate. \\

\noindent Further, let us introduce $\*G_T=\mathrm{diag}(T^{(1+\tau)/2}\*I_{m}, T^{1/2}\*I_{g-m})$. Then, we work out the asymptotic limit of $\*G_T^{-1}\widehat{\*F}^{0\prime}_{M_{\*x}}\widehat{\*F}^0_{M_{\*x}}\*G_T^{-1}$. Particularly, 
\begin{align}\label{GFFG}
\*G_T^{-1}\widehat{\*F}^{0\prime}_{M_{\*x}}\widehat{\*F}^0_{M_{\*x}}\*G_T^{-1}&=\*G_T^{-1}\*F_\*x^{0\prime}\*F_\*x^0\*G_T^{-1} + \*G_T^{-1}\*F_\*x^{0\prime}\overline{\*V}_{M}^0\*G_T^{-1}+\*G_T^{-1}\overline{\*V}_{M}^{0\prime}\*F_\*x^{0}\*G_T^{-1} +\*G_T^{-1}\overline{\*V}_{M}^{0\prime}\overline{\*V}_{M}^0\*G_T^{-1}\notag\\
    &=\*S + O_p(N^{-1/2}T^{-\tau/2}) + O_p(T^{-1/2}),
\end{align}
where $\*S=\mathrm{diag}(T^{-(1+\tau)}\*F_\*x'\*F_\*x, T^{-1}\overline{\*V}_{M_2}^{0\prime}\overline{\*V}_{M_2}^0)\in \mathbb{R}^{g\times g}$. The rate comes from the leading, but vanishing components of the last three terms on the right-hand side of (\ref{GFFG}). Specifically, 
\begin{align}
    &\left\| T^{-1-\tau/2}\overline{\*V}_{M_2}^{0\prime}\overline{\*V}_{M_1}^{0}\right\|\leq T^{-\tau/2}N^{-1/2}\left\|\*q_M \right\|^2\left\|\overline{\*H}_{M_0} \right\|\left\|\overline{\*H}_{M_0^C} \right\|\left\|NT^{-1}\overline{\*V}'\overline{\*V} \right\|=O_p(N^{-1/2}T^{-\tau/2}),\\
    &\left| T^{-1-\tau/2}\*F_\*x'\overline{\*V}_{M_2}^{0}\right\|\leq \sqrt{N}T^{-1/2}\underbrace{\left\|T^{-(1+\tau)/2}\*F_\*x'\overline{\*V} \right\|}_{O_p(N^{-1/2})}\left\|\*q_M \right\|\left\|\overline{\*H}_{M_0^C} \right\|=O_p(T^{-1/2}), \\
    & \left\|T^{-(1+\tau)}\overline{\*V}^{0\prime}_{M_1}\overline{\*V}^{0}_{M_1} \right\|\leq N^{-1}T^{-\tau}\left\| \*q_M\right\|^2\left\| \overline{\*H}_{M_0}\right\|^2\left\|NT^{-1}\overline{\*V}'\overline{\*V} \right\|=O_p(N^{-1}T^{-\tau}),\\
    &\left| T^{-(1+\tau)}\*F_\*x'\overline{\*V}_{M_1}^{0}\right\|\leq N^{-1/2}T^{-(1+\tau)/2}\underbrace{\left\|\sqrt{N}T^{-(1+\tau)/2}\*F_\*x'\overline{\*V} \right\|}_{O_p(1)}\left\|\*q_M \right\|\left\|\overline{\*H}_{M_0^C} \right\|=O_p(N^{-1/2}T^{-(1+\tau)/2})
\end{align}
by Corollary (\ref{CorollaryA1}). By following the approach in \cite{bai2004estimating}, we see that this implies that 
\begin{align}
    \left\|(\*G_T^{-1}\widehat{\*F}^{0\prime}_{M_{\*x}}\widehat{\*F}^{0}_{M_\*x}\*G_T^{-1})^+-\*S^+ \right\|&=\left\|\*S^+(\*S-\*G_T^{-1}\widehat{\*F}^{0\prime}_{M_{\*x}}\widehat{\*F}^{0}_{M_\*x}\*G_T^{-1})  (\*G_T^{-1}\widehat{\*F}^{0\prime}_{M_{\*x}}\widehat{\*F}^{0}_{M_\*x}\*G_T^{-1})^+\right\|\notag\\ 
    &\leq \left\|\*S^+ \right\|\left\|\*G_T^{-1}\widehat{\*F}^{0\prime}_{M_{\*x}}\widehat{\*F}^{0}_{M_\*x}\*G_T^{-1}-\*S \right\| \left\|(\*G_T^{-1}\widehat{\*F}^{0\prime}_{M_{\*x}}\widehat{\*F}^{0}_{M_\*x}\*G_T^{-1})^+ \right\|\notag\\
    &=O_p(N^{-1/2}T^{-\tau/2}) + O_p(T^{-1/2}),
\end{align}
so the rate result applies to the MP inverses, as well.
\\

\noindent In what follows, we will use a slightly different path than \cite{de2024cross} in order to sharpen some rates. Let us use the fact that $\*q_{M_0}$ gives the minimal set of averages that asymptotically span the space of $\*F_\*x$. This means that $\mathrm{rank}(\overline{\+\Gamma}\*q_{M_0})=\mathrm{rank}(\overline{\+\Gamma}_{M_0})=m$ and this matrix is $m\times m$. This means that 
\begin{align}\label{solution_F}
\widehat{\*F}_{M_0}=\overline{\*X}\*q_{M_0}=\*F_\*x\overline{\+\Gamma}_{M_0}+\overline{\*V}_{M_0}\Longleftrightarrow \*F_\*x=(\widehat{\*F}_{M_0}-\overline{\*V}_{M_0})\overline{\+\Gamma}_{M_0}^{-1}.
\end{align}
Next, observe that 
\begin{align}
    \overline{\*Q}_{M_0}&=\frac{1}{NT}\sum_{i=1}^N\*X_i'\*M_{\widehat{\*F}_{M_0}}\*X_i=\frac{1}{NT}\sum_{i=1}^N(\*F_\*x\+\Gamma_i+\*V_i)'\*M_{\widehat{\*F}_{M_0}}(\*F_\*x\+\Gamma_i+\*V_i)\notag\\
    &= \frac{1}{NT}\sum_{i=1}^N(\*V_i-\overline{\*V}_{M_0}\overline{\+\Gamma}_{M_0}^{-1}\+\Gamma_i)'\*M_{\widehat{\*F}_{M_0}}(\*V_i-\overline{\*V}_{M_0}\overline{\+\Gamma}_{M_0}^{-1}\+\Gamma_i),
\end{align}
where we used the fact that $\*M_{\widehat{\*F}_{M_0}}\widehat{\*F}_{M_0}=\*0_{T\times m}$. The trick is to recognize that since $M_0\subset M$, we also have that $\*M_{\widehat{\*F}_{M_{\*x}}}\widehat{\*F}_{M_0}=\*0_{T\times m}$, because, based on (\ref{partial_rep}) and the block-wise formula for a projection matrix, we have
\begin{align}
\*M_{\widehat{\*F}_{M_{\*x}}}\widehat{\*F}_{M_0}&=\left(\*I_T-\*P_{\overline{\*X}\*q_{M_0}}-\*P_{\*M_{\overline{\*X}\*q_{M_0}}\overline{\*X}\*q_{M_0^C}}\right)\widehat{\*F}_{M_0}\notag\\
&=\left(\*M_{\overline{\*X}\*q_{M_0}}-\*P_{\*M_{\overline{\*X}\*q_{M_0}}\overline{\*X}\*q_{M_0^C}}\right)\overline{\*X}\*q_{M_0}\notag\\
&=\*M_{\overline{\*X}\*q_{M_0}}\overline{\*X}\*q_{M_0}-\*M_{\overline{\*X}\*q_{M_0}}\overline{\*X}\*q_{M_0^C}(\*q_{M_0^C}'\overline{\*X}'\*M_{\overline{\*X}\*q_{M_0}}\overline{\*X}\*q_{M_0^C})^+\*q_{M_0^C}'\overline{\*X}'\*M_{\overline{\*X}\*q_{M_0}}\overline{\*X}\*q_{M_0}\notag\\
&=\*0_{T\times m}
\end{align}
and so 
\begin{align}
      \overline{\*Q}_{M}&=\frac{1}{NT}\sum_{i=1}^N\*X_i'\*M_{\widehat{\*F}_{M_{\*x}}}\*X_i=\frac{1}{NT}\sum_{i=1}^N(\*F_\*x\+\Gamma_i+\*V_i)'\*M_{\widehat{\*F}_{M_{\*x}}}(\*F_\*x\+\Gamma_i+\*V_i)\notag\\
      &= \frac{1}{NT}\sum_{i=1}^N(\*V_i-\overline{\*V}_{M_0}\overline{\+\Gamma}_{M_0}^{-1}\+\Gamma_i)'\*M_{\widehat{\*F}_{M_{\*x}}}(\*V_i-\overline{\*V}_{M_0}\overline{\+\Gamma}_{M_0}^{-1}\+\Gamma_i).
\end{align}
This implies that by appropriately adding and subtracting, we can write 
\begin{align}
    \overline{\*Q}_{M}-\overline{\*Q}_{M_0}&= \frac{1}{NT}\sum_{i=1}^N(\*V_i-\overline{\*V}_{M_0}\overline{\+\Gamma}_{M_0}^{-1}\+\Gamma_i)'(\*M_{\widehat{\*F}_{M_{\*x}}}-\*M_{\widehat{\*F}_{M_0}})(\*V_i-\overline{\*V}_{M_0}\overline{\+\Gamma}_{M_0}^{-1}\+\Gamma_i)\notag\\
    &=  \frac{1}{NT}\sum_{i=1}^N(\*V_i-\overline{\*V}_{M_0}\overline{\+\Gamma}_{M_0}^{-1}\+\Gamma_i)'(\*M_{\*F_\*x^0}-\*M_{\widehat{\*F}_{M_0}})(\*V_i-\overline{\*V}_{M_0}\overline{\+\Gamma}_{M_0}^{-1}\+\Gamma_i)\notag\\
    &-\frac{1}{NT}\sum_{i=1}^N(\*V_i-\overline{\*V}_{M_0}\overline{\+\Gamma}_{M_0}^{-1}\+\Gamma_i)'(\*M_{\*F_\*x^0}-\*M_{\widehat{\*F}_{M_{\*x}}})(\*V_i-\overline{\*V}_{M_0}\overline{\+\Gamma}_{M_0}^{-1})\notag\\
    &= \frac{1}{NT}\sum_{i=1}^N(\*V_i-\overline{\*V}_{M_0}\overline{\+\Gamma}_{M_0}^{-1}\+\Gamma_i)'(\*M_{\*F_\*x}-\*M_{\widehat{\*F}_{M_0}})(\*V_i-\overline{\*V}_{M_0}\overline{\+\Gamma}_{M_0}^{-1}\+\Gamma_i)\notag\\
    &-\frac{1}{NT}\sum_{i=1}^N(\*V_i-\overline{\*V}_{M_0}\overline{\+\Gamma}_{M_0}^{-1}\+\Gamma_i)'(\*M_{\*F_\*x^0}-\*M_{\widehat{\*F}_{M_{\*x}}})(\*V_i-\overline{\*V}_{M_0}\overline{\+\Gamma}_{M_0}^{-1}\+\Gamma_i)\notag\\
    &=\*I -\mathbf{II},
\end{align}
where in the second-to-last equality we used the MP inverse properties: 
\begin{align}
    &(\mathbf{F}_{\*x}^{0\prime}\mathbf{F}_{\*x}^0)^+ = \left[\begin{array}{cc} \mathbf{F}_\*x'\mathbf{F}_\*x & \mathbf{0}_{m\times (g-m)}\\
\mathbf{0}_{(g-m)\times m} & \mathbf{0}_{(g-m)}\end{array}\right]^+=\left[\begin{array}{cc}(\mathbf{F}_\*x'\mathbf{F}_\*x)^+ & \mathbf{0}_{m\times (g-m)}\\
\mathbf{0}_{(g-m)\times m} & \mathbf{0}_{(g-m)} \end{array}\right],
\end{align}
leading to 
\begin{align}\label{MP_trick}
   \*P_{\*F^0_{\*x}}&= \*F_{\*x}^{0}(\mathbf{F}_{\*x}^{0\prime}\mathbf{F}_{\*x}^0)^+\*F_{\*x}^{0\prime}=\left[\begin{array}{cc}\*F_\*x, \*0_{T\times (g-m)}\end{array}\right]\left[\begin{array}{cc}(\mathbf{F}_\*x'\mathbf{F}_\*x)^+ & \mathbf{0}_{m\times (g-m)}\\
\mathbf{0}_{(g-m)\times m} & \mathbf{0}_{(g-m)} \end{array}\right]\left[\begin{array}{cc} \*F_\*x' \\ \*0_{(g-m)\times T}\end{array}\right]\notag\\
&=\*F_\*x(\*F_\*x'\*F_\*x)^+ \*F_\*x' =\*P_{\*F_\*x},
\end{align}
where $(\*F_\*x'\*F_\*x)^+=(\*F_\*x'\*F_\*x)^{-1}$, which is bounded in probability for finite $T$. Hence, we now need to obtain the order of $\mathbf{II}$, because it dominates $\*I$ due to the case of $m<g$. To achieve this, we start with the decomposition 
\begin{align}\label{M-M}
    \*M_{\*F_\*x^0}-\*M_{\widehat{\*F}_{M_{\*x}}}&=T^{-1}\overline{\*V}_{M_2}^{0}(T^{-1}\overline{\*V}_{M_2}^{0\prime}\overline{\*V}_{M_2}^{0})^+\overline{\*V}_{M_2}^{0\prime}+T^{-(1+\tau)}\overline{\*V}_{M_1}^{0}(T^{-(1+\tau)}\*F_{\*x}'\*F_\*x)^+\overline{\*V}_{M_1}^{0\prime}\notag\\
    &+ T^{-(1+\tau)}\overline{\*V}_{M_1}^{0}(T^{-(1+\tau)}\*F_{\*x}'\*F_\*x)^+\*F_\*x'+T^{-(1+\tau)}\*F_\*x(T^{-(1+\tau)}\*F_{\*x}'\*F_\*x)^+\overline{\*V}_{M_1}^{0\prime}\notag\\
    &+\widehat{\*F}^{0}_{M_\*x}\*G_T^{-1}((\*G_T^{-1}\widehat{\*F}^{0\prime}_{M_{\*x}}\widehat{\*F}^{0}_{M_\*x}\*G_T^{-1})^+-\*S^+)\*G_T^{-1}\widehat{\*F}^{0\prime}_{M_{\*x}}.
\end{align}
Hence, if we insert (\ref{M-M}) into $\mathbf{II}$, we obtain $\mathbf{II}=\mathbf{II}^1-\mathbf{II}^2-\mathbf{II}^3+\mathbf{II}^4$, and so 
\begin{align}\label{II1}
    \left\| \mathbf{II}^1\right\|&=\left\|\frac{1}{NT}\sum_{i=1}^N\*V_i'(\*M_{\*F_\*x^0}-\*M_{\widehat{\*F}_{M_{\*x}}})\*V_i \right\|\notag\\
    &\leq \frac{1}{N}\sum_{i=1}^N\left\|T^{-1}\*V_i'\overline{\*V}_{M_2}^{0} \right\|^2\left\|(T^{-1}\overline{\*V}_{M_2}^{0\prime}\overline{\*V}_{M_2}^{0})^+ \right\|+T^{-\tau}\frac{1}{N}\sum_{i=1}^N\left\|T^{-1}\*V_i' \overline{\*V}_{M_1}^{0}\right\|^2\left\|(T^{-(1+\tau)}\*F_{\*x}'\*F_\*x)^+ \right\|\notag\\
    &+2\frac{1}{N}\sum_{i=1}^N\left\|T^{-1}\*V_i' \overline{\*V}_{M_1}^{0}\right\|\left\|(T^{-(1+\tau)}\*F_{\*x}'\*F_\*x)^+ \right\|\left\|T^{-(1+\tau)}\*F_\*x'\*V_i \right\|\notag\\
    &+\underbrace{\frac{1}{N}\sum_{i=1}^N\left\|T^{-1/2}\*V_i' \widehat{\*F}^{0}_{M_\*x}\*G_T^{-1}\right\|^2\left\|(\*G_T^{-1}\widehat{\*F}^{0\prime}_{M_{\*x}}\widehat{\*F}^{0}_{M_\*x}\*G_T^{-1})^+-\*S^+ \right\|}_{=o_p(C_{N,T}^{-2})}\notag\\
    &= O_p(N^{-1})+ O_p(T^{-1})\notag\\
    &=O_p(C_{N,T}^{-2}),
\end{align}
which is driven by the first term, because $\left\|T^{-1}\*V_i'\overline{\*V}_{M_2}^{0} \right\|=O_p(N^{-1/2})+O_p(T^{-1/2}) $, whereas 
\begin{align}
    \left\|T^{-1/2}\*V_i' \widehat{\*F}^{0}_{M_\*x}\*G_T^{-1} \right\|&\leq \left\| T^{-1}\*V_i'\overline{\*V}^0_M\right\|\left\|\sqrt{T}\*G_T^{-1} \right\| + T^{-1/2}\left\|T^{-(1+\tau)/2}\*V_i'\*F_\*x \right\|\notag\\
    &=O_p(N^{-1/2})+O_p(T^{-1/2}) =O_p(C_{N,T}^{-1}), 
\end{align}
and so the last term in (\ref{II1}) is $o_p(C_{N,T}^{-2})$. The other terms are clearly dominated. Next, we go to 
\begin{align}\label{II2}
    \left\|\mathbf{II}^2\right\|&=\left\|\frac{1}{NT}\sum_{i=1}^N\+\Gamma_i'\overline{\+\Gamma}_{M_0}^{-1\prime}\overline{\*V}_{M_0}^{0\prime}(\*M_{\*F_\*x^0}-\*M_{\widehat{\*F}_{M_{\*x}}})\*V_i \right\|\notag\\
    &\leq \underbrace{\frac{1}{N}\sum_{i=1}^N \left\|\+\Gamma_i'\overline{\+\Gamma}_{M_0}^{-1\prime} \right\| \left\| T^{-1}\overline{\*V}_{M_0}^{0\prime}\overline{\*V}_{M_2}^{0}\right\|\left\| (T^{-1}\overline{\*V}_{M_2}^{0\prime}\overline{\*V}_{M_2}^{0})^+\right\|\left\|T^{-1}\*V_i'\overline{\*V}_{M_2}^{0} \right\|}_{=O_p(C_{N,T}^{-2})}\notag\\
    &+T^{-\tau}\frac{1}{N}\sum_{i=1}^N\left\|\+\Gamma_i'\overline{\+\Gamma}_{M_0}^{-1\prime} \right\|\left\|T^{-1}\overline{\*V}_{M_0}^{0\prime}\overline{\*V}_{M_1}^{0} \right\|\left\|(T^{-(1+\tau)}\*F_{\*x}'\*F_\*x)^+ \right\|\left\|T^{-1}\*V_i' \overline{\*V}_{M_1}^{0}\right\|\notag\\
    &+\underbrace{\frac{1}{N}\sum_{i=1}^N\left\|\+\Gamma_i'\overline{\+\Gamma}_{M_0}^{-1\prime} \right\|\left\|T^{-1}\overline{\*V}_{M_0}^{0\prime}\overline{\*V}_{M_1}^{0} \right\|\left\|(T^{-(1+\tau)}\*F_{\*x}'\*F_\*x)^+ \right\|\left\| T^{-(1+\tau)}\*F_\*x'\*V_i\right\|}_{=O_p(N^{-1}T^{-(1+\tau)/2})}\notag\\
    &+\frac{1}{N}\sum_{i=1}^N\left\|\+\Gamma_i'\overline{\+\Gamma}_{M_0}^{-1\prime} \right\|\left\|T^{-(1+\tau)}\overline{\*V}_{M_0}^{0\prime} \*F_\*x\right\|\left\|(T^{-(1+\tau)}\*F_{\*x}'\*F_\*x)^+ \right\|\left\|T^{-1}\overline{\*V}_{M_1}^{0\prime}\*V_i \right\|\notag\\
    &+\frac{1}{N}\sum_{i=1}^N\left\|\+\Gamma_i'\overline{\+\Gamma}_{M_0}^{-1\prime} \right\|\left\|T^{-1/2}\overline{\*V}_{M_0}^{0\prime}\widehat{\*F}^{0}_{M_\*x}\*G_T^{-1} \right|\left\| (\*G_T^{-1}\widehat{\*F}^{0\prime}_{M_{\*x}}\widehat{\*F}^{0}_{M_\*x}\*G_T^{-1})^+-\*S^+\right\|\left\|T^{-1/2}\*V_i' \widehat{\*F}^{0}_{M_\*x}\*G_T^{-1}\right\|\notag\\
    &=O_p(C_{N,T}^{-2}), 
\end{align}
since 
\begin{align}
    \left\|T^{-1/2}\overline{\*V}_{M_0}^{0\prime}\widehat{\*F}^{0}_{M_\*x}\*G_T^{-1} \right\|&\leq (NT)^{-1/2}\left\|T^{-(1+\tau)/2}\sqrt{N}\overline{\*V}_{M_0}^{0\prime}\*F_{\*x} \right\|+N^{-1/2}\left\|\sqrt{T}\*G_T^{-1} \right\|\left\|T^{-1}\sqrt{N}\overline{\*V}_{M_0}^{0\prime}\overline{\*V}_{M}^{0} \right\|\notag\\
    &=O_p(N^{-1/2}),
\end{align}
and so the overall order is dominated by the first and third terms in (\ref{II2}). Further, note that $\mathbf{II}^3$ is just a transpose of $\mathbf{II}^2$ and so 
\begin{align}
     \left\|\mathbf{II}^3\right\|&=\left\|\frac{1}{NT}\sum_{i=1}^N\*V_i'(\*M_{\*F_\*x^0}-\*M_{\widehat{\*F}_{M_{\*x}}})\overline{\*V}_{M_0}^{0}\overline{\+\Gamma}_{M_0}^{-1}\+\Gamma_i \right\|=O_p(C_{N,T}^{-2}).
\end{align}
Eventually we move on to 
\begin{align}
      \left\|\mathbf{II}^4\right\|&=\left\|\frac{1}{NT}\sum_{i=1}^N\+\Gamma_i'\overline{\+\Gamma}_{M_0}^{-1\prime}\overline{\*V}_{M_0}^{0\prime}(\*M_{\*F_\*x^0}-\*M_{\widehat{\*F}_{M_{\*x}}})\overline{\*V}_{M_0}^{0}\overline{\+\Gamma}_{M_0}^{-1}\+\Gamma_i \right\|\notag\\
      &\underbrace{\leq \frac{1}{N}\sum_{i=1}^N\left\|\+\Gamma_i'\overline{\+\Gamma}_{M_0}^{-1\prime} \right\|^2\left\| T^{-1}\overline{\*V}_{M_0}^{0\prime}\overline{\*V}_{M_2}^{0}\right\|^2\left\| (T^{-1}\overline{\*V}_{M_2}^{0\prime}\overline{\*V}_{M_2}^{0})^+\right\|}_{=O_p(N^{-1})}\notag\\
      &+T^{-\tau}\frac{1}{N}\sum_{i=1}^N\left\|\+\Gamma_i'\overline{\+\Gamma}_{M_0}^{-1\prime} \right\|^2\left\|T^{-1}\overline{\*V}_{M_0}^{0\prime} \overline{\*V}_{M_1}^{0}\right\|^2\left\|(T^{-(1+\tau)}\*F_{\*x}'\*F_\*x)^+ \right\|\notag\\
      &+2\frac{1}{N}\sum_{i=1}^N\left\|\+\Gamma_i'\overline{\+\Gamma}_{M_0}^{-1\prime} \right\|^2\left\|T^{-1}\overline{\*V}_{M_0}^{0\prime} \overline{\*V}_{M_1}^{0}\right\|\left\|(T^{-(1+\tau)}\*F_{\*x}'\*F_\*x)^+ \right\|\left\|T^{-(1+\tau)}\*F_\*x'\overline{\*V}_{M_0}^{0} \right\|\notag\\
      &+\underbrace{\frac{1}{N}\sum_{i=1}^N\left\|\+\Gamma_i'\overline{\+\Gamma}_{M_0}^{-1\prime} \right\|^2\left\|T^{-1/2}\overline{\*V}_{M_0}^{0\prime}\widehat{\*F}^{0}_{M_\*x}\*G_T^{-1} \right\|^2\left\| (\*G_T^{-1}\widehat{\*F}^{0\prime}_{M_{\*x}}\widehat{\*F}^{0}_{M_\*x}\*G_T^{-1})^+-\*S^+\right\|}_{O_p(N^{-3/2}T^{-\tau/2})+O_p(N^{-1}T^{-1/2})}\notag\\
      &= O_p(N^{-1}),
\end{align}
which is decided by the first term, which contributes to the order of $O_p(C_{N,T}^{-2})$ (note that provided that we assume $TN^{-1}=O(1)$ this term itself can also be seen as $O_p(C_{N,T}^{-2})$). Thus, overall we obtain 
\begin{align}
    \left\|\overline{\*Q}_{M}-\overline{\*Q}_{M_0}\right\|&=O_p(C_{N,T}^{-2}).
\end{align}

\noindent Now, we put the results together in a fashion similar to \cite{de2024cross}. In particular, because of (\ref{solution_F}), we can easily demonstrate that $\left\|\overline{\*Q}_{M_0}^{-1} \right\|=O_p(1)$ (see e.g. the analysis of CCEP denominator in e.g. \cite{stauskas2022complete}, where even more general factors are considered). Then 
\begin{align}
    \mathrm{IC}^{DVS}(M)-\mathrm{IC}^{DVS}(M_0)&= \ln\mathrm{det}\left(\*I_k+\left[\overline{\*Q}_{M}-\overline{\*Q}_{M_0}\right]\overline{\*Q}_{M_0}^{-1} \right)+k(g-m)p_{N,T}\notag\\
    &= \ln\mathrm{det}\left(\*I_k+  O_p(C_{N,T}^{-2})\right) + k(g-m)p_{N,T}\notag\\
    &= \ln\mathrm{det}(\*I_k)+ O_p(C_{N,T}^{-2})+ k(g-m)p_{N,T}\notag\\
    &=O_p(C_{N,T}^{-2})+ k(g-m)p_{N,T},
\end{align}
where the approximation comes from p. 119 in \cite{paulsen1984order}:
\begin{align*}
    \ln\mathrm{det}(\*I+O_p(n^{-1}))=O_p(n^{-1})
\end{align*}
for some integer $n$. However, given that $p_{N,T}^{-1}O_p(C_{N,T}^{-2})=o(1)$, then we have that 
\begin{align}
    p_{N,T}^{-1}(\mathrm{IC}^{DVS}(M)-\mathrm{IC}^{DVS}(M_0))= k(g-m)+o_p(1)
\end{align}
and so 
\begin{align}
    \mathbb{P}\left( \mathrm{IC}^{DVS}(M)-\mathrm{IC}^{DVS}(M_0)<0 \right)= \mathbb{P}\left(  p_{N,T}^{-1}(\mathrm{IC}^{DVS}(M)-\mathrm{IC}^{DVS}(M_0))<0 \right)\to 0,
\end{align}
and so the overspecification risk is eliminated. 
\subsubsection{Proof Changes for IC of \cite{margaritella2023using}}
\paragraph{Homogeneous $\+\beta$}
Note that IC of MW is based on the residuals from the first stage CCE regression under the full set of available $k+1$ CAs (note that $|M|=k+1$ now). In this case, we have $\widehat{\*F}=[\overline{\*y}, \overline{\*X}]=\*F\overline{\*C}+\overline{\*U}$, where now $\overline{\*C}$ is a loading matrix with rank of $m$ even when $N\to \infty$. In particular, let us assume homogeneous $\+\beta$. Then 
\begin{align}\label{upsilonhat}
    \widehat{\+\nu}_i=\*y_i-\*X_i\widehat{\+\beta}=-\*X_i(\widehat{\+\beta}-\+\beta)+\*F\+\gamma_i+\+\varepsilon_i.
\end{align}
We then obtain 
\begin{align}
    \mathrm{IC}^{MW}(M)-\mathrm{IC}^{MW}(M_0)&=\ln \left[1+\left(\frac{1}{NT} \sum_{i=1}^{N} \widehat{\+\nu}_i' (\mathbf{M}_{\widehat{\*F}_M}-\mathbf{M}_{\widehat{\*F}_{M_0}})  \widehat{\+\nu}_i \right)\left(\frac{1}{NT} \sum_{i=1}^{N} \widehat{\+\nu}_i'\mathbf{M}_{\widehat{\*F}_{M_0}} \widehat{\+\nu}_i\right)^{-1}\right]\notag\\
    &+(g-m)\cdot p_{N,T}\notag\\
    &= \ln \left[1+T^\tau\left(\frac{1}{NT^{1+\tau}} \sum_{i=1}^{N} \widehat{\+\nu}_i' (\mathbf{M}_{\widehat{\*F}_M}-\mathbf{M}_{\widehat{\*F}_{M_0}})  \widehat{\+\nu}_i \right)\left(\frac{1}{NT} \sum_{i=1}^{N} \widehat{\+\nu}_i'\mathbf{M}_{\widehat{\*F}_{M_0}} \widehat{\+\nu}_i\right)^{-1}\right]\notag\\
    &+(g-m)\cdot p_{N,T}
\end{align}
where, by inserting (\ref{upsilonhat}) we further analyze 
\begin{align}
    \frac{1}{NT^{1+\tau}} \sum_{i=1}^{N} \widehat{\+\nu}_i' (\mathbf{M}_{\widehat{\*F}_M}-\mathbf{M}_{\widehat{\*F}_{M_0}})  \widehat{\+\nu}_i&=\frac{1}{NT^{1+\tau}}\sum_{i=1}^N(\widehat{\+\beta}-\+\beta)'\*X_i'(\mathbf{M}_{\widehat{\*F}_M}-\mathbf{M}_{\widehat{\*F}_{M_0}})\*X_i(\widehat{\+\beta}-\+\beta)\notag\\
    &-\frac{1}{NT^{1+\tau}}\sum_{i=1}^N(\widehat{\+\beta}-\+\beta)'\*X_i'(\mathbf{M}_{\widehat{\*F}_M}-\mathbf{M}_{\widehat{\*F}_{M_0}})(\*F\+\gamma_i+\+\varepsilon_i)\notag\\
    &-\frac{1}{NT^{1+\tau}}\sum_{i=1}^N(\*F\+\gamma_i+\+\varepsilon_i)'(\mathbf{M}_{\widehat{\*F}_M}-\mathbf{M}_{\widehat{\*F}_{M_0}})\*X_i(\widehat{\+\beta}-\+\beta)\notag\\
    &+\frac{1}{NT^{1+\tau}}\sum_{i=1}^N(\*F\+\gamma_i+\+\varepsilon_i)'(\mathbf{M}_{\widehat{\*F}_M}-\mathbf{M}_{\widehat{\*F}_{M_0}})(\*F\+\gamma_i+\+\varepsilon_i).
\end{align}
Clearly,
\begin{align}
    &\left\|\frac{1}{NT^{1+\tau}}\sum_{i=1}^N(\widehat{\+\beta}-\+\beta)'\*X_i'(\mathbf{M}_{\widehat{\*F}_M}-\mathbf{M}_{\widehat{\*F}_{M_0}})\*X_i(\widehat{\+\beta}-\+\beta) \right\|\\
    &\leq (NT)^{-1}\underbrace{\left\|\sqrt{NT}(\widehat{\+\beta}-\+\beta) \right\|^2}_{O_p(1)}\left\|\frac{1}{NT^{1+\tau}}\sum_{i=1}^N\*X_i'(\mathbf{M}_{\widehat{\*F}_M}-\mathbf{M}_{\widehat{\*F}_{M_0}})\*X_i \right\|\notag\\
    &=\begin{cases}
        O_p((NT)^{-1})=o_p(1),\hspace{1mm}\text{under}\hspace{1mm}M\subset M_0\\
        o_p(C_{N,T}^{-2})=o_p(1), \hspace{1mm}\text{under}\hspace{1mm}M_0\subset M,
    \end{cases}
\end{align}
because the rate of $\left\|\frac{1}{NT^{1+\tau}}\sum_{i=1}^N\*X_i'(\mathbf{M}_{\widehat{\*F}_M}-\mathbf{M}_{\widehat{\*F}_{M_0}})\*X_i \right\| $ was already derived for the IC of DVS under both cases (either $O_p(1)$ or vanishing), and its behavior will not change when we use the full set of averages $\widehat{\*F}=[\overline{\*y}, \overline{\*X}]$. Also, $\left\|\sqrt{NT}(\widehat{\+\beta}-\+\beta)\right\|=O_p(1)$ under general factors (see \citealp{westerlund2018cce}), and thus the $\sqrt{NT}$-consistency will hold under our assumptions, as well. Similarly,
\begin{align}
    \left\|\frac{1}{NT^{1+\tau}}\sum_{i=1}^N(\widehat{\+\beta}-\+\beta)'\*X_i'(\mathbf{M}_{\widehat{\*F}_M}-\mathbf{M}_{\widehat{\*F}_{M_0}})(\*F\+\gamma_i+\+\varepsilon_i)\right\|&\leq (NT)^{-1/2}\left\|\sqrt{NT} (\widehat{\+\beta}-\+\beta)\right\|\notag\\
    &\times\left\| \frac{1}{NT^{1+\tau}}\sum_{i=1}^N\*X_i'(\mathbf{M}_{\widehat{\*F}_M}-\mathbf{M}_{\widehat{\*F}_{M_0}})(\*F\+\gamma_i+\+\varepsilon_i)\right\|\notag\\
    & =\begin{cases}
        O_p((NT)^{-1/2}),\hspace{1mm}\text{under}\hspace{1mm}M\subset M_0\\
        o_p(C_{N,T}^{-2})=o_p(1), \hspace{1mm}\text{under}\hspace{1mm}M_0\subset M,
        \end{cases}
\end{align}
because asymptotically $\*F\+\gamma_i+\+\varepsilon_i$ behaves identically to $\*X_i=\*F\+\Gamma_i+\*V_i$, and thus the dominating component will have the same rate. Eventually, by following the same analysis of $\mathbf{I},\mathbf{II}$ and $\mathbf{III}$ in (\ref{I-II-III}), we have 
\begin{align}
    \left\|\frac{1}{NT^{1+\tau}}\sum_{i=1}^N(\*F\+\gamma_i+\+\varepsilon_i)'(\mathbf{M}_{\widehat{\*F}_M}-\mathbf{M}_{\widehat{\*F}_{M_0}})(\*F\+\gamma_i+\+\varepsilon_i) \right\|=\begin{cases}
O_p(1),\hspace{1mm}\text{under}\hspace{1mm}M\subset M_0\\
O_p(C_{N,T}^{-2})=o_p(1), \hspace{1mm}\text{under}\hspace{1mm}M_0\subset M.
 \end{cases}
\end{align}
In fact, we can demonstrate that in analogy to the analysis of (\ref{I-II-III}), we can decompose and simplify the term:
\begin{align}\label{MW_exp1}
    &\frac{1}{NT^{1+\tau}}\sum_{i=1}^N(\*F\+\gamma_i+\+\varepsilon_i)'(\mathbf{M}_{\widehat{\*F}_M}-\mathbf{M}_{\widehat{\*F}_{M_0}})(\*F\+\gamma_i+\+\varepsilon_i)\notag\\
    & = \frac{1}{NT^{1+\tau}}\sum_{i=1}^N(\*F\+\gamma_i+\+\varepsilon_i)'(\*M_{\widehat{\*F}_{M}}-\*M_{\*F\overline{\+\Gamma}\*q_M})(\*F\+\gamma_i+\+\varepsilon_i)\notag\\
    &-\frac{1}{NT^{1+\tau}}\sum_{i=1}^N(\*F\+\gamma_i+\+\varepsilon_i)'(\*M_{\widehat{\*F}_{M_0}}-\*M_{\*F\overline{\+\Gamma}\*q_{M_0}})(\*F\+\gamma_i+\+\varepsilon_i)\notag\\
    &+\frac{1}{NT^{1+\tau}}\sum_{i=1}^N(\*F\+\gamma_i+\+\varepsilon_i)'(\*M_{\*F\overline{\+\Gamma}\*q_M}-\*M_{\*F\overline{\+\Gamma}\*q_{M_0}})(\*F\+\gamma_i+\+\varepsilon_i)\notag\\
    &=\frac{1}{NT^{1+\tau}}\sum_{i=1}^N(\*F\+\gamma_i+\+\varepsilon_i)'(\*M_{\*F\overline{\+\Gamma}\*q_M}-\*M_{\*F\overline{\+\Gamma}\*q_{M_0}})(\*F\+\gamma_i+\+\varepsilon_i)+o_p(T^{-\tau})\notag\\
    &= \frac{1}{NT^{1+\tau}}\sum_{i=1}^N(\*F\+\gamma_i+\+\varepsilon_i)'\*P_{\*M_{\*F\overline{\+\Gamma}\*q_{M}}\*F\overline{\+\Gamma}\*q_{M^C}}(\*F\+\gamma_i+\+\varepsilon_i) + o_p(T^{-\tau}) \notag\\
    &= \frac{1}{NT^{1+\tau}}\sum_{i=1}^N\+\gamma_i'\*F'\*P_{\*M_{\*F\overline{\+\Gamma}\*q_{M}}\*F\overline{\+\Gamma}\*q_{M^C}}\*F\+\gamma_i + o_p(T^{-\tau}) \notag\\
    &=c_{NT}+o_p(T^{-\tau})
\end{align}
 for a strictly positive $c_{NT}$ with $c_{NT}\to c>0$ under $M\subset M_0$. \\
 
 \noindent To complete, we will assume that $\tau \in (0,0.5)$, in order to isolate $o_p(T^{-\tau})$ terms. Then we have 
 \begin{align}
      \frac{1}{NT^{1+\tau}} \sum_{i=1}^{N} \widehat{\+\nu}_i' (\mathbf{M}_{\widehat{\*F}_M}-\mathbf{M}_{\widehat{\*F}_{M_0}})  \widehat{\+\nu}_i=c_{NT}+o_p(T^{-\tau}),
 \end{align}
and, because we can show that $b_{NT}:=\frac{1}{NT} \sum_{i=1}^{N} \widehat{\+\nu}_i'\mathbf{M}_{\widehat{\*F}_{M_0}} \widehat{\+\nu}_i$ converges to a positive scalar quantity, we can analogously show that 
\begin{align}
      (\mathrm{IC}^{MW}(M)&-\mathrm{IC}^{MW}(M_0))/\ln(T)=\ln(1+T^{\tau}(c_{NT}\times b_{NT}^{-1}+o_p(T^{-\tau}))/\ln(T)+o(1)\notag\\
      &=\ln(1+T^\tau c_{NT}\times b_{NT}^{-1})/\ln(T) + o_p(1)\notag\\
      &=\ln(T^\tau c\times b_{NT}^{-1})/\ln(T)+\left(\ln(1+T^\tau c_{NT}\times b_{NT}^{-1})-\ln(T^\tau c_{NT}\times b_{NT}^{-1})) \right)/\ln(T)\notag\\
      &+o_p(1)\notag\\
      &=\ln(T^\tau c_{NT}\times b_{NT}^{-1})/\ln(T)+\frac{1}{\ln(T)}\ln\left(1+\frac{b_{NT}}{cT^\tau} \right) + o_p(1)\notag\\
      &= \tau + \ln(c_{NT}\times b_{NT}^{-1})/\ln(T)+\frac{1}{\ln(T)}\left(\frac{b_{NT}}{c_{NT}T^\tau}+O(T^{-2\tau}) \right)+o_p(1)\notag\\
      &\to_p \tau >0
\end{align}
as $(N,T)\to \infty$. Therefore, 
\begin{align}
    \mathbb{P}(\mathrm{IC}^{MW}(M)-\mathrm{IC}^{MW}(M_0)<0)=\mathbb{P}\left( \frac{\mathrm{IC}^{MW}(M)-\mathrm{IC}^{MW}(M_0)}{\ln(T)}<0\right)\to 0
\end{align}
as $(N,T)\to \infty$. Under $M_0\subset M$, the proof goes exactly the same as in \cite{margaritella2023using},and so given the rates declared above and the fact that $p_{N,T}^{-1}O_p(C_{N,T}^{-2})=o(1)$, we obtain 
\begin{align}
    p_{N,T}^{-1}(\mathrm{IC}^{MW}(M)-\mathrm{IC}^{MW}(M_0))= (g-m)+o_p(1)
\end{align}
and so 
\begin{align}
    \mathbb{P}\left( \mathrm{IC}^{MW}(M)-\mathrm{IC}^{MW}(M_0)<0 \right)= \mathbb{P}\left(  p_{N,T}^{-1}(\mathrm{IC}^{MW}(M)-\mathrm{IC}^{MW}(M_0))<0) \right)\to 0,
\end{align}
and the overspecification risk is eliminated. 
\paragraph{Heterogeneous $\+\beta$}
\noindent If the parameter vector is heterogeneous, e.g. $\+\beta_i=\+\beta+\+\upsilon_i$, where $\+\upsilon_i$ is IID and mean-zero, then 
\begin{align}
      \widehat{\+\nu}_i=\*y_i-\*X_i\widehat{\+\beta}&=-\*X_i(\widehat{\+\beta}-\+\beta)+\*X_i\+\upsilon_i+\*F\+\gamma_i+\+\varepsilon_i\notag\\
      &=-\*X_i(\widehat{\+\beta}-\+\beta)+\*F\+\Gamma_i\+\upsilon_i+\*V_i\+\upsilon_i+\*F\+\gamma_i+\+\varepsilon_i\notag \\
      &= -\*X_i(\widehat{\+\beta}-\+\beta)+\*F(\+\Gamma_i\+\upsilon_i+\+\gamma_i) + \*V_i\+\upsilon_i+\+\varepsilon_i\notag\\
      &=-\*X_i(\widehat{\+\beta}-\+\beta)+\*F\+\gamma_i^* + \+\varepsilon_i^*,
\end{align}
with the obvious definitions of $\+\gamma_i^*$ and $\+\varepsilon_i^*$. Note that now $\left\|\sqrt{N}(\widehat{\+\beta}-\+\beta)\right\|=O_p(1)$ under general factors (see e.g. \citealp{stauskas2022complete}), which is a considerably lower rate. This means that now, for example,  
\begin{align}
     \left\|\frac{1}{NT^{1+\tau}}\sum_{i=1}^N(\widehat{\+\beta}-\+\beta)'\*X_i'(\mathbf{M}_{\widehat{\*F}_M}-\mathbf{M}_{\widehat{\*F}_{M_0}})\*X_i(\widehat{\+\beta}-\+\beta) \right\|&\leq N^{-1}\underbrace{\left\|\sqrt{N}(\widehat{\+\beta}-\+\beta) \right\|^2}_{O_p(1)}\left\|\frac{1}{NT^{1+\tau}}\sum_{i=1}^N\*X_i'(\mathbf{M}_{\widehat{\*F}_M}-\mathbf{M}_{\widehat{\*F}_{M_0}})\*X_i \right\|\notag\\
    &=\begin{cases}
        O_p(N^{-1}),\hspace{1mm}\text{under}\hspace{1mm}M\subset M_0\\
        o_p(C_{N,T}^{-2})=o_p(1), \hspace{1mm}\text{under}\hspace{1mm}M_0\subset M,
    \end{cases}
\end{align}
and other terms will behave similarly:
\begin{align}
    &\left\| \frac{1}{NT^{1+\tau}}\sum_{i=1}^N(\widehat{\+\beta}-\+\beta)'\*X_i'(\mathbf{M}_{\widehat{\*F}_M}-\mathbf{M}_{\widehat{\*F}_{M_0}})(\*F\+\gamma_i^*+\+\varepsilon_i^*)\right\|\notag\\
    &\leq N^{-1/2}\left\|\sqrt{N}(\widehat{\+\beta}-\+\beta) \right\|\left\|\frac{1}{NT^{1+\tau}}\sum_{i=1}^N\*X_i'(\mathbf{M}_{\widehat{\*F}_M}-\mathbf{M}_{\widehat{\*F}_{M_0}})(\*F\+\gamma_i^*+\+\varepsilon_i^*) \right\|\notag\\
    &=\begin{cases}
        O_p(N^{-1/2}),\hspace{1mm}\text{under}\hspace{1mm}M\subset M_0\\
        o_p(C_{N,T}^{-2})=o_p(1), \hspace{1mm}\text{under}\hspace{1mm}M_0\subset M.
    \end{cases}
\end{align}
Finally, 
\begin{align}
    &  \left\|\frac{1}{NT^{1+\tau}}\sum_{i=1}^N(\*F\+\gamma_i^*+\+\varepsilon_i^*)'(\mathbf{M}_{\widehat{\*F}_M}-\mathbf{M}_{\widehat{\*F}_{M_0}})(\*F\+\gamma_i^*+\+\varepsilon_i^*) \right\|=\begin{cases}
O_p(1),\hspace{1mm}\text{under}\hspace{1mm}M\subset M_0\\
O_p(C_{N,T}^{-2})=o_p(1), \hspace{1mm}\text{under}\hspace{1mm}M_0\subset M,
 \end{cases}
\end{align}
and, in particular, under $M\subset M_0$:
\begin{align}
    &\frac{1}{NT^{1+\tau}}\sum_{i=1}^N(\*F\+\gamma_i^*+\+\varepsilon_i^*)'(\mathbf{M}_{\widehat{\*F}_M}-\mathbf{M}_{\widehat{\*F}_{M_0}})(\*F\+\gamma_i^*+\+\varepsilon_i^*)\notag\\
    &=\frac{1}{NT^{1+\tau}}\sum_{i=1}^N(\*F\+\gamma_i^*+\+\varepsilon_i^*)'(\*M_{\*F\overline{\+\Gamma}\*q_M}-\*M_{\*F\overline{\+\Gamma}\*q_{M_0}})(\*F\+\gamma_i^*+\+\varepsilon_i^*)+o_p(T^{-\tau})\notag\\
    &= \frac{1}{NT^{1+\tau}}\sum_{i=1}^N(\*F\+\gamma_i^*+\+\varepsilon_i^*)'\*P_{\*M_{\*F\overline{\+\Gamma}\*q_{M}}\*F\overline{\+\Gamma}\*q_{M^C}}(\*F\+\gamma_i^*+\+\varepsilon_i^*) + o_p(T^{-\tau}) \notag\\
    &= \frac{1}{NT^{1+\tau}}\sum_{i=1}^N\+\gamma_i^{*\prime}\*F'\*P_{\*M_{\*F\overline{\+\Gamma}\*q_{M}}\*F\overline{\+\Gamma}\*q_{M^C}}\*F\+\gamma_i^* + o_p(T^{-\tau}) \notag\\
    &=d_{NT}+o_p(T^{-\tau})
\end{align}
with $d_{NT}\to_pd>0$ in analogy to the analysis in (\ref{MW_exp1}). Now, we need to assume $T^{\tau}N^{-1/2}\to 0$ in addition to $\tau \in (0,0.5)$ to isolate the $o_p(T^{-\tau})$ terms. Then in the under-specification case, we have
\begin{align}
     \frac{1}{NT^{1+\tau}} \sum_{i=1}^{N} \widehat{\+\nu}_i' (\mathbf{M}_{\widehat{\*F}_M}-\mathbf{M}_{\widehat{\*F}_{M_0}})  \widehat{\+\nu}_i=d_{NT}+o_p(T^{-\tau}),
\end{align}
and, since again we can show that $e_{NT}:=\frac{1}{NT} \sum_{i=1}^{N} \widehat{\+\nu}_i'\mathbf{M}_{\widehat{\*F}_{M_0}} \widehat{\+\nu}_i$ converges to a positive scalar quantity, we can demonstrate that 
\begin{align}
      (\mathrm{IC}^{MW}(M)&-\mathrm{IC}^{MW}(M_0))/\ln(T)=\ln(1+T^{\tau}(d_{NT}\times e_{NT}^{-1}+o_p(T^{-\tau}))/\ln(T)+o(1)\notag\\
      &=\ln(1+T^\tau d_{NT}\times e_{NT}^{-1})/\ln(T) + o_p(1)\notag\\
      &\to_p \tau>0
      \end{align}
      as $(N,T)\to \infty$. Therefore, 
\begin{align}
    \mathbb{P}(\mathrm{IC}^{MW}(M)-\mathrm{IC}^{MW}(M_0)<0)=\mathbb{P}\left( \frac{\mathrm{IC}^{MW}(M)-\mathrm{IC}^{MW}(M_0)}{\ln(T)}<0\right)\to 0
\end{align}
as $(N,T)\to \infty$. Under $M_0\subset M$, the proof once again the same as in \cite{margaritella2023using},and so given the rates declared above and the fact that $p_{N,T}^{-1}O_p(C_{N,T}^{-2})=o(1)$, we get 
\begin{align}
    p_{N,T}^{-1}(\mathrm{IC}^{MW}(M)-\mathrm{IC}^{MW}(M_0))= (g-m)+o_p(1)
\end{align}
and so 
\begin{align}
    \mathbb{P}\left( \mathrm{IC}^{MW}(M)-\mathrm{IC}^{MW}(M_0)<0 \right)= \mathbb{P}\left(  p_{N,T}^{-1}(\mathrm{IC}^{MW}(M)-\mathrm{IC}^{MW}(M_0))<0) \right)\to 0,
\end{align}
as expected. Note that the restrictions on $\tau$ discussed here are just sufficient conditions, since we do not see such need in the simulationd exercises.
\subsection{Additional Monte Carlo Results}
\subsubsection{Idiosyncratics Independent Over Time}

\begin{table}[H]
    \centering
    \resizebox{0.8\columnwidth}{!}{%
    \begin{tabular}{cc|cccc|cccc|cccc}
    \hline\hline
    & & \multicolumn{4}{c}{Stationary $\*F$} & \multicolumn{8}{c}{Non-Stationary $\*F$} \\
    & & \multicolumn{4}{c}{$\tau = 0$} & \multicolumn{4}{c}{$\tau = 0.4$} & \multicolumn{4}{c}{$\tau = 0.9$} \\
    $N$ & $T$ & $IC^{MW}_1$ & $IC^{MW}_2$ & $IC^{DVS}_1$ & $IC^{DVS}_2$ &$IC^{MW}_1$ & $IC^{MW}_2$ & $IC^{DVS}_1$ & $IC^{DVS}_2$&$IC^{MW}_1$ & $IC^{MW}_2$ & $IC^{DVS}_1$ & $IC^{DVS}_2$  \\ \hline %
    \multicolumn{14}{c}{Correct Selection Frequency for $g$} \\ \hline
    50 & 50 & 31.00 & 15.90 & 36.40 & 16.10 & 7.30 & 3.60 & 3.50 & 0.30 & 0.10 & 0.00 & 0.00 & 0.00 \\
100 & 50 & 97.30 & 95.00 & 99.20 & 96.40 & 91.50 & 82.80 & 95.10 & 85.90 & 11.50 & 8.60 & 13.40 & 7.50 \\
200 & 50 & 95.50 & 94.30 & 99.70 & 98.80 & 48.60 & 45.50 & 77.70 & 69.20 & 0.20 & 0.20 & 5.50 & 4.10 \\
300 & 50 & 100.00 & 100.00 & 99.60 & 99.50 & 81.30 & 78.20 & 69.90 & 67.70 & 4.30 & 3.70 & 1.80 & 1.60 \\
500 & 50 & 100.00 & 100.00 & 100.00 & 100.00 & 99.60 & 99.40 & 96.30 & 95.10 & 21.20 & 20.10 & 11.50 & 10.10 \\ \hline
50 & 100 & 97.10 & 89.30 & 100.00 & 100.00 & 68.60 & 55.30 & 99.30 & 94.50 & 7.00 & 5.90 & 18.20 & 10.90 \\
100 & 100 & 100.00 & 99.90 & 99.90 & 99.00 & 62.50 & 49.40 & 56.30 & 39.70 & 2.70 & 1.30 & 1.70 & 0.60 \\
200 & 100 & 100.00 & 100.00 & 100.00 & 100.00 & 99.90 & 99.60 & 99.90 & 99.70 & 24.50 & 19.70 & 26.30 & 18.80 \\
300 & 100 & 100.00 & 100.00 & 100.00 & 100.00 & 100.00 & 100.00 & 100.00 & 100.00 & 39.10 & 34.30 & 46.60 & 39.30 \\
500 & 100 & 100.00 & 100.00 & 100.00 & 100.00 & 99.70 & 99.70 & 95.90 & 94.80 & 16.30 & 15.30 & 8.50 & 7.60 \\ \hline
50 & 200 & 99.70 & 99.90 & 99.90 & 99.80 & 97.70 & 96.70 & 89.00 & 79.90 & 9.80 & 8.70 & 5.10 & 3.50 \\
100 & 200 & 100.00 & 100.00 & 100.00 & 100.00 & 93.00 & 88.80 & 99.50 & 98.60 & 5.30 & 4.70 & 12.80 & 7.70 \\
200 & 200 & 100.00 & 100.00 & 100.00 & 100.00 & 100.00 & 100.00 & 100.00 & 100.00 & 40.80 & 32.70 & 60.20 & 44.50 \\
300 & 200 & 100.00 & 100.00 & 100.00 & 100.00 & 100.00 & 100.00 & 100.00 & 100.00 & 22.40 & 18.90 & 81.40 & 73.20 \\
500 & 200 & 100.00 & 100.00 & 100.00 & 100.00 & 100.00 & 100.00 & 100.00 & 100.00 & 54.10 & 49.30 & 80.80 & 75.50 \\ \hline
50 & 300 & 100.00 & 100.00 & 100.00 & 100.00 & 88.60 & 86.40 & 99.90 & 99.40 & 7.40 & 6.20 & 15.20 & 11.00 \\
100 & 300 & 98.90 & 99.80 & 100.00 & 100.00 & 98.70 & 99.40 & 100.00 & 100.00 & 21.30 & 18.60 & 27.20 & 22.40 \\
200 & 300 & 100.00 & 100.00 & 100.00 & 100.00 & 100.00 & 100.00 & 100.00 & 100.00 & 56.00 & 49.10 & 52.60 & 45.60 \\
300 & 300 & 100.00 & 100.00 & 100.00 & 100.00 & 100.00 & 100.00 & 100.00 & 100.00 & 89.30 & 82.90 & 94.40 & 89.10 \\
500 & 300 & 100.00 & 100.00 & 100.00 & 100.00 & 100.00 & 100.00 & 100.00 & 100.00 & 80.10 & 75.70 & 93.70 & 91.90 \\ \hline
50 & 500 & 100.00 & 100.00 & 100.00 & 100.00 & 90.10 & 89.00 & 100.00 & 100.00 & 3.40 & 3.10 & 11.20 & 10.20 \\
100 & 500 & 100.00 & 100.00 & 100.00 & 100.00 & 100.00 & 100.00 & 100.00 & 100.00 & 38.50 & 37.10 & 53.70 & 50.80 \\
200 & 500 & 100.00 & 100.00 & 100.00 & 100.00 & 100.00 & 100.00 & 100.00 & 100.00 & 84.10 & 81.30 & 97.30 & 95.80 \\
300 & 500 & 100.00 & 100.00 & 100.00 & 100.00 & 100.00 & 100.00 & 100.00 & 100.00 & 88.30 & 85.90 & 97.00 & 95.60 \\
500 & 500 & 100.00 & 100.00 & 100.00 & 100.00 & 100.00 & 100.00 & 100.00 & 100.00 & 92.70 & 88.20 & 99.50 & 98.70 \\ \hline \multicolumn{14}{c}{Average of $ g $} \\ \hline
50 & 50 & 3.16 & 2.48 & 3.14 & 1.96 & 2.41 & 2.02 & 2.27 & 1.68 & 1.51 & 1.38 & 1.40 & 1.23 \\
100 & 50 & 3.97 & 3.94 & 3.99 & 3.96 & 3.88 & 3.71 & 3.95 & 3.82 & 2.11 & 1.96 & 2.23 & 1.96 \\
200 & 50 & 3.96 & 3.94 & 3.99 & 3.98 & 3.43 & 3.38 & 3.56 & 3.38 & 1.86 & 1.82 & 1.61 & 1.54 \\
300 & 50 & 4.00 & 4.00 & 4.00 & 4.00 & 3.80 & 3.77 & 3.70 & 3.67 & 2.21 & 2.17 & 2.14 & 2.07 \\
500 & 50 & 4.00 & 4.00 & 4.00 & 4.00 & 4.00 & 3.99 & 3.96 & 3.95 & 2.65 & 2.61 & 2.37 & 2.32 \\ \hline
50 & 100 & 3.96 & 3.86 & 4.00 & 4.00 & 3.65 & 3.34 & 3.99 & 3.92 & 2.08 & 1.95 & 2.31 & 2.03 \\
100 & 100 & 4.00 & 4.00 & 4.00 & 3.99 & 3.45 & 3.19 & 3.51 & 3.21 & 1.45 & 1.33 & 1.73 & 1.52 \\
200 & 100 & 4.00 & 4.00 & 4.00 & 4.00 & 4.00 & 4.00 & 4.00 & 4.00 & 2.69 & 2.54 & 2.62 & 2.41 \\
300 & 100 & 4.00 & 4.00 & 4.00 & 4.00 & 4.00 & 4.00 & 4.00 & 4.00 & 3.14 & 3.05 & 3.21 & 3.06 \\
500 & 100 & 4.00 & 4.00 & 4.00 & 4.00 & 4.00 & 4.00 & 3.96 & 3.95 & 2.63 & 2.58 & 2.76 & 2.70 \\ \hline
50 & 200 & 4.00 & 4.00 & 4.00 & 4.00 & 3.98 & 3.96 & 3.84 & 3.70 & 1.93 & 1.86 & 1.58 & 1.50 \\
100 & 200 & 4.00 & 4.00 & 4.00 & 4.00 & 3.92 & 3.87 & 3.99 & 3.98 & 2.09 & 2.02 & 1.95 & 1.77 \\
200 & 200 & 4.00 & 4.00 & 4.00 & 4.00 & 4.00 & 4.00 & 4.00 & 4.00 & 3.16 & 2.98 & 3.36 & 3.04 \\
300 & 200 & 4.00 & 4.00 & 4.00 & 4.00 & 4.00 & 4.00 & 4.00 & 4.00 & 3.01 & 2.92 & 3.61 & 3.44 \\
500 & 200 & 4.00 & 4.00 & 4.00 & 4.00 & 4.00 & 4.00 & 4.00 & 4.00 & 3.21 & 3.11 & 3.69 & 3.56 \\ \hline
50 & 300 & 4.00 & 4.00 & 4.00 & 4.00 & 3.89 & 3.86 & 4.00 & 3.99 & 2.14 & 2.09 & 2.24 & 2.11 \\
100 & 300 & 4.01 & 4.00 & 4.00 & 4.00 & 4.01 & 4.00 & 4.00 & 4.00 & 2.83 & 2.75 & 2.69 & 2.57 \\
200 & 300 & 4.00 & 4.00 & 4.00 & 4.00 & 4.00 & 4.00 & 4.00 & 4.00 & 3.50 & 3.38 & 3.48 & 3.38 \\
300 & 300 & 4.00 & 4.00 & 4.00 & 4.00 & 4.00 & 4.00 & 4.00 & 4.00 & 3.85 & 3.74 & 3.94 & 3.87 \\
500 & 300 & 4.00 & 4.00 & 4.00 & 4.00 & 4.00 & 4.00 & 4.00 & 4.00 & 3.80 & 3.75 & 3.93 & 3.91 \\ \hline
50 & 500 & 4.00 & 4.00 & 4.00 & 4.00 & 3.90 & 3.89 & 4.00 & 4.00 & 1.92 & 1.89 & 1.98 & 1.94 \\
100 & 500 & 4.00 & 4.00 & 4.00 & 4.00 & 4.00 & 4.00 & 4.00 & 4.00 & 3.08 & 3.04 & 3.29 & 3.22 \\
200 & 500 & 4.00 & 4.00 & 4.00 & 4.00 & 4.00 & 4.00 & 4.00 & 4.00 & 3.81 & 3.77 & 3.97 & 3.95 \\
300 & 500 & 4.00 & 4.00 & 4.00 & 4.00 & 4.00 & 4.00 & 4.00 & 4.00 & 3.87 & 3.83 & 3.97 & 3.96 \\
500 & 500 & 4.00 & 4.00 & 4.00 & 4.00 & 4.00 & 4.00 & 4.00 & 4.00 & 3.90 & 3.83 & 3.99 & 3.98 \\
\hline\hline
    \end{tabular}}
    \caption{IC with Cross-Sectional Averages -Average of $g$ with $m = 4$ and $K \in \{8,9\}$. DVS criteria from \cite{de2024cross}, MW from \cite{margaritella2023using}, see (2.1) and (2.2) in the main text. For $DVS_1$ and $MW_1$, $p_{N,T} = \frac{N+T}{NT}\ln(\frac{NT}{N+T})$; for  $DVS_2$ and $MW_2$, $p_{N,T} = \frac{N+T}{NT}\ln(C^2_{N,T})$ with $C_{N,T} = \min(\sqrt{N},\sqrt{T})$. Idiosyncratics in $\*x_{i,t}$, $\*v_{i,t}$, and $\varepsilon_{i,t}$ are uncorrelated over time, but weakly correlated across units, see (3.2) in the main text.}
    \label{tab:my_label}
\end{table}

\begin{table}[H]
     \centering
    \resizebox{0.8\columnwidth}{!}{%
    \begin{tabular}{cc| cccc| cccc| cccc}
    \hline\hline
    & & \multicolumn{4}{c}{Stationary $\*F$} & \multicolumn{8}{c}{Non-Stationary $\*F$} \\
    & & \multicolumn{4}{c}{$\tau = 0$} & \multicolumn{4}{c}{$\tau = 0.4$} & \multicolumn{4}{c}{$\tau = 0.9$} \\
    $N$ & $T$ & $MW_1$ & $MW_2$ & $DVS_1$ & $DVS_2$ &$MW_1$ & $MW_2$ & $DVS_1$ & $DVS_2$&$MW_1$ & $MW_2$ & $DVS_1$ & $DVS_2$  \\\hline
     \multicolumn{14}{c}{Correct Selection Frequency for $g$} \\ \hline
    50 & 50 & 82.60 & 74.70 & 37.60 & 16.30 & 26.30 & 17.90 & 2.60 & 0.40 & 0.10 & 0.00 & 0.00 & 0.00 \\
100 & 50 & 100.00 & 100.00 & 99.20 & 95.20 & 99.90 & 99.70 & 94.10 & 83.60 & 42.00 & 33.80 & 7.20 & 4.30 \\
200 & 50 & 100.00 & 100.00 & 99.60 & 98.70 & 99.90 & 99.90 & 75.70 & 67.40 & 28.20 & 24.80 & 2.30 & 1.40 \\
300 & 50 & 100.00 & 100.00 & 99.60 & 99.50 & 93.90 & 92.80 & 69.70 & 67.00 & 8.80 & 7.40 & 1.50 & 1.10 \\
500 & 50 & 100.00 & 100.00 & 100.00 & 100.00 & 100.00 & 100.00 & 96.20 & 94.90 & 49.80 & 47.70 & 9.70 & 8.50 \\ \hline
50 & 100 & 100.00 & 100.00 & 100.00 & 99.40 & 100.00 & 100.00 & 97.20 & 89.80 & 42.40 & 32.10 & 7.50 & 3.90 \\
100 & 100 & 100.00 & 100.00 & 99.90 & 98.40 & 90.50 & 79.90 & 47.90 & 31.80 & 2.80 & 1.70 & 0.40 & 0.20 \\
200 & 100 & 100.00 & 100.00 & 100.00 & 100.00 & 100.00 & 100.00 & 99.90 & 99.70 & 78.90 & 72.30 & 19.30 & 12.90 \\
300 & 100 & 100.00 & 100.00 & 100.00 & 100.00 & 100.00 & 100.00 & 100.00 & 100.00 & 85.50 & 83.20 & 41.80 & 36.10 \\
500 & 100 & 100.00 & 100.00 & 100.00 & 100.00 & 100.00 & 99.80 & 96.00 & 94.70 & 27.10 & 25.20 & 7.40 & 6.80 \\ \hline
50 & 200 & 100.00 & 100.00 & 100.00 & 99.60 & 100.00 & 100.00 & 85.50 & 73.00 & 59.30 & 53.30 & 0.60 & 0.50 \\
100 & 200 & 100.00 & 100.00 & 100.00 & 100.00 & 100.00 & 100.00 & 99.50 & 98.40 & 49.80 & 42.90 & 5.50 & 3.80 \\
200 & 200 & 100.00 & 100.00 & 100.00 & 100.00 & 100.00 & 100.00 & 100.00 & 100.00 & 95.20 & 92.20 & 49.70 & 36.30 \\
300 & 200 & 100.00 & 100.00 & 100.00 & 100.00 & 100.00 & 100.00 & 100.00 & 100.00 & 98.80 & 96.90 & 77.10 & 67.70 \\
500 & 200 & 100.00 & 100.00 & 100.00 & 100.00 & 100.00 & 100.00 & 100.00 & 100.00 & 99.80 & 99.70 & 76.40 & 69.70 \\ \hline
50 & 300 & 100.00 & 100.00 & 100.00 & 100.00 & 100.00 & 100.00 & 99.50 & 98.90 & 58.50 & 55.40 & 4.80 & 4.00 \\
100 & 300 & 100.00 & 100.00 & 100.00 & 100.00 & 100.00 & 100.00 & 100.00 & 99.90 & 85.50 & 82.80 & 16.70 & 12.50 \\
200 & 300 & 100.00 & 100.00 & 100.00 & 100.00 & 100.00 & 100.00 & 100.00 & 100.00 & 91.50 & 88.50 & 48.10 & 40.50 \\
300 & 300 & 100.00 & 100.00 & 100.00 & 100.00 & 100.00 & 100.00 & 100.00 & 100.00 & 99.90 & 99.70 & 91.40 & 85.60 \\
500 & 300 & 100.00 & 100.00 & 100.00 & 100.00 & 100.00 & 100.00 & 100.00 & 100.00 & 99.90 & 99.70 & 93.10 & 90.90 \\ \hline
50 & 500 & 100.00 & 100.00 & 100.00 & 100.00 & 100.00 & 100.00 & 99.80 & 99.80 & 17.40 & 16.00 & 3.00 & 2.30 \\
100 & 500 & 100.00 & 100.00 & 100.00 & 100.00 & 100.00 & 100.00 & 100.00 & 100.00 & 95.70 & 95.00 & 40.70 & 37.10 \\
200 & 500 & 100.00 & 100.00 & 100.00 & 100.00 & 100.00 & 100.00 & 100.00 & 100.00 & 99.90 & 99.90 & 95.70 & 94.20 \\
300 & 500 & 100.00 & 100.00 & 100.00 & 100.00 & 100.00 & 100.00 & 100.00 & 100.00 & 100.00 & 100.00 & 95.80 & 94.20 \\
500 & 500 & 100.00 & 100.00 & 100.00 & 100.00 & 100.00 & 100.00 & 100.00 & 100.00 & 100.00 & 100.00 & 99.30 & 97.80 \\ \hline \multicolumn{14}{c}{Average of $ g $} \\ \hline
50 & 50 & 3.83 & 3.74 & 3.07 & 1.87 & 3.15 & 2.97 & 2.19 & 1.63 & 1.85 & 1.71 & 1.32 & 1.20 \\
100 & 50 & 4.00 & 4.00 & 3.99 & 3.94 & 4.00 & 4.00 & 3.93 & 3.78 & 3.11 & 2.90 & 2.02 & 1.80 \\
200 & 50 & 4.00 & 4.00 & 3.99 & 3.97 & 4.00 & 4.00 & 3.52 & 3.35 & 2.48 & 2.38 & 1.58 & 1.52 \\
300 & 50 & 4.00 & 4.00 & 4.00 & 4.00 & 3.94 & 3.93 & 3.70 & 3.67 & 2.82 & 2.77 & 2.09 & 2.03 \\
500 & 50 & 4.00 & 4.00 & 4.00 & 4.00 & 4.00 & 4.00 & 3.96 & 3.95 & 3.38 & 3.35 & 2.33 & 2.27 \\ \hline
50 & 100 & 4.00 & 4.00 & 4.00 & 3.99 & 4.00 & 4.00 & 3.97 & 3.85 & 3.02 & 2.77 & 2.07 & 1.86 \\
100 & 100 & 4.00 & 4.00 & 4.00 & 3.98 & 3.90 & 3.79 & 3.40 & 3.07 & 1.85 & 1.70 & 1.45 & 1.33 \\
200 & 100 & 4.00 & 4.00 & 4.00 & 4.00 & 4.00 & 4.00 & 4.00 & 4.00 & 3.69 & 3.58 & 2.48 & 2.27 \\
300 & 100 & 4.00 & 4.00 & 4.00 & 4.00 & 4.00 & 4.00 & 4.00 & 4.00 & 3.84 & 3.81 & 3.15 & 3.01 \\
500 & 100 & 4.00 & 4.00 & 4.00 & 4.00 & 4.00 & 4.00 & 3.96 & 3.95 & 3.20 & 3.17 & 2.69 & 2.63 \\ \hline
50 & 200 & 4.00 & 4.00 & 4.00 & 4.00 & 4.00 & 4.00 & 3.77 & 3.55 & 3.20 & 3.08 & 1.45 & 1.41 \\
100 & 200 & 4.00 & 4.00 & 4.00 & 4.00 & 4.00 & 4.00 & 3.99 & 3.98 & 3.10 & 2.93 & 1.82 & 1.71 \\
200 & 200 & 4.00 & 4.00 & 4.00 & 4.00 & 4.00 & 4.00 & 4.00 & 4.00 & 3.94 & 3.89 & 3.19 & 2.90 \\
300 & 200 & 4.00 & 4.00 & 4.00 & 4.00 & 4.00 & 4.00 & 4.00 & 4.00 & 3.98 & 3.94 & 3.54 & 3.35 \\
500 & 200 & 4.00 & 4.00 & 4.00 & 4.00 & 4.00 & 4.00 & 4.00 & 4.00 & 4.00 & 4.00 & 3.62 & 3.48 \\ \hline
50 & 300 & 4.00 & 4.00 & 4.00 & 4.00 & 4.00 & 4.00 & 4.00 & 3.99 & 3.40 & 3.33 & 1.96 & 1.90 \\
100 & 300 & 4.00 & 4.00 & 4.00 & 4.00 & 4.00 & 4.00 & 4.00 & 4.00 & 3.78 & 3.73 & 2.53 & 2.42 \\
200 & 300 & 4.00 & 4.00 & 4.00 & 4.00 & 4.00 & 4.00 & 4.00 & 4.00 & 3.92 & 3.88 & 3.43 & 3.32 \\
300 & 300 & 4.00 & 4.00 & 4.00 & 4.00 & 4.00 & 4.00 & 4.00 & 4.00 & 4.00 & 4.00 & 3.91 & 3.83 \\
500 & 300 & 4.00 & 4.00 & 4.00 & 4.00 & 4.00 & 4.00 & 4.00 & 4.00 & 4.00 & 4.00 & 3.92 & 3.89 \\ \hline
50 & 500 & 4.00 & 4.00 & 4.00 & 4.00 & 4.00 & 4.00 & 4.00 & 4.00 & 2.42 & 2.38 & 1.75 & 1.71 \\
100 & 500 & 4.00 & 4.00 & 4.00 & 4.00 & 4.00 & 4.00 & 4.00 & 4.00 & 3.95 & 3.94 & 3.04 & 2.95 \\
200 & 500 & 4.00 & 4.00 & 4.00 & 4.00 & 4.00 & 4.00 & 4.00 & 4.00 & 4.00 & 4.00 & 3.95 & 3.93 \\
300 & 500 & 4.00 & 4.00 & 4.00 & 4.00 & 4.00 & 4.00 & 4.00 & 4.00 & 4.00 & 4.00 & 3.96 & 3.94 \\
500 & 500 & 4.00 & 4.00 & 4.00 & 4.00 & 4.00 & 4.00 & 4.00 & 4.00 & 4.00 & 4.00 & 3.99 & 3.97 \\
\hline\hline
    \end{tabular}}
    \caption{IC with Cross-Sectional Averages -Average of $g$ with $m = 4$ and $K \in \{8,9\}$ with generated factors, $\overline{\mathbf{Z}}_M = \mathbf{F}\*q_M$. DVS criteria from \cite{de2024cross}, MW from \cite{margaritella2023using}, see (2.1) and (2.2) in the main text. For $DVS_1$ and $MW_1$, $p_{N,T} = \frac{N+T}{NT}\ln(\frac{NT}{N+T})$; for  $DVS_2$ and $MW_2$, $p_{N,T} = \frac{N+T}{NT}\ln(C^2_{N,T})$ with $C_{N,T} = \min(\sqrt{N},\sqrt{T})$. Idiosyncratics in $\*x_{i,t}$, $\*v_{i,t}$, and $\varepsilon_{i,t}$ are uncorrelated over time, but weakly correlated across units, see (3.2) in the main text. }
    \label{tab:my_label}
\end{table}

\clearpage
\subsubsection{Non-Stationary Idiosyncratics in $\*X$}
\begin{table}[H]
    \centering
        \resizebox{0.8\columnwidth}{!}{%
    \begin{tabular}{cc| cccc| cccc| cccc}
    \hline\hline
    & & \multicolumn{4}{c}{Stationary $\*F$} & \multicolumn{8}{c}{Non-Stationary $\*F$} \\
    & & \multicolumn{4}{c}{$\tau = 0$} & \multicolumn{4}{c}{$\tau = 0.4$} & \multicolumn{4}{c}{$\tau = 0.9$} \\
    $N$ & $T$ & $MW_1$ & $MW_2$ & $DVS_1$ & $DVS_2$ &$MW_1$ & $MW_2$ & $DVS_1$ & $DVS_2$&$MW_1$ & $MW_2$ & $DVS_1$ & $DVS_2$  \\\hline
     \multicolumn{14}{c}{Correct Selection Frequency for $g$} \\ \hline
    50 & 50 & 31.00 & 15.90 & 36.40 & 16.10 & 7.40 & 3.60 & 3.50 & 0.30 & 0.10 & 0.00 & 0.00 & 0.00 \\
100 & 50 & 97.30 & 95.00 & 99.20 & 96.40 & 91.50 & 82.40 & 95.10 & 86.20 & 12.20 & 9.10 & 13.90 & 7.60 \\
200 & 50 & 95.50 & 94.30 & 99.70 & 98.80 & 48.70 & 45.60 & 78.30 & 69.70 & 0.20 & 0.20 & 5.80 & 4.30 \\
300 & 50 & 100.00 & 100.00 & 99.60 & 99.50 & 81.20 & 78.20 & 69.70 & 67.50 & 3.90 & 3.40 & 1.60 & 1.40 \\
500 & 50 & 100.00 & 100.00 & 100.00 & 100.00 & 99.60 & 99.40 & 96.00 & 94.60 & 21.80 & 20.50 & 11.60 & 9.90 \\ \hline
50 & 100 & 97.10 & 89.30 & 100.00 & 100.00 & 68.20 & 54.70 & 99.40 & 94.40 & 7.00 & 5.70 & 18.30 & 10.80 \\
100 & 100 & 100.00 & 99.90 & 99.90 & 99.00 & 62.90 & 49.40 & 55.80 & 38.80 & 2.50 & 1.20 & 1.60 & 0.40 \\
200 & 100 & 100.00 & 100.00 & 100.00 & 100.00 & 99.90 & 99.60 & 99.90 & 99.70 & 24.20 & 19.20 & 25.80 & 18.60 \\
300 & 100 & 100.00 & 100.00 & 100.00 & 100.00 & 100.00 & 100.00 & 100.00 & 100.00 & 39.50 & 34.70 & 46.90 & 39.00 \\
500 & 100 & 100.00 & 100.00 & 100.00 & 100.00 & 99.70 & 99.70 & 95.50 & 94.50 & 16.40 & 15.20 & 8.40 & 7.40 \\ \hline
50 & 200 & 99.70 & 99.90 & 99.90 & 99.80 & 97.80 & 96.70 & 88.70 & 79.60 & 9.90 & 8.80 & 5.50 & 3.70 \\
100 & 200 & 100.00 & 100.00 & 100.00 & 100.00 & 93.10 & 88.70 & 99.60 & 98.80 & 5.10 & 4.60 & 12.70 & 7.70 \\
200 & 200 & 100.00 & 100.00 & 100.00 & 100.00 & 100.00 & 100.00 & 100.00 & 100.00 & 41.00 & 33.00 & 60.40 & 44.50 \\
300 & 200 & 100.00 & 100.00 & 100.00 & 100.00 & 100.00 & 100.00 & 100.00 & 100.00 & 22.50 & 19.20 & 81.00 & 72.40 \\
500 & 200 & 100.00 & 100.00 & 100.00 & 100.00 & 100.00 & 100.00 & 100.00 & 100.00 & 54.10 & 49.20 & 80.70 & 75.40 \\ \hline
50 & 300 & 100.00 & 100.00 & 100.00 & 100.00 & 88.90 & 86.70 & 100.00 & 99.50 & 7.10 & 6.00 & 14.10 & 10.20 \\
100 & 300 & 98.90 & 99.80 & 100.00 & 100.00 & 98.80 & 99.50 & 100.00 & 100.00 & 21.10 & 18.40 & 26.90 & 22.10 \\
200 & 300 & 100.00 & 100.00 & 100.00 & 100.00 & 100.00 & 100.00 & 100.00 & 100.00 & 55.90 & 49.00 & 52.10 & 45.30 \\
300 & 300 & 100.00 & 100.00 & 100.00 & 100.00 & 100.00 & 100.00 & 100.00 & 100.00 & 88.50 & 82.10 & 94.20 & 88.90 \\
500 & 300 & 100.00 & 100.00 & 100.00 & 100.00 & 100.00 & 100.00 & 100.00 & 100.00 & 79.90 & 75.40 & 94.20 & 92.20 \\ \hline
50 & 500 & 100.00 & 100.00 & 100.00 & 100.00 & 89.70 & 88.70 & 100.00 & 100.00 & 3.40 & 3.00 & 10.60 & 9.40 \\
100 & 500 & 100.00 & 100.00 & 100.00 & 100.00 & 100.00 & 100.00 & 100.00 & 100.00 & 39.30 & 37.80 & 53.00 & 50.00 \\
200 & 500 & 100.00 & 100.00 & 100.00 & 100.00 & 100.00 & 100.00 & 100.00 & 100.00 & 84.00 & 81.00 & 97.40 & 96.00 \\
300 & 500 & 100.00 & 100.00 & 100.00 & 100.00 & 100.00 & 100.00 & 100.00 & 100.00 & 89.00 & 86.50 & 97.20 & 95.90 \\
500 & 500 & 100.00 & 100.00 & 100.00 & 100.00 & 100.00 & 100.00 & 100.00 & 100.00 & 92.50 & 87.90 & 99.40 & 98.40 \\ \hline \multicolumn{14}{c}{Average of $ g $} \\ \hline
50 & 50 & 3.16 & 2.48 & 3.14 & 1.96 & 2.43 & 2.04 & 2.28 & 1.70 & 1.50 & 1.37 & 1.39 & 1.23 \\
100 & 50 & 3.97 & 3.94 & 3.99 & 3.96 & 3.88 & 3.70 & 3.95 & 3.82 & 2.13 & 1.96 & 2.25 & 1.96 \\
200 & 50 & 3.96 & 3.94 & 3.99 & 3.98 & 3.44 & 3.38 & 3.57 & 3.39 & 1.88 & 1.84 & 1.62 & 1.55 \\
300 & 50 & 4.00 & 4.00 & 4.00 & 4.00 & 3.80 & 3.77 & 3.70 & 3.67 & 2.19 & 2.15 & 2.12 & 2.06 \\
500 & 50 & 4.00 & 4.00 & 4.00 & 4.00 & 4.00 & 3.99 & 3.96 & 3.94 & 2.65 & 2.61 & 2.38 & 2.32 \\ \hline
50 & 100 & 3.96 & 3.86 & 4.00 & 4.00 & 3.64 & 3.33 & 3.99 & 3.92 & 2.06 & 1.94 & 2.30 & 2.02 \\
100 & 100 & 4.00 & 4.00 & 4.00 & 3.99 & 3.46 & 3.19 & 3.50 & 3.20 & 1.44 & 1.32 & 1.73 & 1.52 \\
200 & 100 & 4.00 & 4.00 & 4.00 & 4.00 & 4.00 & 4.00 & 4.00 & 4.00 & 2.69 & 2.54 & 2.62 & 2.42 \\
300 & 100 & 4.00 & 4.00 & 4.00 & 4.00 & 4.00 & 4.00 & 4.00 & 4.00 & 3.15 & 3.06 & 3.22 & 3.06 \\
500 & 100 & 4.00 & 4.00 & 4.00 & 4.00 & 4.00 & 4.00 & 3.96 & 3.94 & 2.63 & 2.58 & 2.75 & 2.69 \\ \hline
50 & 200 & 4.00 & 4.00 & 4.00 & 4.00 & 3.98 & 3.96 & 3.84 & 3.69 & 1.94 & 1.86 & 1.60 & 1.51 \\
100 & 200 & 4.00 & 4.00 & 4.00 & 4.00 & 3.92 & 3.87 & 3.99 & 3.98 & 2.09 & 2.01 & 1.95 & 1.76 \\
200 & 200 & 4.00 & 4.00 & 4.00 & 4.00 & 4.00 & 4.00 & 4.00 & 4.00 & 3.16 & 2.98 & 3.37 & 3.05 \\
300 & 200 & 4.00 & 4.00 & 4.00 & 4.00 & 4.00 & 4.00 & 4.00 & 4.00 & 3.01 & 2.92 & 3.60 & 3.42 \\
500 & 200 & 4.00 & 4.00 & 4.00 & 4.00 & 4.00 & 4.00 & 4.00 & 4.00 & 3.22 & 3.11 & 3.69 & 3.56 \\ \hline
50 & 300 & 4.00 & 4.00 & 4.00 & 4.00 & 3.89 & 3.87 & 4.00 & 3.99 & 2.12 & 2.07 & 2.21 & 2.09 \\
100 & 300 & 4.01 & 4.00 & 4.00 & 4.00 & 4.01 & 4.00 & 4.00 & 4.00 & 2.83 & 2.75 & 2.68 & 2.56 \\
200 & 300 & 4.00 & 4.00 & 4.00 & 4.00 & 4.00 & 4.00 & 4.00 & 4.00 & 3.49 & 3.38 & 3.48 & 3.38 \\
300 & 300 & 4.00 & 4.00 & 4.00 & 4.00 & 4.00 & 4.00 & 4.00 & 4.00 & 3.83 & 3.73 & 3.94 & 3.87 \\
500 & 300 & 4.00 & 4.00 & 4.00 & 4.00 & 4.00 & 4.00 & 4.00 & 4.00 & 3.79 & 3.75 & 3.93 & 3.91 \\ \hline
50 & 500 & 4.00 & 4.00 & 4.00 & 4.00 & 3.90 & 3.88 & 4.00 & 4.00 & 1.93 & 1.89 & 1.97 & 1.92 \\
100 & 500 & 4.00 & 4.00 & 4.00 & 4.00 & 4.00 & 4.00 & 4.00 & 4.00 & 3.09 & 3.05 & 3.29 & 3.22 \\
200 & 500 & 4.00 & 4.00 & 4.00 & 4.00 & 4.00 & 4.00 & 4.00 & 4.00 & 3.81 & 3.77 & 3.97 & 3.95 \\
300 & 500 & 4.00 & 4.00 & 4.00 & 4.00 & 4.00 & 4.00 & 4.00 & 4.00 & 3.87 & 3.84 & 3.97 & 3.96 \\
500 & 500 & 4.00 & 4.00 & 4.00 & 4.00 & 4.00 & 4.00 & 4.00 & 4.00 & 3.90 & 3.83 & 3.99 & 3.98 \\
\hline\hline
    \end{tabular}}
    \caption{IC with Cross-Sectional Averages -Average of $g$ with $m = 4$ and $K \in \{8,9\}$. DVS criteria from \cite{de2024cross}, MW from \cite{margaritella2023using}, see (2.1) and (2.2) in the main text. For $DVS_1$ and $MW_1$, $p_{N,T} = \frac{N+T}{NT}\ln(\frac{NT}{N+T})$; for  $DVS_2$ and $MW_2$, $p_{N,T} = \frac{N+T}{NT}\ln(C^2_{N,T})$ with $C_{N,T} = \min(\sqrt{N},\sqrt{T})$. Idiosyncratics in $\*x_{i,t}$, $\*v_{i,t}$ are generated same as factors. $\varepsilon_{i,t}$ are uncorrelated over time, but weakly correlated across units, see (3.2) in the main text. }
    \label{tab:my_label}
\end{table}

\begin{table}[H]
    \centering
      \resizebox{0.8\columnwidth}{!}{%
    \begin{tabular}{cc| cccc| cccc| cccc}
    \hline\hline
    & & \multicolumn{4}{c}{Stationary $\*F$} & \multicolumn{8}{c}{Non-Stationary $\*F$} \\
    & & \multicolumn{4}{c}{$\tau = 0$} & \multicolumn{4}{c}{$\tau = 0.4$} & \multicolumn{4}{c}{$\tau = 0.9$} \\
    $N$ & $T$ & $MW_1$ & $MW_2$ & $DVS_1$ & $DVS_2$ &$MW_1$ & $MW_2$ & $DVS_1$ & $DVS_2$&$MW_1$ & $MW_2$ & $DVS_1$ & $DVS_2$  \\\hline
     \multicolumn{14}{c}{Correct Selection Frequency for $g$} \\ \hline
    50 & 50 & 82.60 & 74.70 & 37.60 & 16.30 & 26.70 & 18.10 & 2.60 & 0.40 & 0.10 & 0.00 & 0.00 & 0.00 \\
100 & 50 & 100.00 & 100.00 & 99.20 & 95.20 & 99.90 & 99.70 & 93.90 & 83.80 & 42.20 & 33.70 & 7.40 & 4.20 \\
200 & 50 & 100.00 & 100.00 & 99.60 & 98.70 & 99.90 & 99.90 & 76.20 & 68.00 & 29.00 & 25.90 & 2.60 & 1.60 \\
300 & 50 & 100.00 & 100.00 & 99.60 & 99.50 & 93.40 & 92.50 & 69.50 & 66.80 & 8.60 & 7.30 & 1.30 & 1.00 \\
500 & 50 & 100.00 & 100.00 & 100.00 & 100.00 & 100.00 & 100.00 & 95.80 & 94.50 & 50.70 & 48.70 & 9.80 & 8.50 \\ \hline
50 & 100 & 100.00 & 100.00 & 100.00 & 99.40 & 100.00 & 100.00 & 97.20 & 89.40 & 42.50 & 32.60 & 7.70 & 3.90 \\
100 & 100 & 100.00 & 100.00 & 99.90 & 98.40 & 90.30 & 80.00 & 47.40 & 31.50 & 2.50 & 1.40 & 0.20 & 0.10 \\
200 & 100 & 100.00 & 100.00 & 100.00 & 100.00 & 100.00 & 100.00 & 99.90 & 99.70 & 79.20 & 72.40 & 18.80 & 12.80 \\
300 & 100 & 100.00 & 100.00 & 100.00 & 100.00 & 100.00 & 100.00 & 100.00 & 100.00 & 85.30 & 83.00 & 42.00 & 36.00 \\
500 & 100 & 100.00 & 100.00 & 100.00 & 100.00 & 100.00 & 99.80 & 95.60 & 94.40 & 27.30 & 25.30 & 7.20 & 6.80 \\ \hline
50 & 200 & 100.00 & 100.00 & 100.00 & 99.60 & 100.00 & 100.00 & 85.20 & 72.80 & 59.80 & 53.80 & 0.60 & 0.50 \\
100 & 200 & 100.00 & 100.00 & 100.00 & 100.00 & 100.00 & 100.00 & 99.60 & 98.50 & 50.20 & 43.00 & 5.40 & 3.80 \\
200 & 200 & 100.00 & 100.00 & 100.00 & 100.00 & 100.00 & 100.00 & 100.00 & 100.00 & 94.90 & 92.00 & 49.50 & 36.40 \\
300 & 200 & 100.00 & 100.00 & 100.00 & 100.00 & 100.00 & 100.00 & 100.00 & 100.00 & 98.70 & 97.00 & 76.60 & 66.70 \\
500 & 200 & 100.00 & 100.00 & 100.00 & 100.00 & 100.00 & 100.00 & 100.00 & 100.00 & 99.80 & 99.70 & 76.10 & 69.60 \\ \hline
50 & 300 & 100.00 & 100.00 & 100.00 & 100.00 & 100.00 & 100.00 & 99.60 & 99.00 & 57.90 & 54.80 & 4.60 & 3.90 \\
100 & 300 & 100.00 & 100.00 & 100.00 & 100.00 & 100.00 & 100.00 & 100.00 & 99.90 & 84.50 & 81.80 & 16.30 & 12.00 \\
200 & 300 & 100.00 & 100.00 & 100.00 & 100.00 & 100.00 & 100.00 & 100.00 & 100.00 & 91.50 & 88.60 & 47.70 & 40.60 \\
300 & 300 & 100.00 & 100.00 & 100.00 & 100.00 & 100.00 & 100.00 & 100.00 & 100.00 & 99.80 & 99.60 & 91.40 & 85.30 \\
500 & 300 & 100.00 & 100.00 & 100.00 & 100.00 & 100.00 & 100.00 & 100.00 & 100.00 & 99.90 & 99.60 & 93.60 & 91.10 \\ \hline
50 & 500 & 100.00 & 100.00 & 100.00 & 100.00 & 100.00 & 100.00 & 99.80 & 99.80 & 17.00 & 15.60 & 2.70 & 2.10 \\
100 & 500 & 100.00 & 100.00 & 100.00 & 100.00 & 100.00 & 100.00 & 100.00 & 100.00 & 95.70 & 95.00 & 39.80 & 36.10 \\
200 & 500 & 100.00 & 100.00 & 100.00 & 100.00 & 100.00 & 100.00 & 100.00 & 100.00 & 99.90 & 99.90 & 95.90 & 94.40 \\
300 & 500 & 100.00 & 100.00 & 100.00 & 100.00 & 100.00 & 100.00 & 100.00 & 100.00 & 100.00 & 100.00 & 96.20 & 94.70 \\
500 & 500 & 100.00 & 100.00 & 100.00 & 100.00 & 100.00 & 100.00 & 100.00 & 100.00 & 100.00 & 100.00 & 99.20 & 97.70 \\ \hline \multicolumn{14}{c}{Average of $ g $} \\ \hline
50 & 50 & 3.83 & 3.74 & 3.07 & 1.87 & 3.15 & 2.98 & 2.20 & 1.64 & 1.85 & 1.70 & 1.31 & 1.20 \\
100 & 50 & 4.00 & 4.00 & 3.99 & 3.94 & 4.00 & 4.00 & 3.93 & 3.78 & 3.12 & 2.90 & 2.03 & 1.79 \\
200 & 50 & 4.00 & 4.00 & 3.99 & 3.97 & 4.00 & 4.00 & 3.53 & 3.36 & 2.49 & 2.41 & 1.60 & 1.53 \\
300 & 50 & 4.00 & 4.00 & 4.00 & 4.00 & 3.93 & 3.92 & 3.69 & 3.66 & 2.82 & 2.77 & 2.08 & 2.02 \\
500 & 50 & 4.00 & 4.00 & 4.00 & 4.00 & 4.00 & 4.00 & 3.96 & 3.94 & 3.39 & 3.35 & 2.34 & 2.28 \\ \hline
50 & 100 & 4.00 & 4.00 & 4.00 & 3.99 & 4.00 & 4.00 & 3.97 & 3.85 & 3.02 & 2.77 & 2.07 & 1.85 \\
100 & 100 & 4.00 & 4.00 & 4.00 & 3.98 & 3.90 & 3.79 & 3.40 & 3.07 & 1.85 & 1.69 & 1.44 & 1.32 \\
200 & 100 & 4.00 & 4.00 & 4.00 & 4.00 & 4.00 & 4.00 & 4.00 & 4.00 & 3.70 & 3.59 & 2.48 & 2.27 \\
300 & 100 & 4.00 & 4.00 & 4.00 & 4.00 & 4.00 & 4.00 & 4.00 & 4.00 & 3.84 & 3.81 & 3.16 & 3.02 \\
500 & 100 & 4.00 & 4.00 & 4.00 & 4.00 & 4.00 & 4.00 & 3.96 & 3.94 & 3.21 & 3.17 & 2.68 & 2.62 \\ \hline
50 & 200 & 4.00 & 4.00 & 4.00 & 4.00 & 4.00 & 4.00 & 3.76 & 3.55 & 3.21 & 3.09 & 1.45 & 1.41 \\
100 & 200 & 4.00 & 4.00 & 4.00 & 4.00 & 4.00 & 4.00 & 3.99 & 3.98 & 3.11 & 2.94 & 1.82 & 1.72 \\
200 & 200 & 4.00 & 4.00 & 4.00 & 4.00 & 4.00 & 4.00 & 4.00 & 4.00 & 3.93 & 3.89 & 3.19 & 2.91 \\
300 & 200 & 4.00 & 4.00 & 4.00 & 4.00 & 4.00 & 4.00 & 4.00 & 4.00 & 3.98 & 3.94 & 3.53 & 3.33 \\
500 & 200 & 4.00 & 4.00 & 4.00 & 4.00 & 4.00 & 4.00 & 4.00 & 4.00 & 4.00 & 4.00 & 3.61 & 3.48 \\ \hline
50 & 300 & 4.00 & 4.00 & 4.00 & 4.00 & 4.00 & 4.00 & 4.00 & 3.99 & 3.38 & 3.32 & 1.95 & 1.89 \\
100 & 300 & 4.00 & 4.00 & 4.00 & 4.00 & 4.00 & 4.00 & 4.00 & 4.00 & 3.76 & 3.72 & 2.52 & 2.40 \\
200 & 300 & 4.00 & 4.00 & 4.00 & 4.00 & 4.00 & 4.00 & 4.00 & 4.00 & 3.92 & 3.88 & 3.43 & 3.32 \\
300 & 300 & 4.00 & 4.00 & 4.00 & 4.00 & 4.00 & 4.00 & 4.00 & 4.00 & 4.00 & 4.00 & 3.91 & 3.83 \\
500 & 300 & 4.00 & 4.00 & 4.00 & 4.00 & 4.00 & 4.00 & 4.00 & 4.00 & 4.00 & 4.00 & 3.93 & 3.89 \\ \hline
50 & 500 & 4.00 & 4.00 & 4.00 & 4.00 & 4.00 & 4.00 & 4.00 & 4.00 & 2.40 & 2.36 & 1.74 & 1.71 \\
100 & 500 & 4.00 & 4.00 & 4.00 & 4.00 & 4.00 & 4.00 & 4.00 & 4.00 & 3.95 & 3.94 & 3.03 & 2.94 \\
200 & 500 & 4.00 & 4.00 & 4.00 & 4.00 & 4.00 & 4.00 & 4.00 & 4.00 & 4.00 & 4.00 & 3.95 & 3.93 \\
300 & 500 & 4.00 & 4.00 & 4.00 & 4.00 & 4.00 & 4.00 & 4.00 & 4.00 & 4.00 & 4.00 & 3.96 & 3.95 \\
500 & 500 & 4.00 & 4.00 & 4.00 & 4.00 & 4.00 & 4.00 & 4.00 & 4.00 & 4.00 & 4.00 & 3.99 & 3.97 \\
\hline\hline
    \end{tabular}}
    \caption{IC with Cross-Sectional Averages -Average of $g$ with $m = 4$ and $K \in \{8,9\}$ with generated factors, $\overline{\mathbf{Z}}_M = \mathbf{F}\*q_M$. DVS criteria from \cite{de2024cross}, MW from \cite{margaritella2023using}, see (2.1) and (2.2) in the main text. For $DVS_1$ and $MW_1$, $p_{N,T} = \frac{N+T}{NT}\ln(\frac{NT}{N+T})$; for  $DVS_2$ and $MW_2$, $p_{N,T} = \frac{N+T}{NT}\ln(C^2_{N,T})$ with $C_{N,T} = \min(\sqrt{N},\sqrt{T})$. Idiosyncratics in $\*x_{i,t}$, $\*v_{i,t}$ are generated same as factors. $\varepsilon_{i,t}$  are uncorrelated over time, but weakly correlated across units, see (3.2) in the main text.}
    \label{tab:my_label}
\end{table}

\clearpage
\subsubsection{Penalty by \cite{bai2004estimating}}
\begin{table}[H]
    \centering
    \resizebox{0.8\columnwidth}{!}{%
    \begin{tabular}{cc|cccc|cccc|cccc}
    \hline\hline
    & & \multicolumn{4}{c}{Stationary $\*F$} & \multicolumn{8}{c}{Non-Stationary $\*F$} \\
    & & \multicolumn{4}{c}{$\tau = 0$} & \multicolumn{4}{c}{$\tau = 0.4$} & \multicolumn{4}{c}{$\tau = 0.9$} \\
    $N$ & $T$ & $IC^{MW}_1$ & $IC^{MW}_2$ & $IC^{DVS}_1$ & $IC^{DVS}_2$ &$IC^{MW}_1$ & $IC^{MW}_2$ & $IC^{DVS}_1$ & $IC^{DVS}_2$&$IC^{MW}_1$ & $IC^{MW}_2$ & $IC^{DVS}_1$ & $IC^{DVS}_2$  \\\hline
     \multicolumn{14}{c}{Correct Selection Frequency for $g$} \\ \hline
    50 & 50 & 0.00 & 0.00 & 0.00 & 0.00 & 0.00 & 0.00 & 0.00 & 0.00 & 0.00 & 0.00 & 0.00 & 0.00 \\
100 & 50 & 0.00 & 0.00 & 0.00 & 0.00 & 0.00 & 0.00 & 0.00 & 0.00 & 0.00 & 0.00 & 0.00 & 0.00 \\
200 & 50 & 0.00 & 0.00 & 0.00 & 0.00 & 0.00 & 0.00 & 0.00 & 0.00 & 0.00 & 0.00 & 0.00 & 0.00 \\
300 & 50 & 0.00 & 0.00 & 0.00 & 0.00 & 0.00 & 0.00 & 0.00 & 0.00 & 0.00 & 0.00 & 0.00 & 0.00 \\
500 & 50 & 0.00 & 0.00 & 0.00 & 0.00 & 0.10 & 0.10 & 0.00 & 0.00 & 0.00 & 0.00 & 0.00 & 0.00 \\ \hline
50 & 100 & 0.00 & 0.00 & 0.00 & 0.00 & 0.00 & 0.00 & 0.00 & 0.00 & 0.00 & 0.00 & 0.00 & 0.00 \\
100 & 100 & 0.00 & 0.00 & 0.00 & 0.00 & 0.00 & 0.00 & 0.00 & 0.00 & 0.00 & 0.00 & 0.00 & 0.00 \\
200 & 100 & 0.00 & 0.00 & 0.00 & 0.00 & 0.00 & 0.00 & 0.00 & 0.00 & 0.00 & 0.00 & 0.00 & 0.00 \\
300 & 100 & 0.00 & 0.00 & 0.00 & 0.00 & 0.10 & 0.10 & 0.00 & 0.00 & 0.00 & 0.00 & 0.00 & 0.00 \\
500 & 100 & 0.10 & 0.00 & 0.00 & 0.00 & 1.20 & 0.40 & 0.00 & 0.00 & 0.10 & 0.10 & 0.00 & 0.00 \\ \hline
50 & 200 & 0.00 & 0.00 & 0.00 & 0.00 & 0.00 & 0.00 & 0.00 & 0.00 & 0.00 & 0.00 & 0.00 & 0.00 \\
100 & 200 & 0.00 & 0.00 & 0.00 & 0.00 & 0.00 & 0.00 & 0.00 & 0.00 & 0.00 & 0.00 & 0.00 & 0.00 \\
200 & 200 & 0.00 & 0.00 & 0.00 & 0.00 & 0.40 & 0.00 & 0.00 & 0.00 & 0.00 & 0.00 & 0.00 & 0.00 \\
300 & 200 & 2.40 & 0.00 & 0.00 & 0.00 & 8.80 & 2.70 & 0.00 & 0.00 & 0.10 & 0.00 & 0.00 & 0.00 \\
500 & 200 & 26.10 & 4.70 & 0.00 & 0.00 & 12.90 & 7.20 & 0.00 & 0.00 & 0.40 & 0.10 & 0.00 & 0.00 \\ \hline
50 & 300 & 0.00 & 0.00 & 0.00 & 0.00 & 0.00 & 0.00 & 0.00 & 0.00 & 0.00 & 0.00 & 0.00 & 0.00 \\
100 & 300 & 0.00 & 0.00 & 0.00 & 0.00 & 0.00 & 0.00 & 0.00 & 0.00 & 0.00 & 0.00 & 0.00 & 0.00 \\
200 & 300 & 0.00 & 0.00 & 0.00 & 0.00 & 0.00 & 0.00 & 0.00 & 0.00 & 0.10 & 0.10 & 0.00 & 0.00 \\
300 & 300 & 19.30 & 0.00 & 0.00 & 0.00 & 16.00 & 1.90 & 0.00 & 0.00 & 0.90 & 0.70 & 0.00 & 0.00 \\
500 & 300 & 99.90 & 92.90 & 0.00 & 0.00 & 72.70 & 53.60 & 1.20 & 0.00 & 2.70 & 1.30 & 0.00 & 0.00 \\ \hline
50 & 500 & 0.00 & 0.00 & 0.00 & 0.00 & 0.00 & 0.00 & 0.00 & 0.00 & 0.00 & 0.00 & 0.00 & 0.00 \\
100 & 500 & 0.00 & 0.00 & 0.00 & 0.00 & 0.00 & 0.00 & 0.00 & 0.00 & 0.00 & 0.00 & 0.00 & 0.00 \\
200 & 500 & 0.00 & 0.00 & 0.00 & 0.00 & 0.30 & 0.00 & 0.00 & 0.00 & 0.80 & 0.50 & 0.00 & 0.00 \\
300 & 500 & 15.80 & 0.10 & 0.00 & 0.00 & 14.90 & 3.10 & 0.00 & 0.00 & 0.90 & 0.70 & 0.00 & 0.00 \\
500 & 500 & 100.00 & 100.00 & 77.40 & 0.20 & 94.50 & 81.20 & 31.50 & 2.00 & 2.50 & 1.70 & 0.30 & 0.00 \\ \hline \multicolumn{14}{c}{Average of $ g $} \\ \hline
50 & 50 & 1.00 & 1.00 & 1.00 & 1.00 & 1.00 & 1.00 & 1.00 & 1.00 & 1.02 & 1.01 & 1.00 & 1.00 \\
100 & 50 & 1.00 & 1.00 & 1.00 & 1.00 & 1.00 & 1.00 & 1.00 & 1.00 & 1.07 & 1.05 & 1.01 & 1.00 \\
200 & 50 & 1.00 & 1.00 & 1.00 & 1.00 & 1.03 & 1.02 & 1.00 & 1.00 & 1.09 & 1.08 & 1.00 & 1.00 \\
300 & 50 & 1.00 & 1.00 & 1.00 & 1.00 & 1.05 & 1.04 & 1.00 & 1.00 & 1.12 & 1.10 & 1.02 & 1.02 \\
500 & 50 & 1.00 & 1.00 & 1.00 & 1.00 & 1.03 & 1.02 & 1.00 & 1.00 & 1.20 & 1.18 & 1.04 & 1.04 \\ \hline
50 & 100 & 1.00 & 1.00 & 1.00 & 1.00 & 1.00 & 1.00 & 1.00 & 1.00 & 1.01 & 1.01 & 1.00 & 1.00 \\
100 & 100 & 1.00 & 1.00 & 1.00 & 1.00 & 1.07 & 1.04 & 1.00 & 1.00 & 1.01 & 1.00 & 1.00 & 1.00 \\
200 & 100 & 1.00 & 1.00 & 1.00 & 1.00 & 1.09 & 1.06 & 1.00 & 1.00 & 1.12 & 1.10 & 1.01 & 1.00 \\
300 & 100 & 1.00 & 1.00 & 1.00 & 1.00 & 1.05 & 1.03 & 1.00 & 1.00 & 1.21 & 1.18 & 1.06 & 1.04 \\
500 & 100 & 1.01 & 1.01 & 1.00 & 1.00 & 1.26 & 1.21 & 1.01 & 1.00 & 1.16 & 1.15 & 1.06 & 1.05 \\ \hline
50 & 200 & 1.00 & 1.00 & 1.00 & 1.00 & 1.00 & 1.00 & 1.00 & 1.00 & 1.01 & 1.01 & 1.00 & 1.00 \\
100 & 200 & 1.00 & 1.00 & 1.00 & 1.00 & 1.00 & 1.00 & 1.00 & 1.00 & 1.07 & 1.05 & 1.00 & 1.00 \\
200 & 200 & 1.00 & 1.00 & 1.00 & 1.00 & 1.23 & 1.11 & 1.00 & 1.00 & 1.23 & 1.17 & 1.08 & 1.04 \\
300 & 200 & 1.07 & 1.00 & 1.00 & 1.00 & 1.47 & 1.17 & 1.01 & 1.00 & 1.31 & 1.25 & 1.07 & 1.05 \\
500 & 200 & 1.80 & 1.15 & 1.00 & 1.00 & 1.80 & 1.55 & 1.16 & 1.11 & 1.25 & 1.23 & 1.06 & 1.06 \\ \hline
50 & 300 & 1.00 & 1.00 & 1.00 & 1.00 & 1.00 & 1.00 & 1.00 & 1.00 & 1.01 & 1.01 & 1.00 & 1.00 \\
100 & 300 & 1.00 & 1.00 & 1.00 & 1.00 & 1.00 & 1.00 & 1.00 & 1.00 & 1.07 & 1.05 & 1.01 & 1.00 \\
200 & 300 & 1.00 & 1.00 & 1.00 & 1.00 & 1.04 & 1.02 & 1.00 & 1.00 & 1.26 & 1.19 & 1.04 & 1.02 \\
300 & 300 & 1.61 & 1.00 & 1.00 & 1.00 & 1.82 & 1.24 & 1.02 & 1.00 & 1.39 & 1.31 & 1.07 & 1.04 \\
500 & 300 & 4.00 & 3.80 & 1.02 & 1.00 & 3.58 & 3.05 & 1.21 & 1.03 & 1.73 & 1.61 & 1.31 & 1.23 \\ \hline
50 & 500 & 1.00 & 1.00 & 1.00 & 1.00 & 1.00 & 1.00 & 1.00 & 1.00 & 1.01 & 1.01 & 1.00 & 1.00 \\
100 & 500 & 1.00 & 1.00 & 1.00 & 1.00 & 1.00 & 1.00 & 1.00 & 1.00 & 1.08 & 1.07 & 1.00 & 1.00 \\
200 & 500 & 1.09 & 1.06 & 1.00 & 1.00 & 1.01 & 1.00 & 1.00 & 1.00 & 1.30 & 1.25 & 1.08 & 1.06 \\
300 & 500 & 1.58 & 1.05 & 1.00 & 1.00 & 1.78 & 1.22 & 1.05 & 1.01 & 1.47 & 1.40 & 1.18 & 1.13 \\
500 & 500 & 4.00 & 4.00 & 3.65 & 1.06 & 3.90 & 3.60 & 2.52 & 1.52 & 1.67 & 1.56 & 1.28 & 1.19 \\
\hline\hline
    \end{tabular}}
    \caption{IC with Cross-Sectional Averages -Average of $g$ with $m = 4$ and $K \in \{8,9\}$. DVS criteria from \cite{de2024cross}, MW from \cite{margaritella2023using}, see (2.1) and (2.2) in the main text. For $IC^{DVS}_1$ and $IC^{MW}_1$, $\tilde{p}_{N,T} = \frac{N+T}{NT}\ln(\frac{NT}{N+T}) \ln(T)$; for  $IC^{DVS}_2$ and $IC^{MW}_2$, $\widetilde{p}_{N,T}=\ln(T)p_{N,T} $ with $C_{N,T} = \min(\sqrt{N},\sqrt{T})$. Idiosyncratics in $\*x_{i,t}$, $\*v_{i,t}$, and $\varepsilon_{i,t}$  uncorrelated over time, but weakly correlated across units, see (3.2) in the main text. }
    \label{tab:my_label}
\end{table}

\clearpage
\subsubsection{Idiosyncratics Correlated Over Time}
\begin{table}[H]
    \centering
      \resizebox{0.8\columnwidth}{!}{%
    \begin{tabular}{cc| cccc| cccc| cccc}
    \hline\hline
    & & \multicolumn{4}{c}{Stationary $\*F$} & \multicolumn{8}{c}{Non-Stationary $\*F$} \\
    & & \multicolumn{4}{c}{$\tau = 0$} & \multicolumn{4}{c}{$\tau = 0.4$} & \multicolumn{4}{c}{$\tau = 0.9$} \\
    $N$ & $T$ & $MW_1$ & $MW_2$ & $DVS_1$ & $DVS_2$ &$MW_1$ & $MW_2$ & $DVS_1$ & $DVS_2$&$MW_1$ & $MW_2$ & $DVS_1$ & $DVS_2$  \\\hline
     \multicolumn{14}{c}{Correct Selection Frequency for $g$} \\ \hline
    50 & 50 & 74.10 & 28.50 & 98.70 & 72.70 & 11.70 & 6.40 & 12.30 & 4.30 & 0.60 & 0.10 & 0.00 & 0.00 \\
100 & 50 & 94.80 & 93.30 & 99.40 & 97.10 & 94.00 & 89.00 & 99.30 & 97.20 & 23.50 & 15.90 & 43.20 & 26.90 \\
200 & 50 & 96.80 & 95.20 & 99.90 & 99.50 & 64.50 & 61.40 & 97.90 & 95.30 & 1.60 & 1.30 & 24.30 & 19.20 \\
300 & 50 & 100.00 & 100.00 & 100.00 & 100.00 & 90.00 & 88.30 & 91.00 & 89.60 & 10.30 & 8.80 & 12.30 & 10.20 \\
500 & 50 & 100.00 & 100.00 & 100.00 & 100.00 & 99.80 & 99.80 & 99.70 & 99.40 & 47.10 & 45.10 & 50.10 & 47.10 \\ \hline
50 & 100 & 77.30 & 86.90 & 98.80 & 96.10 & 69.10 & 58.00 & 99.90 & 98.70 & 7.70 & 5.80 & 33.00 & 21.60 \\
100 & 100 & 98.90 & 94.90 & 99.60 & 96.80 & 69.20 & 56.10 & 78.20 & 59.70 & 3.00 & 1.20 & 4.00 & 1.10 \\
200 & 100 & 99.70 & 100.00 & 100.00 & 100.00 & 99.90 & 99.70 & 99.90 & 99.90 & 34.70 & 27.80 & 52.90 & 40.20 \\
300 & 100 & 100.00 & 100.00 & 100.00 & 100.00 & 100.00 & 100.00 & 100.00 & 100.00 & 54.60 & 48.80 & 77.30 & 70.90 \\
500 & 100 & 100.00 & 100.00 & 100.00 & 100.00 & 99.90 & 99.90 & 99.20 & 98.90 & 30.40 & 27.40 & 23.50 & 20.50 \\ \hline
50 & 200 & 99.30 & 99.00 & 100.00 & 100.00 & 94.50 & 94.10 & 95.70 & 89.80 & 9.80 & 8.50 & 8.70 & 5.60 \\
100 & 200 & 95.90 & 96.40 & 100.00 & 100.00 & 92.30 & 88.30 & 99.80 & 99.70 & 5.40 & 4.60 & 21.40 & 13.70 \\
200 & 200 & 100.00 & 100.00 & 100.00 & 100.00 & 100.00 & 100.00 & 100.00 & 100.00 & 45.40 & 36.40 & 76.80 & 63.60 \\
300 & 200 & 100.00 & 100.00 & 100.00 & 100.00 & 100.00 & 100.00 & 100.00 & 100.00 & 27.60 & 22.60 & 94.70 & 88.30 \\
500 & 200 & 100.00 & 100.00 & 100.00 & 100.00 & 100.00 & 100.00 & 100.00 & 100.00 & 66.00 & 59.80 & 95.40 & 91.50 \\ \hline
50 & 300 & 99.10 & 99.60 & 100.00 & 100.00 & 85.40 & 82.60 & 100.00 & 99.90 & 6.70 & 5.60 & 18.70 & 15.70 \\
100 & 300 & 100.00 & 100.00 & 100.00 & 100.00 & 95.50 & 98.50 & 100.00 & 100.00 & 19.10 & 17.50 & 36.50 & 29.90 \\
200 & 300 & 100.00 & 100.00 & 100.00 & 100.00 & 100.00 & 100.00 & 100.00 & 100.00 & 56.80 & 50.80 & 63.10 & 55.10 \\
300 & 300 & 100.00 & 100.00 & 100.00 & 100.00 & 100.00 & 100.00 & 100.00 & 100.00 & 90.50 & 84.10 & 98.30 & 95.10 \\
500 & 300 & 100.00 & 100.00 & 100.00 & 100.00 & 100.00 & 100.00 & 100.00 & 100.00 & 85.00 & 79.90 & 98.00 & 96.70 \\ \hline
50 & 500 & 100.00 & 100.00 & 100.00 & 100.00 & 85.10 & 83.10 & 100.00 & 100.00 & 2.80 & 2.30 & 12.30 & 11.30 \\
100 & 500 & 100.00 & 100.00 & 100.00 & 100.00 & 100.00 & 100.00 & 100.00 & 100.00 & 36.80 & 34.40 & 60.60 & 55.60 \\
200 & 500 & 100.00 & 100.00 & 100.00 & 100.00 & 100.00 & 100.00 & 100.00 & 100.00 & 83.10 & 80.80 & 98.60 & 97.80 \\
300 & 500 & 100.00 & 100.00 & 100.00 & 100.00 & 100.00 & 100.00 & 100.00 & 100.00 & 88.60 & 84.70 & 99.10 & 98.10 \\
500 & 500 & 100.00 & 100.00 & 100.00 & 100.00 & 100.00 & 100.00 & 100.00 & 100.00 & 93.60 & 88.70 & 99.90 & 99.70 \\ \hline \multicolumn{14}{c}{Average of $ g $} \\ \hline
50 & 50 & 3.57 & 2.11 & 3.99 & 3.53 & 2.65 & 2.21 & 2.76 & 2.24 & 1.63 & 1.47 & 1.65 & 1.43 \\
100 & 50 & 4.00 & 3.91 & 3.99 & 3.95 & 3.94 & 3.84 & 3.99 & 3.97 & 2.52 & 2.26 & 3.10 & 2.71 \\
200 & 50 & 3.98 & 3.96 & 4.00 & 3.99 & 3.62 & 3.58 & 3.96 & 3.91 & 2.19 & 2.12 & 2.30 & 2.14 \\
300 & 50 & 4.00 & 4.00 & 4.00 & 4.00 & 3.90 & 3.88 & 3.91 & 3.90 & 2.62 & 2.55 & 2.81 & 2.72 \\
500 & 50 & 4.00 & 4.00 & 4.00 & 4.00 & 4.00 & 4.00 & 4.00 & 3.99 & 3.27 & 3.23 & 3.36 & 3.30 \\ \hline
50 & 100 & 4.22 & 4.01 & 3.99 & 3.95 & 3.68 & 3.39 & 4.00 & 3.98 & 2.17 & 2.00 & 2.72 & 2.40 \\
100 & 100 & 3.99 & 3.92 & 4.00 & 3.96 & 3.60 & 3.33 & 3.78 & 3.54 & 1.51 & 1.35 & 1.99 & 1.73 \\
200 & 100 & 4.00 & 4.00 & 4.00 & 4.00 & 4.00 & 4.00 & 4.00 & 4.00 & 2.98 & 2.80 & 3.26 & 2.98 \\
300 & 100 & 4.00 & 4.00 & 4.00 & 4.00 & 4.00 & 4.00 & 4.00 & 4.00 & 3.43 & 3.33 & 3.74 & 3.64 \\
500 & 100 & 4.00 & 4.00 & 4.00 & 4.00 & 4.00 & 4.00 & 3.99 & 3.99 & 3.03 & 2.94 & 3.15 & 3.10 \\ \hline
50 & 200 & 4.00 & 3.99 & 4.00 & 4.00 & 3.98 & 3.95 & 3.94 & 3.86 & 1.93 & 1.85 & 1.74 & 1.62 \\
100 & 200 & 4.00 & 3.97 & 4.00 & 4.00 & 3.92 & 3.87 & 4.00 & 4.00 & 2.11 & 2.03 & 2.21 & 1.98 \\
200 & 200 & 4.00 & 4.00 & 4.00 & 4.00 & 4.00 & 4.00 & 4.00 & 4.00 & 3.25 & 3.05 & 3.66 & 3.42 \\
300 & 200 & 4.00 & 4.00 & 4.00 & 4.00 & 4.00 & 4.00 & 4.00 & 4.00 & 3.11 & 3.01 & 3.89 & 3.75 \\
500 & 200 & 4.00 & 4.00 & 4.00 & 4.00 & 4.00 & 4.00 & 4.00 & 4.00 & 3.43 & 3.32 & 3.93 & 3.88 \\ \hline
50 & 300 & 4.01 & 4.00 & 4.00 & 4.00 & 3.85 & 3.83 & 4.00 & 4.00 & 2.08 & 2.02 & 2.36 & 2.26 \\
100 & 300 & 4.00 & 4.00 & 4.00 & 4.00 & 4.04 & 4.01 & 4.00 & 4.00 & 2.81 & 2.73 & 2.89 & 2.74 \\
200 & 300 & 4.00 & 4.00 & 4.00 & 4.00 & 4.00 & 4.00 & 4.00 & 4.00 & 3.51 & 3.41 & 3.62 & 3.51 \\
300 & 300 & 4.00 & 4.00 & 4.00 & 4.00 & 4.00 & 4.00 & 4.00 & 4.00 & 3.87 & 3.77 & 3.98 & 3.95 \\
500 & 300 & 4.00 & 4.00 & 4.00 & 4.00 & 4.00 & 4.00 & 4.00 & 4.00 & 3.85 & 3.79 & 3.98 & 3.96 \\ \hline
50 & 500 & 4.00 & 4.00 & 4.00 & 4.00 & 3.84 & 3.82 & 4.00 & 4.00 & 1.85 & 1.81 & 2.04 & 1.99 \\
100 & 500 & 4.00 & 4.00 & 4.00 & 4.00 & 4.00 & 4.00 & 4.00 & 4.00 & 3.03 & 2.97 & 3.43 & 3.33 \\
200 & 500 & 4.00 & 4.00 & 4.00 & 4.00 & 4.00 & 4.00 & 4.00 & 4.00 & 3.80 & 3.76 & 3.98 & 3.97 \\
300 & 500 & 4.00 & 4.00 & 4.00 & 4.00 & 4.00 & 4.00 & 4.00 & 4.00 & 3.87 & 3.82 & 3.99 & 3.98 \\
500 & 500 & 4.00 & 4.00 & 4.00 & 4.00 & 4.00 & 4.00 & 4.00 & 4.00 & 3.92 & 3.84 & 4.00 & 4.00 \\
\hline\hline
    \end{tabular}}
    \caption{IC with Cross-Sectional Averages -Average of $g$ with $m = 4$ and $K \in \{8,9\}$. DVS criteria from \cite{de2024cross}, MW from \cite{margaritella2023using}, see (2.1) and (2.2) in the main text. For $DVS_1$ and $MW_1$, $p_{N,T} = \frac{N+T}{NT}\ln(\frac{NT}{N+T})$; for  $DVS_2$ and $MW_2$, $p_{N,T} = \frac{N+T}{NT}\ln(C^2_{N,T})$ with $C_{N,T} = \min(\sqrt{N},\sqrt{T})$. Idiosyncratics in $\*x_{i,t}$, $\*v_{i,t}$, and $\varepsilon_{i,t}$ are weakly correlated time and units units. }
    \label{tab:my_label}
\end{table}

\begin{table}[H]
    \centering
      \resizebox{0.8\columnwidth}{!}{%
    \begin{tabular}{cc| cccc| cccc| cccc}
    \hline\hline
    & & \multicolumn{4}{c}{Stationary $\*F$} & \multicolumn{8}{c}{Non-Stationary $\*F$} \\
    & & \multicolumn{4}{c}{$\tau = 0$} & \multicolumn{4}{c}{$\tau = 0.4$} & \multicolumn{4}{c}{$\tau = 0.9$} \\
    $N$ & $T$ & $MW_1$ & $MW_2$ & $DVS_1$ & $DVS_2$ &$MW_1$ & $MW_2$ & $DVS_1$ & $DVS_2$&$MW_1$ & $MW_2$ & $DVS_1$ & $DVS_2$  \\\hline
     \multicolumn{14}{c}{Correct Selection Frequency for $g$} \\ \hline
    50 & 50 & 100.00 & 99.80 & 97.20 & 61.60 & 36.40 & 26.90 & 13.90 & 4.60 & 0.50 & 0.10 & 0.00 & 0.00 \\
100 & 50 & 100.00 & 99.80 & 99.30 & 96.40 & 99.80 & 99.70 & 99.10 & 96.30 & 63.70 & 53.20 & 38.00 & 23.50 \\
200 & 50 & 100.00 & 100.00 & 99.80 & 99.40 & 100.00 & 100.00 & 98.50 & 95.80 & 56.70 & 50.60 & 24.50 & 18.70 \\
300 & 50 & 100.00 & 100.00 & 100.00 & 100.00 & 97.20 & 96.70 & 91.80 & 89.70 & 29.60 & 26.10 & 12.50 & 10.70 \\
500 & 50 & 100.00 & 100.00 & 100.00 & 100.00 & 100.00 & 100.00 & 99.60 & 99.60 & 77.20 & 74.90 & 50.40 & 46.50 \\ \hline
50 & 100 & 100.00 & 100.00 & 97.10 & 90.80 & 100.00 & 99.90 & 99.70 & 97.10 & 50.10 & 38.70 & 19.60 & 11.20 \\
100 & 100 & 100.00 & 100.00 & 99.50 & 96.50 & 93.10 & 85.10 & 72.10 & 54.20 & 4.60 & 2.40 & 0.70 & 0.30 \\
200 & 100 & 100.00 & 100.00 & 100.00 & 100.00 & 100.00 & 100.00 & 99.90 & 99.90 & 89.00 & 84.60 & 49.20 & 36.50 \\
300 & 100 & 100.00 & 100.00 & 100.00 & 100.00 & 100.00 & 100.00 & 100.00 & 100.00 & 93.80 & 91.10 & 77.30 & 70.00 \\
500 & 100 & 100.00 & 100.00 & 100.00 & 100.00 & 100.00 & 100.00 & 99.20 & 98.90 & 48.90 & 44.30 & 23.30 & 20.40 \\ \hline
50 & 200 & 100.00 & 100.00 & 100.00 & 99.90 & 100.00 & 100.00 & 94.00 & 87.50 & 57.70 & 52.20 & 2.40 & 1.80 \\
100 & 200 & 100.00 & 100.00 & 100.00 & 100.00 & 100.00 & 100.00 & 99.80 & 99.80 & 53.50 & 45.20 & 12.40 & 7.70 \\
200 & 200 & 100.00 & 100.00 & 100.00 & 100.00 & 100.00 & 100.00 & 100.00 & 100.00 & 97.40 & 93.40 & 72.50 & 56.60 \\
300 & 200 & 100.00 & 100.00 & 100.00 & 100.00 & 100.00 & 100.00 & 100.00 & 100.00 & 99.40 & 98.70 & 93.10 & 87.20 \\
500 & 200 & 100.00 & 100.00 & 100.00 & 100.00 & 100.00 & 100.00 & 100.00 & 100.00 & 99.70 & 99.70 & 94.90 & 91.00 \\ \hline
50 & 300 & 100.00 & 100.00 & 100.00 & 100.00 & 100.00 & 100.00 & 100.00 & 99.60 & 54.70 & 50.90 & 7.60 & 6.20 \\
100 & 300 & 100.00 & 100.00 & 100.00 & 100.00 & 100.00 & 100.00 & 100.00 & 100.00 & 84.30 & 80.70 & 26.10 & 19.80 \\
200 & 300 & 100.00 & 100.00 & 100.00 & 100.00 & 100.00 & 100.00 & 100.00 & 100.00 & 92.20 & 89.30 & 61.30 & 52.60 \\
300 & 300 & 100.00 & 100.00 & 100.00 & 100.00 & 100.00 & 100.00 & 100.00 & 100.00 & 99.90 & 99.70 & 97.20 & 93.60 \\
500 & 300 & 100.00 & 100.00 & 100.00 & 100.00 & 100.00 & 100.00 & 100.00 & 100.00 & 100.00 & 99.80 & 97.90 & 96.30 \\ \hline
50 & 500 & 100.00 & 100.00 & 100.00 & 100.00 & 100.00 & 100.00 & 99.90 & 99.90 & 14.70 & 13.60 & 4.00 & 3.50 \\
100 & 500 & 100.00 & 100.00 & 100.00 & 100.00 & 100.00 & 100.00 & 100.00 & 100.00 & 93.90 & 93.20 & 47.60 & 43.90 \\
200 & 500 & 100.00 & 100.00 & 100.00 & 100.00 & 100.00 & 100.00 & 100.00 & 100.00 & 99.90 & 99.90 & 98.20 & 96.60 \\
300 & 500 & 100.00 & 100.00 & 100.00 & 100.00 & 100.00 & 100.00 & 100.00 & 100.00 & 100.00 & 100.00 & 98.50 & 97.00 \\
500 & 500 & 100.00 & 100.00 & 100.00 & 100.00 & 100.00 & 100.00 & 100.00 & 100.00 & 100.00 & 100.00 & 99.90 & 99.50 \\ \hline \multicolumn{14}{c}{Average of $ g $} \\ \hline
50 & 50 & 4.00 & 4.00 & 3.97 & 3.33 & 3.29 & 3.11 & 2.77 & 2.23 & 2.02 & 1.83 & 1.59 & 1.37 \\
100 & 50 & 4.00 & 4.00 & 3.99 & 3.94 & 4.00 & 4.00 & 3.99 & 3.96 & 3.51 & 3.32 & 3.00 & 2.61 \\
200 & 50 & 4.00 & 4.00 & 4.00 & 3.99 & 4.00 & 4.00 & 3.97 & 3.93 & 3.17 & 3.03 & 2.41 & 2.23 \\
300 & 50 & 4.00 & 4.00 & 4.00 & 4.00 & 3.97 & 3.97 & 3.92 & 3.90 & 3.22 & 3.17 & 2.83 & 2.75 \\
500 & 50 & 4.00 & 4.00 & 4.00 & 4.00 & 4.00 & 4.00 & 4.00 & 4.00 & 3.75 & 3.72 & 3.37 & 3.30 \\ \hline
50 & 100 & 4.00 & 4.00 & 3.97 & 3.88 & 4.00 & 4.00 & 4.00 & 3.97 & 3.18 & 2.93 & 2.50 & 2.20 \\
100 & 100 & 4.00 & 4.00 & 4.00 & 3.96 & 3.93 & 3.85 & 3.71 & 3.49 & 2.03 & 1.80 & 1.70 & 1.50 \\
200 & 100 & 4.00 & 4.00 & 4.00 & 4.00 & 4.00 & 4.00 & 4.00 & 4.00 & 3.86 & 3.79 & 3.21 & 2.94 \\
300 & 100 & 4.00 & 4.00 & 4.00 & 4.00 & 4.00 & 4.00 & 4.00 & 4.00 & 3.93 & 3.90 & 3.74 & 3.63 \\
500 & 100 & 4.00 & 4.00 & 4.00 & 4.00 & 4.00 & 4.00 & 3.99 & 3.99 & 3.48 & 3.42 & 3.15 & 3.09 \\ \hline
50 & 200 & 4.00 & 4.00 & 4.00 & 4.00 & 4.00 & 4.00 & 3.91 & 3.81 & 3.17 & 3.04 & 1.56 & 1.51 \\
100 & 200 & 4.00 & 4.00 & 4.00 & 4.00 & 4.00 & 4.00 & 4.00 & 4.00 & 3.18 & 2.99 & 2.10 & 1.92 \\
200 & 200 & 4.00 & 4.00 & 4.00 & 4.00 & 4.00 & 4.00 & 4.00 & 4.00 & 3.97 & 3.91 & 3.60 & 3.31 \\
300 & 200 & 4.00 & 4.00 & 4.00 & 4.00 & 4.00 & 4.00 & 4.00 & 4.00 & 3.99 & 3.98 & 3.87 & 3.75 \\
500 & 200 & 4.00 & 4.00 & 4.00 & 4.00 & 4.00 & 4.00 & 4.00 & 4.00 & 4.00 & 4.00 & 3.94 & 3.88 \\ \hline
50 & 300 & 4.00 & 4.00 & 4.00 & 4.00 & 4.00 & 4.00 & 4.00 & 4.00 & 3.32 & 3.24 & 2.08 & 2.01 \\
100 & 300 & 4.00 & 4.00 & 4.00 & 4.00 & 4.00 & 4.00 & 4.00 & 4.00 & 3.76 & 3.70 & 2.75 & 2.61 \\
200 & 300 & 4.00 & 4.00 & 4.00 & 4.00 & 4.00 & 4.00 & 4.00 & 4.00 & 3.92 & 3.89 & 3.60 & 3.49 \\
300 & 300 & 4.00 & 4.00 & 4.00 & 4.00 & 4.00 & 4.00 & 4.00 & 4.00 & 4.00 & 4.00 & 3.97 & 3.93 \\
500 & 300 & 4.00 & 4.00 & 4.00 & 4.00 & 4.00 & 4.00 & 4.00 & 4.00 & 4.00 & 4.00 & 3.98 & 3.96 \\ \hline
50 & 500 & 4.00 & 4.00 & 4.00 & 4.00 & 4.00 & 4.00 & 4.00 & 4.00 & 2.31 & 2.27 & 1.81 & 1.78 \\
100 & 500 & 4.00 & 4.00 & 4.00 & 4.00 & 4.00 & 4.00 & 4.00 & 4.00 & 3.93 & 3.92 & 3.20 & 3.12 \\
200 & 500 & 4.00 & 4.00 & 4.00 & 4.00 & 4.00 & 4.00 & 4.00 & 4.00 & 4.00 & 4.00 & 3.98 & 3.96 \\
300 & 500 & 4.00 & 4.00 & 4.00 & 4.00 & 4.00 & 4.00 & 4.00 & 4.00 & 4.00 & 4.00 & 3.98 & 3.97 \\
500 & 500 & 4.00 & 4.00 & 4.00 & 4.00 & 4.00 & 4.00 & 4.00 & 4.00 & 4.00 & 4.00 & 4.00 & 3.99 \\
\hline\hline
    \end{tabular}}
    \caption{IC with Cross-Sectional Averages -Average of $g$ with $m = 4$ and $K \in \{8,9\}$ with generated factors, $\overline{\mathbf{Z}}_M = \mathbf{F}\*q_M$. DVS criteria from \cite{de2024cross}, MW from \cite{margaritella2023using}, see (2.1) and (2.2) in the main text. For $DVS_1$ and $MW_1$, $p_{N,T} = \frac{N+T}{NT}\ln(\frac{NT}{N+T})$; for  $DVS_2$ and $MW_2$, $p_{N,T} = \frac{N+T}{NT}\ln(C^2_{N,T})$ with $C_{N,T} = \min(\sqrt{N},\sqrt{T})$.  Idiosyncratics in $\*x_{i,t}$, $\*v_{i,t}$, and $\varepsilon_{i,t}$ are weakly correlated time and units units.  }
    \label{tab:my_label}
\end{table}

\clearpage
\subsubsection{Eigenvalue Ratio}
\begin{table}[H]
    \centering
    \resizebox{0.8\columnwidth}{!}{%
    \begin{tabular}{cc| cccc| cccc| cccc}
    \hline\hline
    & & \multicolumn{4}{c}{Stationary $\*F$} & \multicolumn{8}{c}{Non-Stationary $\*F$} \\
    & & \multicolumn{4}{c}{$\tau = 0$} & \multicolumn{4}{c}{$\tau = 0.4$} & \multicolumn{4}{c}{$\tau = 0.9$} \\
    $N$ & $T$ & $ER(\*X)$ & $ER(\widetilde{\*X})$ & $ER(\*Z)$ & $ER(\widetilde{\*Z})$ & $ER(\*X)$ & $ER(\widetilde{\*X})$ & $ER(\*Z)$ & $ER(\widetilde{\*Z})$ & $ER(\*X)$ & $ER(\widetilde{\*X})$ & $ER(\*Z)$ & $ER(\widetilde{\*Z})$  \\\hline
     \multicolumn{14}{c}{Correct Selection Frequency for $g$} \\ \hline
    50 & 50 & 66.80 & 66.80 & 14.20 & 14.20 & 5.60 & 5.60 & 3.10 & 3.10 & 0.00 & 0.00 & 0.00 & 0.00 \\
100 & 50 & 99.80 & 99.80 & 51.10 & 51.10 & 95.80 & 95.80 & 37.80 & 37.80 & 0.80 & 0.80 & 1.20 & 1.20 \\
200 & 50 & 97.40 & 97.40 & 24.20 & 24.20 & 52.10 & 52.10 & 11.00 & 11.00 & 0.20 & 0.20 & 0.10 & 0.10 \\
300 & 50 & 100.00 & 100.00 & 17.40 & 17.40 & 80.90 & 80.90 & 14.90 & 14.90 & 1.70 & 1.70 & 0.90 & 0.90 \\
500 & 50 & 99.20 & 99.20 & 5.90 & 5.90 & 83.10 & 83.10 & 7.80 & 7.80 & 7.90 & 7.90 & 3.30 & 3.30 \\ \hline
50 & 100 & 100.00 & 100.00 & 52.70 & 52.70 & 98.70 & 98.70 & 37.40 & 37.40 & 1.00 & 1.00 & 1.10 & 1.10 \\
100 & 100 & 99.60 & 99.60 & 48.80 & 48.80 & 24.50 & 24.50 & 7.70 & 7.70 & 0.00 & 0.00 & 0.00 & 0.00 \\
200 & 100 & 100.00 & 100.00 & 42.10 & 42.10 & 98.00 & 98.00 & 25.70 & 25.70 & 2.10 & 2.10 & 2.00 & 2.00 \\
300 & 100 & 100.00 & 100.00 & 8.40 & 8.40 & 97.70 & 97.70 & 14.40 & 14.40 & 15.30 & 15.30 & 4.70 & 4.70 \\
500 & 100 & 99.70 & 99.70 & 0.40 & 0.40 & 68.20 & 68.20 & 4.80 & 4.80 & 3.10 & 3.10 & 1.50 & 1.50 \\ \hline
50 & 200 & 100.00 & 100.00 & 32.70 & 32.70 & 93.60 & 93.60 & 27.30 & 27.30 & 0.00 & 0.00 & 0.00 & 0.00 \\
100 & 200 & 100.00 & 100.00 & 21.20 & 21.20 & 94.50 & 94.50 & 25.80 & 25.80 & 0.00 & 0.00 & 0.00 & 0.00 \\
200 & 200 & 100.00 & 100.00 & 38.50 & 38.50 & 98.70 & 98.70 & 25.70 & 25.70 & 5.30 & 5.30 & 2.70 & 2.70 \\
300 & 200 & 100.00 & 100.00 & 1.70 & 1.70 & 93.50 & 93.50 & 7.60 & 7.60 & 7.00 & 7.00 & 2.20 & 2.20 \\
500 & 200 & 100.00 & 100.00 & 0.00 & 0.00 & 80.40 & 80.40 & 2.20 & 2.20 & 7.70 & 7.70 & 2.30 & 2.30 \\ \hline
50 & 300 & 100.00 & 100.00 & 40.70 & 40.70 & 99.60 & 99.60 & 32.30 & 32.30 & 0.00 & 0.00 & 0.90 & 0.90 \\
100 & 300 & 100.00 & 100.00 & 34.20 & 34.20 & 99.70 & 99.70 & 31.00 & 31.00 & 2.40 & 2.40 & 1.40 & 1.40 \\
200 & 300 & 100.00 & 100.00 & 29.70 & 29.70 & 100.00 & 100.00 & 26.40 & 26.40 & 19.40 & 19.40 & 5.40 & 5.40 \\
300 & 300 & 100.00 & 100.00 & 2.70 & 2.70 & 99.80 & 99.80 & 10.70 & 10.70 & 17.80 & 17.80 & 3.80 & 3.80 \\
500 & 300 & 100.00 & 100.00 & 0.00 & 0.00 & 97.50 & 97.50 & 0.30 & 0.30 & 26.90 & 26.90 & 5.90 & 5.90 \\ \hline
50 & 500 & 100.00 & 100.00 & 82.80 & 82.80 & 100.00 & 100.00 & 46.40 & 46.40 & 0.10 & 0.10 & 0.30 & 0.30 \\
100 & 500 & 100.00 & 100.00 & 28.30 & 28.30 & 100.00 & 100.00 & 36.20 & 36.20 & 6.80 & 6.80 & 3.80 & 3.80 \\
200 & 500 & 99.10 & 99.10 & 16.60 & 16.60 & 100.00 & 100.00 & 26.10 & 26.10 & 37.40 & 37.40 & 7.40 & 7.40 \\
300 & 500 & 100.00 & 100.00 & 0.70 & 0.70 & 100.00 & 100.00 & 6.30 & 6.30 & 32.80 & 32.80 & 8.00 & 8.00 \\
500 & 500 & 100.00 & 100.00 & 0.00 & 0.00 & 94.70 & 94.70 & 0.20 & 0.20 & 17.80 & 17.80 & 3.10 & 3.10 \\ \hline \multicolumn{14}{c}{Average of $ g $} \\ \hline
50 & 50 & 3.78 & 3.78 & 1.97 & 1.97 & 3.02 & 3.02 & 2.05 & 2.05 & 2.64 & 2.64 & 1.45 & 1.45 \\
100 & 50 & 4.00 & 4.00 & 2.98 & 2.98 & 3.94 & 3.94 & 2.92 & 2.92 & 1.49 & 1.49 & 1.65 & 1.65 \\
200 & 50 & 3.97 & 3.97 & 3.20 & 3.20 & 3.40 & 3.40 & 2.49 & 2.49 & 1.49 & 1.49 & 1.48 & 1.48 \\
300 & 50 & 4.00 & 4.00 & 4.73 & 4.73 & 3.92 & 3.92 & 4.23 & 4.23 & 1.50 & 1.50 & 1.58 & 1.58 \\
500 & 50 & 4.01 & 4.01 & 4.94 & 4.94 & 4.16 & 4.16 & 4.89 & 4.89 & 1.72 & 1.72 & 1.74 & 1.74 \\ \hline
50 & 100 & 4.00 & 4.00 & 2.69 & 2.69 & 3.99 & 3.99 & 2.71 & 2.71 & 1.98 & 1.98 & 1.66 & 1.66 \\
100 & 100 & 4.00 & 4.00 & 3.08 & 3.08 & 2.30 & 2.30 & 2.16 & 2.16 & 2.05 & 2.05 & 1.38 & 1.38 \\
200 & 100 & 4.00 & 4.00 & 3.06 & 3.06 & 3.98 & 3.98 & 3.22 & 3.22 & 1.50 & 1.50 & 1.54 & 1.54 \\
300 & 100 & 4.00 & 4.00 & 4.89 & 4.89 & 4.02 & 4.02 & 4.73 & 4.73 & 1.90 & 1.90 & 1.81 & 1.81 \\
500 & 100 & 4.00 & 4.00 & 5.00 & 5.00 & 4.21 & 4.21 & 4.91 & 4.91 & 1.59 & 1.59 & 1.54 & 1.54 \\ \hline
50 & 200 & 4.00 & 4.00 & 1.98 & 1.98 & 3.94 & 3.94 & 2.26 & 2.26 & 3.40 & 3.40 & 1.54 & 1.54 \\
100 & 200 & 4.00 & 4.00 & 1.64 & 1.64 & 3.91 & 3.91 & 2.21 & 2.21 & 2.04 & 2.04 & 1.45 & 1.45 \\
200 & 200 & 4.00 & 4.00 & 3.44 & 3.44 & 4.00 & 4.00 & 3.81 & 3.81 & 1.54 & 1.54 & 1.66 & 1.66 \\
300 & 200 & 4.00 & 4.00 & 4.98 & 4.98 & 4.03 & 4.03 & 4.90 & 4.90 & 1.97 & 1.97 & 1.72 & 1.72 \\
500 & 200 & 4.00 & 4.00 & 5.00 & 5.00 & 4.18 & 4.18 & 4.97 & 4.97 & 1.77 & 1.77 & 1.70 & 1.70 \\ \hline
50 & 300 & 4.00 & 4.00 & 2.25 & 2.25 & 3.99 & 3.99 & 2.40 & 2.40 & 2.56 & 2.56 & 1.62 & 1.62 \\
100 & 300 & 4.00 & 4.00 & 2.03 & 2.03 & 4.00 & 4.00 & 2.23 & 2.23 & 1.91 & 1.91 & 1.75 & 1.75 \\
200 & 300 & 4.00 & 4.00 & 3.12 & 3.12 & 4.00 & 4.00 & 3.56 & 3.56 & 2.03 & 2.03 & 1.78 & 1.78 \\
300 & 300 & 4.00 & 4.00 & 4.97 & 4.97 & 4.00 & 4.00 & 4.88 & 4.88 & 2.18 & 2.18 & 1.96 & 1.96 \\
500 & 300 & 4.00 & 4.00 & 5.00 & 5.00 & 4.03 & 4.03 & 5.00 & 5.00 & 3.00 & 3.00 & 2.93 & 2.93 \\ \hline
50 & 500 & 4.00 & 4.00 & 3.48 & 3.48 & 4.00 & 4.00 & 2.90 & 2.90 & 2.90 & 2.90 & 1.60 & 1.60 \\
100 & 500 & 4.00 & 4.00 & 1.85 & 1.85 & 4.00 & 4.00 & 2.12 & 2.12 & 2.00 & 2.00 & 1.81 & 1.81 \\
200 & 500 & 3.99 & 3.99 & 4.25 & 4.25 & 4.00 & 4.00 & 4.36 & 4.36 & 2.59 & 2.59 & 2.03 & 2.03 \\
300 & 500 & 4.00 & 4.00 & 4.99 & 4.99 & 4.00 & 4.00 & 4.94 & 4.94 & 2.68 & 2.68 & 2.40 & 2.40 \\
500 & 500 & 4.00 & 4.00 & 5.00 & 5.00 & 4.05 & 4.05 & 5.00 & 5.00 & 2.42 & 2.42 & 2.25 & 2.25 \\
\hline\hline
    \end{tabular}}
    \caption{IC with Cross-Sectional Averages -Average of $g$ with $m = 4$ and $K \in \{8,9\}$. $ER(\*X)$ selects CA from $\overline{
    \mathbf{X}}$, $ER(\*Z)$ from $(\overline{\*y},\overline{\*X})$. In case of  $\widetilde{\*Z}$ and $\widetilde{\*X}$, the data are scaled by $\widehat{\+\Sigma}_\*Z^{-1/2}$, where $\widehat{\+\Sigma}_\*Z=\frac{1}{NT}\sum_{i=1}^N(\*Z_i-\overline{\*Z})'(\*Z_i-\overline{\*Z})$, see \cite{Juodis2022CCER}. Idiosyncratics in $\*x_{i,t}$, $\*v_{i,t}$, and $\varepsilon_{i,t}$ are uncorrelated over time, but weakly correlated across units, see (3.2) in the main text. }
    \label{tab:my_label}
\end{table}

\clearpage
\subsubsection{Mis-selection with Varying $\tau$}
\begin{figure}[H]
\begin{subfigure}[b]{0.5\textwidth}
       \centering
    \includegraphics[width=0.99\linewidth]{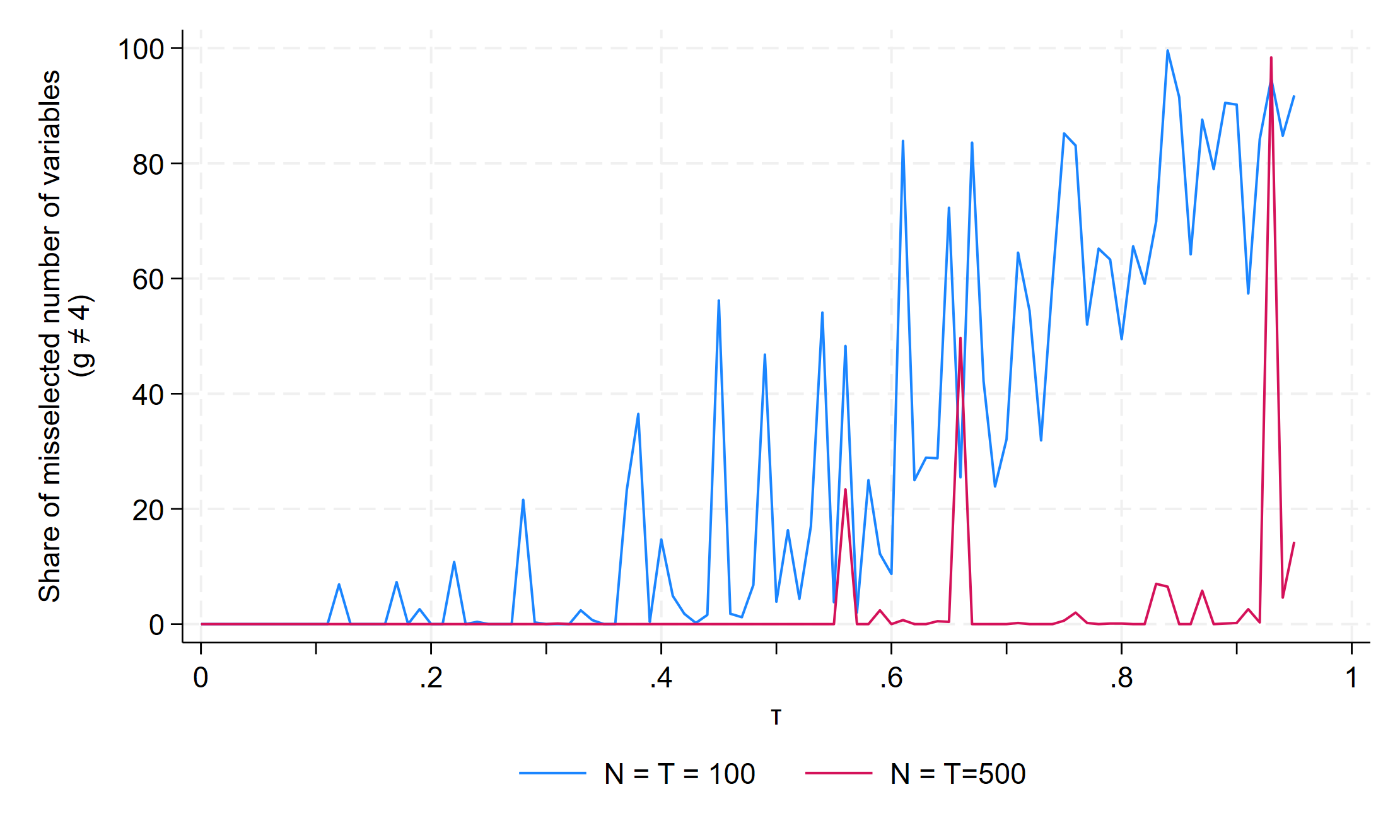}
    \caption{$IC^{DVS}_1$}
\end{subfigure}
\begin{subfigure}[b]{0.5\textwidth}
       \centering
    \includegraphics[width=0.99\linewidth]{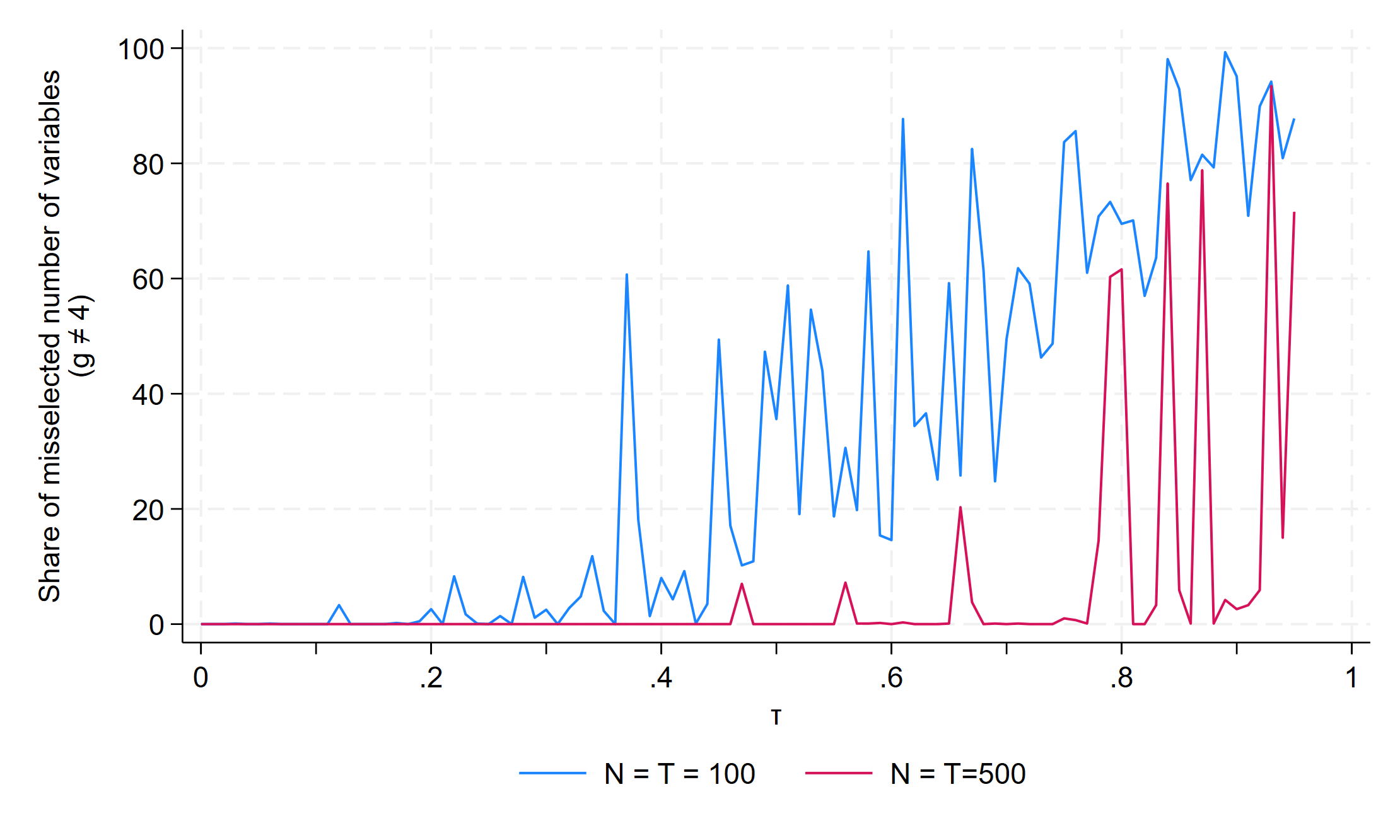}
    \caption{$IC^{MW}_1$}
\end{subfigure}
  \caption{Share of incorrect selected cross-sectional averages with increasing $\tau$. DVS criteria from \cite{de2024cross}, MW from \cite{margaritella2023using}. Idiosyncratics in $\*x_{i,t}$, $\*v_{i,t}$  and $\varepsilon_{i,t}$ uncorrelated over time but weakly correlated across units.}
    \label{fig:Fig1}
\end{figure}

\begin{figure}[H]
\begin{subfigure}[b]{0.5\textwidth}
       \centering
    \includegraphics[width=0.99\linewidth]{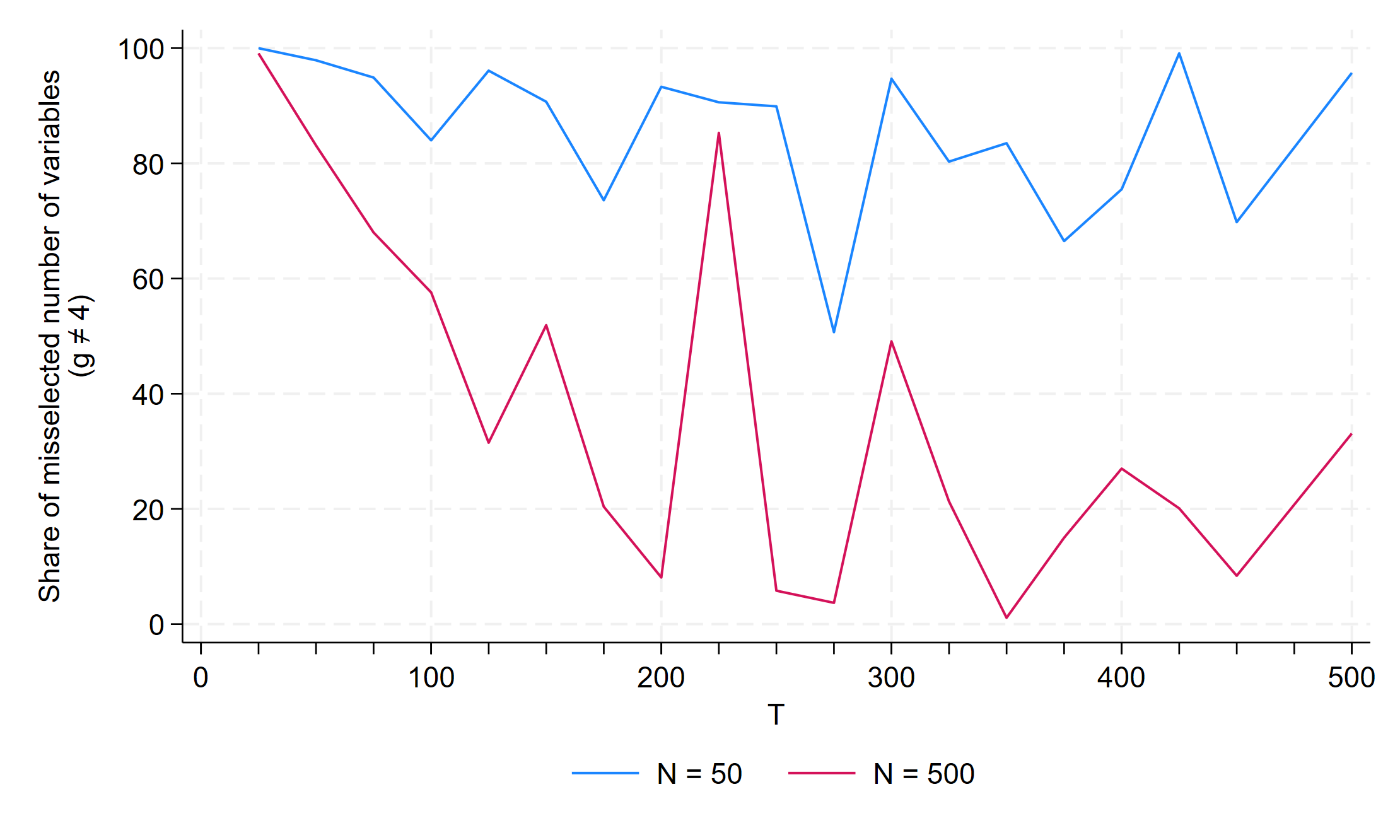}
    \caption{$IC^{DVS}_1$}
\end{subfigure}
\begin{subfigure}[b]{0.5\textwidth}
       \centering
    \includegraphics[width=0.99\linewidth]{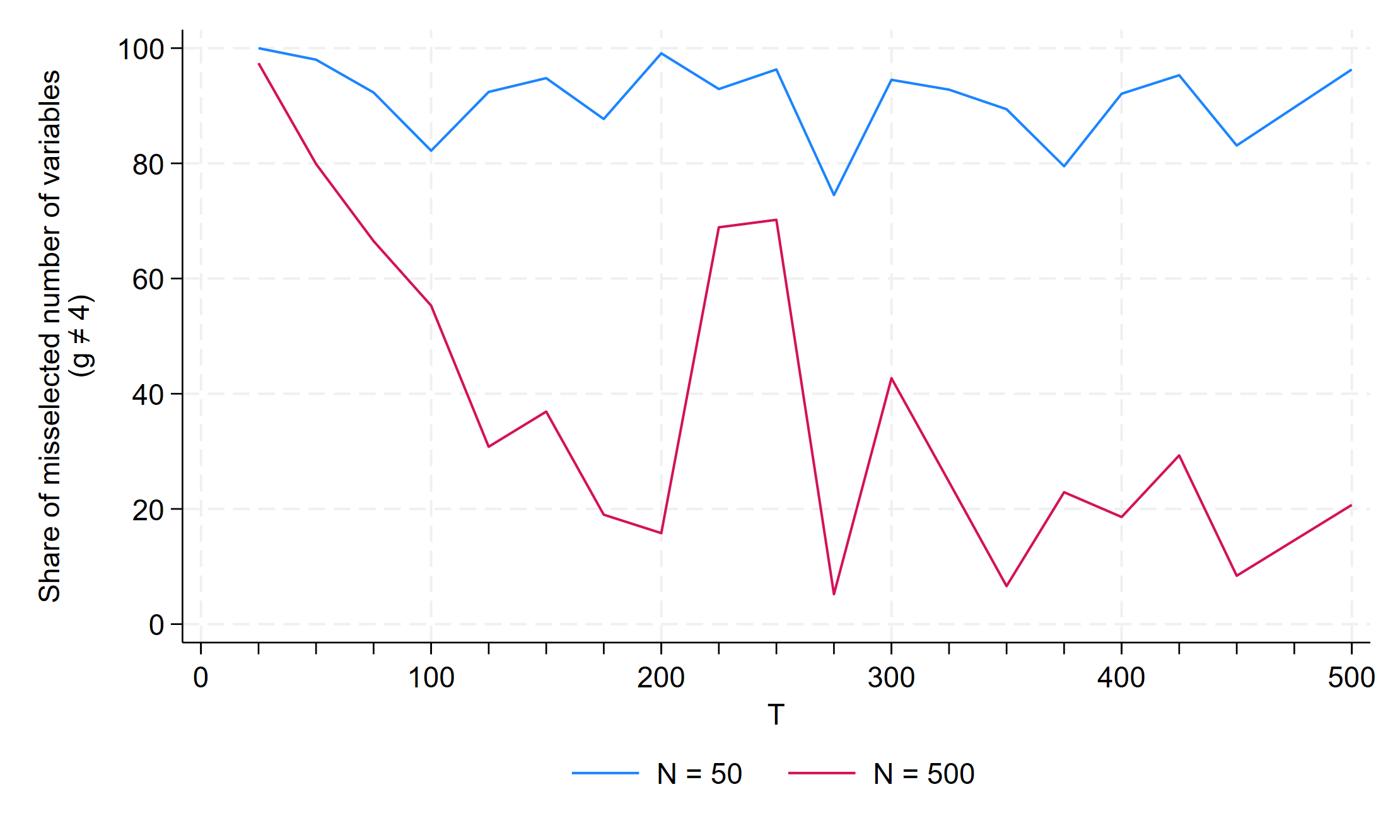}
    \caption{$IC^{MW}_1$}
\end{subfigure}
  \caption{Share of incorrect selected cross-sectional averages with increasing $T$, $\tau = 0.9$. DVS criteria from \cite{de2024cross}, MW from \cite{margaritella2023using}. Idiosyncratics in $\*x_{i,t}$, $\*v_{i,t}$  and $\varepsilon_{i,t}$ uncorrelated over time but weakly correlated across units.}
    \label{fig:Fig2}
\end{figure}

\noindent Figure \ref{fig:Fig1} presents the share of mis-selected number of cross-sectional averages using $IC^{MW}_1$ and $IC^{DVS}_1$ and $N=T=100$ ("small sample") and $N=T=500$ ("large sample"), where we increase $\tau$ in $0.01$ steps from $0$ to $0.95$. As in Table 1 of the manuscript the  idiosyncratics $\*v_{i,t}$ and $\varepsilon_{i,t}$ are uncorrelated over time, but weakly correlated across units. For the small sample, the share of mis-selected number of CAs remains relatively flat and below $20\%$ until $\tau$ reaches levels of around $0.5$, which is the midpoint. Then, it increases to almost $100\%$ with swings showcasing that the criteria become more unstable as $\tau$ increases. As expected, the degree of non-stationarity is much less of a problem for the large sample, and we observe misselection and instability of the criteria only for a high degree of non-stationarity. To shed more light on the importance of the sample size, we present in Figure \ref{fig:Fig2} the share of misselected number of CAs with varying number of time periods from 25 to 500, $\tau=0.9$ and $N=[100,500]$. The share of misselection remains high for $N=100$ for both ICs, but declines for $N=500$. Again, this presents further evidence of the consistency of the ICs, but large samples are required to use them effectively. 

\newpage
\bibliographystyle{apalike}
\bibliography{biblio}
\end{document}